\documentclass[twocolumn,secnumarabic,amssymb,nobibnotes,aps,prd,superscriptaddress,nofootinbib,longbibliography]{revtex4-2}
\pdfoutput=1
\usepackage[utf8]{inputenc}
\usepackage[T1]{fontenc}
\usepackage[british]{babel}
\usepackage[british,cleanlook,printdayon]{isodate}
\usepackage{mathtools}
\usepackage[protrusion,tracking,kerning,spacing,final,babel]{microtype}
\usepackage[svgnames]{xcolor}
\usepackage[final]{hyperref}
\usepackage{lmodern}
\hypersetup{
	colorlinks,
	linkcolor={red!60!black},
	citecolor={green!50!black},
	urlcolor={blue!80!black}
}
\usepackage{cleveref}
\usepackage{aas_macros}
\usepackage{bm}
\usepackage{amsmath}
\usepackage{mathrsfs}
\usepackage[mathscr]{euscript}
\usepackage{array}
\usepackage{ulem}
\usepackage{comment}
\usepackage{booktabs}

\def\0{{(0)}}
\def\sig0{\dot{\sigma}_0}

\def\ph0{\dot{\phi}_0}

\begin{document}
\normalem
\title{%
Can gravitational-wave memory help constrain binary black-hole parameters?\\A LISA case study
}

\author{Silvia Gasparotto}
\email{sgasparotto@ifae.es}
\affiliation{Institut de Fisica d’Altes Energies (IFAE), The Barcelona Institute of Science and Technology, Campus UAB, 08193 Bellaterra (Barcelona), Spain}

\author{Rodrigo Vicente}
\affiliation{Institut de Fisica d’Altes Energies (IFAE), The Barcelona Institute of Science and Technology, Campus UAB, 08193 Bellaterra (Barcelona), Spain}

\author{Diego Blas}
\affiliation{Institut de Fisica d’Altes Energies (IFAE), The Barcelona Institute of Science and Technology, Campus UAB, 08193 Bellaterra (Barcelona), Spain}
\affiliation{Grup de F\'{i}sica Te\`{o}rica, Departament de F\'{i}sica, Universitat Aut\`{o}noma de Barcelona, 08193 Bellaterra, Spain}
    
\author{Alexander C. Jenkins}
\affiliation{Department of Physics and Astronomy, University College London, London WC1E 6BT, United Kingdom}

\author{Enrico Barausse}
\affiliation{SISSA, Via Bonomea 265, 34136 Trieste, Italy and INFN Sezione di Trieste}
\affiliation{IFPU - Institute for Fundamental Physics of the Universe, Via Beirut 2, 34014 Trieste, Italy}

\begin{abstract}
    Besides the transient effect, the passage of a gravitational wave also causes a persistent displacement in the relative position of an interferometer's test masses through the \emph{nonlinear memory effect}.
    This effect is generated by the gravitational backreaction of the waves themselves, and encodes additional information about the source.
    In this work, we explore the implications of using this information for the parameter estimation of massive binary black holes with LISA.
    Based on a Fisher analysis for nonprecessing black hole binaries, our results show that the memory can help to reduce the degeneracy between the luminosity distance and the inclination for binaries observed only for a short time ($\sim$~few hours) before merger. 
    To assess how many such short signals will be detected, we utilized state-of-the-art predictions for the population of massive black hole binaries and models for the gaps expected in the LISA data. 
    We forecast from tens to few hundreds of binaries with observable memory, but only~$\sim \mathcal{O}(0.1)$ events in 4 years for which the memory helps to reduce the degeneracy between distance and inclination.
    Based on this, we conclude that the new information from the nonlinear memory, while promising for testing general relativity in the strong field regime, has probably a limited impact on further constraining the uncertainty on  massive black hole binary parameters with LISA. 
\end{abstract}

\maketitle

\flushbottom

\section{Introduction}\label{sec:intro}

The direct detection of gravitational waves (GWs), predicted by Einstein in 1916~\cite{Einstein1916}, is one of the greatest accomplishments in modern physics, showing (at present) a spectacular agreement with the theory of general relativity (GR)~\cite{LIGOScientific:2019fpa,LIGOScientific:2020tif,LIGOScientific:2021sio}.
By now, almost a hundred GW signals have been observed and interpreted as resulting from the coalescence of compact binaries by LIGO/Virgo~\cite{LIGOScientific:2018mvr,LIGOScientific:2020ibl,LIGOScientific:2021djp}. As the sensitivity of current detectors improves and new detectors become available, it will be possible to estimate the binary parameters more accurately and to find GWs from new types of sources.
This will allow us to not only better test GR in its strong-field regime, but also to probe astrophysics, cosmology and fundamental physics~\cite{Berti:2015itd,Barack:2018yly,manifesto}.
The future space-borne detector Laser Interferometer Space Antenna (LISA)~\cite{Amaro-Seoane2017} will  play a key role in this quest, due to both its expected high signal-to-noise ratio (SNR) measurements and the rich population of sources expected to inhabit its frequency band (from 0.1~mHz to 0.1~Hz)~\cite{eLISA:2013xep,Baibhav:2019rsa,LISA:2022kgy}.

Progress may still be hindered by the fact that some binary parameters --- such as the luminosity distance~$d_\mathrm{L}$ and inclination of the orbital plane with respect to the line of sight~$\iota$ --- may be highly correlated in GW signals, limiting our ability to accurately estimate them~\cite{Cutler:1994ys,Usman:2018imj,Chassande-Mottin:2019nnz}.
This is especially important in the context of \emph{standard sirens}~\cite{Schutz:1986gp,Holz:2005df}, where the precision with which one can estimate the present-day Hubble parameter~$H_0$ depends primarily on how accurately one can measure~$d_\mathrm{L}$~\cite{Nissanke:2009kt,Tamanini:2016zlh,Chassande-Mottin:2019nnz}.
Indeed, this was the main contribution to the large uncertainty on the first estimate of~$H_0$ from~\texttt{GW170817} in Ref.~\cite{LIGOScientific:2017adf}.
The distance-inclination degeneracy can be simply understood by noting that, at leading (Newtonian) order, an inspiralling binary sources the two GW polarizations~$h_+\propto (1+\cos^2 \iota)/d_\mathrm{L}$ and~$h_\times \propto \cos \iota/d_\mathrm{L}$~\cite{poisson_will_2014}; if the detector network is mostly sensitive to one particular combination of~$h_+$ and~$h_\times$, the luminosity distance and inclination are therefore degenerate~\cite{Cutler:1994ys,Chassande-Mottin:2019nnz} (c.f. Appendix~\ref{sec:AppendixFis}).
The degree of this degeneracy depends on the sky location of the binary and the specific detector network~\cite{Chassande-Mottin:2019nnz}, and can be greatly reduced by the observation of the afterglow light curve of an electromagnetic counterpart (which critically depends on~$\iota$)~\cite{Hotokezaka:2018dfi,Chen:2018omi}. 
Interestingly, the degeneracy may also be mitigated by using subleading effects in the waveform.
Examples include the effect of spins misaligned with the orbital angular momentum~\cite{Lang:2006bsg,Apostolatos:1994mx,Lang:2011je} (which lead to the precession of the orbital plane), of higher multipole modes (HMs)~\cite{Arun:2007qv,Porter:2008kn,Trias:2007fp,Klein:2009gza,London:2017bcn,Borhanian:2020vyr,Yang:2022tig,Yang:2022iwn,Yang:2022fgp} (in particular, for unequal component masses or eccentric orbits), or using binary Love relations~\cite{Xie:2022brn} (for neutron star binaries).\footnote{Nevertheless, in the $\sim100$ GW signals observed to date there is only limited evidence for higher multipole content (with no evidence at all beyond~$\ell=3$)~\cite{Hoy:2021dqg} and only one measurement of strong-field precession has been claimed~\cite{Hannam:2021pit} (though some doubt has been cast on this claim due to data-quality issues~\cite{Payne:2022spz,Ng:2022vbz}).}

Another subleading effect in the GW signal with the potential to break the distance-inclination degeneracy is the nonlinear (Christodoulou) GW \emph{memory}~\cite{Christodoulou:1991cr}.
This is a well-grounded prediction of GR which originates from a change in the radiative multipole moments of the gravitational field sourced by the flux of gravitational radiation itself, resulting in a permanent displacement of free-falling test masses upon the passage of GWs~\cite{Payne:1983rrr,Christodoulou:1991cr,Wiseman:1991ss,Blanchet:1992br,Favata:2008yd,Blanchet:2008je,Favata:2010zu}.\footnote{A similar effect sourced by the flux of matter or non-gravitational radiation was actually discovered first and is called the \emph{linear} GW memory~\cite{Zeldovich:1974gvh,Braginsky:1985vlg,Braginsky:1987}. Hereafter, we use ``memory'' to refer exclusively to Christodoulou's GW memory.}
While essentially any source of GWs will generate nonlinear memory,\footnote{See, e.g., Refs.~\cite{Aurrekoetxea:2020tuw,Jenkins:2021kcj}, which studied the nonlinear memory generated by cosmic string loops.} our focus here (and in much of the relevant literature) is on binary black holes (BBHs).
The reason for this is twofold: first, binaries are the one source of \emph{detectable} GWs we definitively know to exist; and second, the amplitude of the memory scales with the total GW energy radiated, which for binaries is $\sim 1\%-10\%$ of the total mass~\cite{Barausse:2012qz}, favoring BBHs over lighter binaries containing neutron stars or white dwarfs.

The gravitational wave memory modifies the BBH waveform by introducing a slowly-growing offset of the oscillations that builds over the whole coalescence and whose time evolution follows that of the instantaneous orbital frequency. This shift rises over the radiation-reaction timescale during the inspiral~\cite{Favata:2008yd,Blanchet:2008je,Favata:2011qi} and accumulates rapidly during the merger before saturating to its final value during the ringdown~\cite{Favata:2009ii}.
Although the memory arises from a 2.5~PN nonlinear interaction in a post-Newtonian (PN) expansion of Einstein equations, because it accumulates over the whole coalescence, it affects the gravitational waveform at leading (Newtonian) order~\cite{Favata:2009ii}, increasing the prospects of observing it in the near future.

Several searches for memory from BBHs have been performed using LIGO/Virgo data, returning only null results thus far~\cite{Hubner:2019sly,Ebersold:2020zah,Zhao:2021hmx,Hubner:2021amk}.
This is in agreement with forecasts for LIGO/Virgo, which show that the detection of memory from a single event would require a much more massive and nearby binary than any yet observed, and that to find collective evidence of memory in the total population of observed binaries one would need $\sim 5$ yr of collected data~\cite{Lasky:2016knh,Yang:2018ceq,Boersma:2020gxx} (or $\sim2.5$ yr for the LIGO/Virgo/KAGRA network, taking into account the expected improvement of the network sensitivities~\cite{Grant:2022bla}).
The difficulty in detecting the memory with current ground-based interferometers resides mostly in the fact that, besides being responsible for only a small amplitude offset, its power is larger at lower frequencies where the detectors sensitivity is limited by several sources of noise~\cite{Buikema:2020}.
However, the prospects for memory detection in single events are considerably better for third-generation ground-based detectors (e.g. Einstein Telescope~\cite{Punturo:2010zz} and Cosmic Explorer~\cite{Reitze:2019iox}) and for the future space-based detectors LISA and TianQin, due to their better sensitivity and low frequency coverage~\cite{Favata:2009ii,Johnson:2018xly,Islo:2019qht,Islam:2021old,Sun:2022pvh,Grant:2022bla}.\footnote{Memory from the merger of supermassive BHs is also a target of Pulsar Timing Arrays (PTAs)~\cite{Seto:2009nv,Cordes:2012zz}, but searches in PTA data thus far have returned only null results~\cite{Wang:2014zls,NANOGrav:2019vto}.}

In this work we investigate the impact of the nonlinear memory on parameter estimation via a Fisher matrix analysis. 
In particular, we focus on how the memory signal breaks the distance-inclination degeneracy, which is crucially important if these binaries are to be used as standard sirens.
We found that the information of the memory can indeed reduce the uncertainty on the luminosity distance by reducing its correlation with the inclination angle, whereas it has almost no impact on the uncertainty of the other parameters. 
For LISA sources its greatest effect involves cases where (\textit{i}) the constituent BHs are light enough [$M_\mathrm{BH}\lesssim10^5\,M_\odot/(1+z)$, with~$z$ the source's cosmological redshift] that the merger takes place near the upper edge of the LISA band, and (\textit{ii}) the information from the primary waveform is limited to a few cycles before the merger. 

The presence of gaps in the data stream and confusion noise from other sources will reduce the effective duration of usable LISA data, and thus the observed number of cycles for BBHs~\cite{Seoane:2021kkk}, making the memory potentially useful for the distance estimation of some BBH events. Therefore, using the state-of-art astrophysical BBH population models described in Refs.~\cite{Barausse:2020gbp, Barausse:2020mdt} (and based upon previous work presented in~\cite{EB12,Sesana:2014bea,Antonini:2015sza}), we assess quantitatively the impact of the memory in the distance estimation of LISA sources, taking into account the presence of gaps in the data stream. Considering the particular gap model used in Ref.~\cite{Dey:2021dem} we did \emph{not} find any significant enhancement in the distance estimation by the inclusion of memory on the BBH waveforms. We do however find a greater number of events with detectable memory at LISA as compared to previous forecasts~\cite{Sun:2022pvh,Islo:2019qht}, especially in our models with heavy BH seeds.

This paper is organized as follows.
In Sec.~\ref{sec:GW_memory} we review the computation of the memory and describe the phenomenology of the signal.
In Sec.~\ref{sec:Parestimation} we describe our Fisher forecasting analysis.
In Sec.~\ref{sec:dist-inc} we present our results for the distance-inclination inference of individual BBHs, and discuss the impact of the binary parameters and signal duration.
In Sec.~\ref{sec:pop} we perform population-level forecasts for LISA using synthetic BBH catalogues, and assess the impact of including the memory on the luminosity distance estimation in the presence of gaps in the data stream.
We conclude in Sec.~\ref{sec:concl}.
Some technical material is discussed in the appendices.
We use geometric units throughout ($c=G=1$).

\section{GW memory waveform} \label{sec:GW_memory}

\subsection{Computation scheme: Thorne's formula} \label{sec:comput}

The most direct way to compute the memory contribution to waveforms would be to extract it directly from numerical relativity simulations.
However, most simulations to date have struggled to accurately capture this information for a number of reasons (see e.g.~\cite{Favata:2008yd}).
Some exceptions are Ref.~\cite{Pollney:2010hs}, where the dominant memory mode~$(\ell,m)=(2,0)$ was first resolved, and the recent work of Ref.~\cite{Mitman:2020pbt}, which used a Cauchy-characteristic extraction (CCE) technique to extract the waveform.
Alternatively, the Bondi, van der Burg, Metzner, and Sachs (BMS) \emph{balance laws}~\cite{Ashtekar:2019viz,Compere:2019gft} have recently been used to add the memory to waveforms~\cite{Khera:2020mcz,Mitman:2020bjf} (see also~\cite{Zhao:2021hmx,Zhao:2021hmx,Sun:2022pvh}). 

Instead, in this work we use a perturbative approach to evaluate the memory~\cite{Favata:2010zu,Favata:2008yd}, which we now briefly review.
A GW strain~$h_0$ (which we call the ``primary'' signal) sources an additional memory strain~$\delta h$, which can be expressed in the transverse-traceless (TT) gauge using Thorne's formula~\cite{Thorne:1992sdb}, 
\begin{equation}
\label{eq:GWmem1}
    \delta h_{ij}^{\mathrm{TT}}(u)= \frac{4}{R}\int_{-\infty}^{u} \mathrm{d}u'\int_R \mathrm{d}\Omega\, \frac{\mathrm{d}^2E_\mathrm{GW}}{\mathrm{d}u'\mathrm{d}\Omega}\left[\frac{n_i n_j}{1-n_k N^k}\right]^{\mathrm{TT}},
\end{equation}
where the angular integral is over the solid angle~$\mathrm{d}\Omega$ of a (large) sphere of radius~$R$ surrounding the source, and~$n^i$ and~$N^i$ are the unit radial vectors pointing, respectively, to~$\mathrm{d}\Omega$ and the detector.
The~TT superscript represents a TT projection with respect to the direction of the detector.
The time integral is over the entire history of the source up to retarded time $u$, which shows that the memory is a hereditary effect.
The GW energy flux carried by the primary GW is~\cite{Favata:2008yd}
\begin{equation}
    \frac{\mathrm{d}^2E_\mathrm{GW}}{\mathrm{d}t\mathrm{d}\Omega}=\frac{R^2}{16 \pi }(\dot{h}_{0,+}^2+\dot{h}_{0,\times}^2),
\end{equation}
where~$\dot{h} \equiv \mathrm{d} h/\mathrm{d}t$ and~$h_{+,\times}\equiv h_{ij}^\mathrm{TT} e^{ij}_{+,\times}$.
We use the same choice of TT-polarization tensors~$e^{ij}_{+,\times}$ as Ref.~\cite{Kidder:2007rt}.
From the spin-weighted spherical harmonic decomposition
\begin{equation}
\label{eq:mode_decomp}
    h_+-ih_\times\equiv \sum_{\ell\geq 2}\,\sum_{|m|\leq \ell} h^{\ell m}(u,r)\, {}_{-2}Y_{\ell m}(\iota,\phi),
\end{equation}
it is possible to show that the sourced memory can be expressed as~\cite{Ebersold:2020zah}
\begin{align}
\label{eq:memorymodes}
    &\delta h^{\ell m}(u)=-R \sum_{\ell',\ell''\geq 2}\, \sum_{m',m''}\sqrt{\frac{(\ell-2)!}{(\ell+2)!}}  \nonumber \\
    &\times \int \mathrm{d}\Omega\, Y_{\ell m}^*\, {}_{-2}Y^*_{\ell'm'} \, {}_{-2}Y_{\ell''m''}\int_{-\infty}^u \mathrm{d}u'\dot{h}_0^{*\ell' m'} \dot{h}_0^{\ell'' m''},
\end{align}
which allows us to straightforwardly compute the  memory modes~$\delta h^{\ell m}$ from the primary waveform modes~$h_0^{\ell m}$.
We can then reconstruct~$\delta h_+$ and~$\delta h_\times$ from Eq.~\eqref{eq:mode_decomp} to give the total strain~$h\approx h_0+\delta h$.
In principle, this process should be iterated to give higher-order contributions (the ``memory of the memory''~\cite{Lasky:2016knh}).
In practice, these extra terms are subleading, and it is sufficient for our purposes to consider just the leading-order memory effect, $\delta h$.

To generate our primary waveforms, we use the surrogate \texttt{NRHybSur3dq8} model~\cite{Varma:2018mmi}, which has only aligned-spin waveforms and includes all higher spherical harmonic modes up to~$(\ell,|m|)=(4,4)$ and is considerably more accurate than other often-used phenomenological models in modeling the merger stage of BBH coalescence~\cite{Boersma:2020gxx}.
We use the publicly available \texttt{GWmemory} package~\cite{Talbot:2018sgr} to implement the calculation scheme described above for the corresponding memory signal.

Equation~\eqref{eq:memorymodes} is valid on a background Minkowski spacetime. It can be extended to a spatially flat Friedmann-Lemaître-Robertson-Walker (FLRW) spacetime using the fact that, for sources at the same luminosity distance~$d_\mathrm{L}$, the memory amplitude in FLRW is enhanced over the Minkowski case by the redshift factor~$(1+z)$~\cite{Tolish:2016ggo,Bieri:2017vni,Jokela:2023suz}. Additionally, we shall use the time at the detector~$t\equiv t_\mathrm{peak}- (1+z)(u_\mathrm{peak}-u)$, where~$t_\mathrm{peak}$ is the instant when the primary strain reaches its peak amplitude. Summarizing, in this work we use
\begin{equation}
\delta h^{\ell m}_\mathrm{FLRW}(t)=(1+z) \delta h^{\ell m}_\mathrm{Mink}\big(u(t)\big)_{R\to d_\mathrm{L}}.
\end{equation}
This can be shown to be equivalent to using redshifted component masses~$M_{i,z}\equiv(1+z) M_i$, with~$i\in\{1,2\}$, and luminosity distance~$d_\mathrm{L}$ to generate the primary signal (e.g., with \texttt{NRHybSur3dq8}), plugging this primary directly into Eq.~\eqref{eq:memorymodes}, and identifying~$(R,u_\mathrm{peak}-u)\to (d_\mathrm{L},t_\mathrm{peak}-t)$. We omit the subscript ``FLRW'' throughout, but a spatially flat FLRW is implicitly assumed in all our expressions.

\subsection{Phenomenology of the signal} \label{sec:phenom}

\begin{figure*}[t!]
    \centering
    \includegraphics[width=0.75\textwidth]{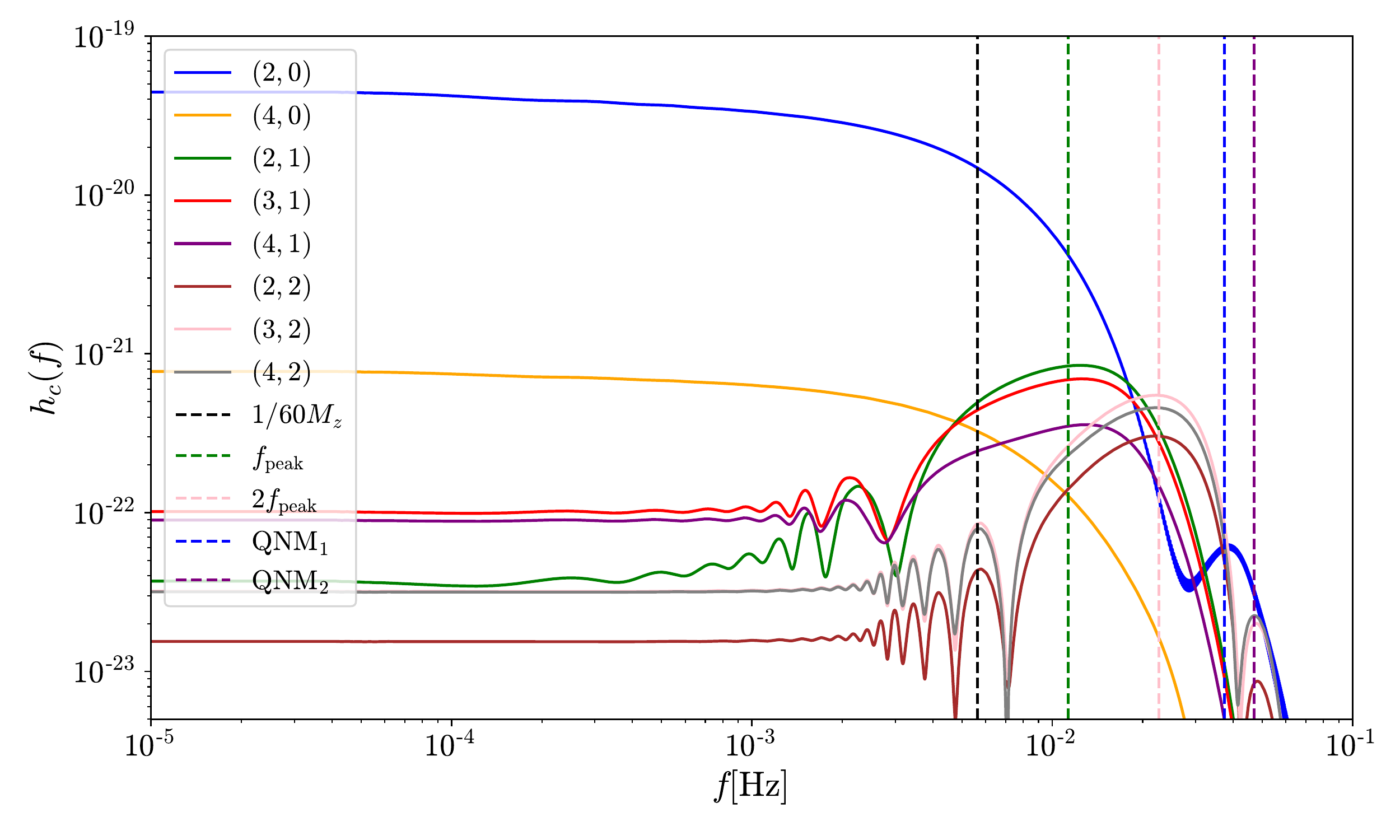}
    \caption{Memory characteristic strain~$h_{\mathrm{c}}^{\ell m}(f)\equiv2f|\widetilde{\delta h}{}^{\ell m}|$ of the most important modes computed from Eq.~\eqref{eq:memorymodes}. We consider the non-spinning ``heavy'' BBH studied in Figs.~\ref{fig:strains} and~\ref{fig:Ellipses}, with total mass~$M=2\times10^5\,M_\odot$, mass ratio~$q=1.2$ and redshift~$z=2$.
    The non-oscillatory~$(\ell,m)=(2,0)$ mode dominates at low frequencies, but is suppressed at~$f\gtrsim 1/60M_z$, where oscillatory~$m\neq0$ modes start becoming important (in particular, at their maxima~$f\sim m f_\mathrm{peak}$).
    We can also see the presence of the ringdown in the memory spectrum in the form of high-frequency peaks [$\mathrm{QNM}_1$ is at~$f=(\tau_{221}^{-1}+\tau_{222}^{-1})/2\pi M_z$, and~$\mathrm{QNM}_2$ at~$f=\tau_{222}^{-1}/\pi M_z$]. (QNMs were computed using the Python package~\texttt{qnm}~\cite{Stein:2019mop}, and the final mass and spin of the remnant BH via~\texttt{surfinBH}~\cite{vijay}, whose fitting procedure is described in~\cite{Varma:2018aht}.)}
    \label{fig:FTmemHM}
\end{figure*}

Equation~\eqref{eq:memorymodes} shows how the memory modes are sourced by pairs of primary modes. For a BBH coalescence occurring in the $x$-$y$ plane, the primary modes have the form~$h_0^{\ell m}\propto e^{-i m \varphi(t)}$ with~$\varphi(t)$ the orbital phase. So, from the time integral in Eq.~\eqref{eq:memorymodes} it is clear that the leading contribution to the memory modes at low frequency (i.e.,~$u \to +\infty$) comes from the nonoscillatory (DC) terms with~$m'-m''=0$ which accumulate in time~\cite{Favata:2008yd,Blanchet:2008je}; these source~$m=0$ memory modes (from the angular integral in Eq.~\eqref{eq:memorymodes}). Note, however, that oscillatory $m\neq 0$ memory modes do become dominant at high frequencies (cf.  Fig.~\ref{fig:FTmemHM}).

The memory sourced in the quasi-circular inspiral of non-spinning BBHs is known analytically up to 3~PN~\cite{Favata:2008yd,Blanchet:2008je}.
Due to the accumulation over the inspiral, the non-oscillatory contribution to the memory enters at Newtonian (0~PN) order in the waveform,
\begin{align}
\label{eq:OPNres}
    \delta h^{(0\mathrm{PN})}_+&=\frac{\eta M_z}{48\, d_\mathrm{L}}[M \omega(t)]^{\frac{2}{3}} \sin^2\iota\,(17+\cos^2\iota),
\end{align}
and $\delta h^{(0\mathrm{PN})}_\times=0$, with the conventional choice of polarization triad~\cite{Favata:2008yd}. The orbital frequency is~$\omega(t)\equiv \dot{\varphi}(t)$ and the symmetric mass ratio~$\eta \equiv M_1 M_2/M^2$, where~$M\equiv M_1+ M_2$ is the total mass and~$M_z\equiv(1+z)M$ is the total redshifted mass.
The memory has the same scaling as the Newtonian primary waveform (cf.  Eq.~\eqref{eq:0PNosc_+}), but rather than the main time dependence coming from the oscillatory term, its time evolution is captured by the instantaneous orbital frequency $\omega(t)$. This explains the typical step shape of the memory, which has a steep increase in the merger-plunge phase and a saturation during the ringdown. Moreover, the memory is characterized by a \emph{different} dependence on the inclination angle~$\iota$ and an overall amplitude~$\sim 20$ times weaker than the primary.
In particular, the two signals have opposite monotonic dependence on the inclination angle, and while the primary signal is maximized for face-on binaries ($\iota=0$), the memory is instead maximized for edge-on binaries ($\iota=\pi/2$).
This behavior is maintained when using primary waveforms generated by \texttt{NRHybSur3dq8}. The different dependence on~$\iota$ is what makes the memory helpful in reducing the ($\iota,d_\mathrm{L}$) correlation (c.f. Fig.~\ref{fig:FM}).


Figure~\ref{fig:FTmemHM} shows the spectral shape of memory mode characteristic strains~$h^{\ell m}_\mathrm{c}(f)\equiv 2f|\widetilde{\delta h}{}^{\ell m}(f)|$, with~$\widetilde{\delta h}(f)\equiv \int \mathrm{d}t\, e^{-i 2\pi f t} \delta h(t)$ the Fourier transform (FT) of~$\delta h$. All modes exhibit a plateau at low frequencies, but the spectral content of the non-oscillatory ($m=0$) modes is clearly distinct from the oscillatory ($m \neq 0$) modes. The plateau of the~$m=0$ modes is easily understood from the approximation~$\delta h^{\ell 0}\approx \Delta h^{\ell0}\mathcal{H}(t-t_\mathrm{peak})$, with~$\mathcal{H}$ the Heaviside step function and where~$\Delta h^{\ell0}$ scales with the fraction of radiated~$E_\mathrm{GW}$ that sources~$\delta h^{\ell 0}$; this results in a constant~$h_\mathrm{c}\approx \Delta h^{\ell0}/\pi$~\cite{Favata:2011qi}. 

Taking into account that the memory growth is not instantaneous, but happens in~$\tau\sim 60 M_z$ (the timescale over which most of~$E_\mathrm{GW}$ is radiated~\cite{Buonanno:2006ui}), one can understand the suppression at~$f\gtrsim 1/60 M_z$ in Fig.~\ref{fig:FTmemHM}. For the~$m\neq 0$ modes the low-frequency plateau has a similar origin, but its value is always subdominant because, due to the oscillations in the integral in Eq.~\eqref{eq:memorymodes}, the memory does not accumulate, averaging out to a net small value that depends strongly on the value of the orbital phase at which the BHs merge; this is also expected from PN results~\cite{Favata:2008yd,Blanchet:2008je}. We note that for~$m\neq0$ the maximum of the strain scales as~$m f_\mathrm{peak}$ where the peak frequency~$f_\mathrm{peak}\sim 0.1/\pi M_z$ roughly corresponds to the moment in which most of the energy is radiated~\cite{Buonanno:2006ui}. The oscillations at the left of these maxima are numerically stable (in particular, they are not artifacts of our FT) and come from interference between different $\ell$-modes in Eq.~\eqref{eq:memorymodes}. On the other hand, the high-frequency peaks seen in all modes are associated with the ringdown stage and are located at $f \approx (\tau_{\ell'm'n'}^{-1}+\tau_{\ell''m''n''}^{-1})/2\pi M_z$, where the complex quasi-normal modes are~$\sigma_{\ell m n}\equiv (\omega_{\ell m n}+ i \tau_{\ell m n}^{-1})/M_z$~\cite{Berti:2005ys}. 
It is interesting to note that these peaks occur at a frequency set by the QNM \emph{decay rate} $\tau_{\ell mn}^{-1}$ (i.e. imaginary frequency), not the oscillatory part $\omega_{\ell mn}$ (i.e. real frequency).
This can be confirmed analytically using Favata's minimal waveform model (MWM; Eq.~(14) of~\cite{Favata:2009ii}).

\begin{figure}[t!]
    \centering
    \includegraphics[width=0.49\textwidth]{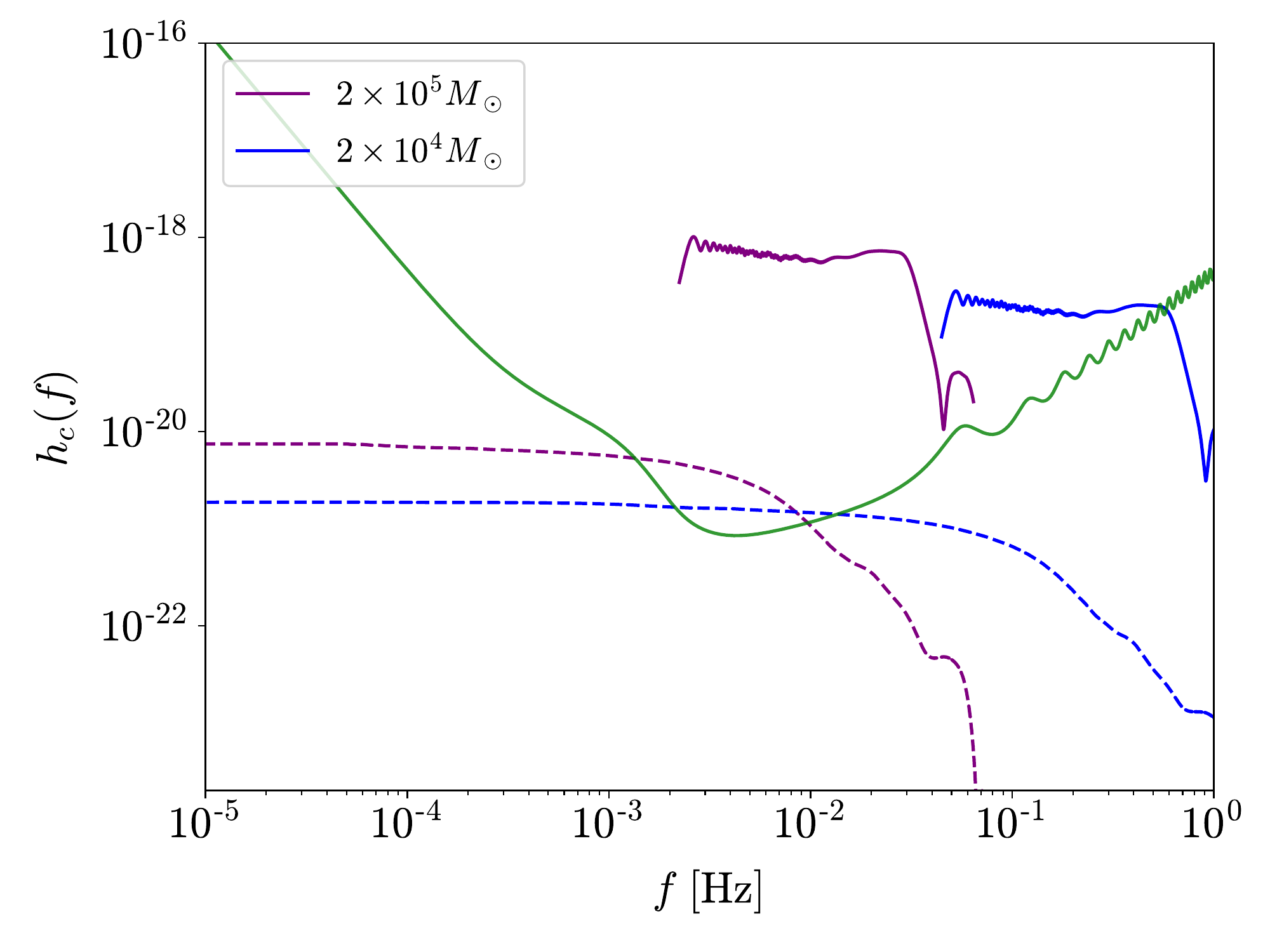}
    \caption{Characteristic strain~$h_\mathrm{c}(f,\iota,\varphi_c)\equiv 2 f |\widetilde{h}_+-i \widetilde{h}_\times|$ of the primary (solid curve) and memory (dashed curve) computed from Eqs.~\eqref{eq:mode_decomp} and~\eqref{eq:memorymodes}, seen from a fixed direction~$\iota=40\, \mathrm{deg}$ and~$\varphi_c=0$. We consider two fiducial non-spinning BBHs, which will also be used in the following sections: a ``light'' binary (in violet) with total mass~$M=2\times10^4M_\odot$ at redshift~$z=0.5$, and an ``heavy'' binary (in blue) with~$M=2\times 10^5M_\odot$ at~$z=2$. Both BBH have mass ratio~$q=1.2$. We consider the last~$25$ cycles before merger for both BBHs (which corresponds to~$\sim6$ minutes for the ``light'' source, and~$\sim2$ hours for the ``heavy'').}
    \label{fig:strains}
\end{figure}

Figure~\ref{fig:strains} shows the characteristic strain~$h_\mathrm{c}(f,\iota,\varphi_c)\equiv 2 f |\widetilde{h}_+-i \widetilde{h}_\times|$ of the primary and memory -- containing all modes up to~$(\ell,|m|)=(4,4)$ -- for two binaries with different total masses and redshifts, seen from a fixed direction~$\iota=40\, \mathrm{deg}$ and~$\varphi_c=0$. At this particular direction, the primary characteristic strain is~$\mathcal{O}(10^2)$ greater than the memory. Thus, the memory adds information to parameter estimation only if the number of cycles that can be observed during inspiral is limited (c.f. Fig.~\ref{fig:Mass&Ndep}); this may happen due to gaps in the data stream and/or confusion noise from other sources. Indeed, whereas truncating the primary waveform at some minimum~$f_\mathrm{in}$ (related to the time/cycles prior to merger) significantly reduces the SNR of the primary, the SNR of the memory is almost unchanged (as also noted in Refs.~\cite{Sun:2022pvh,Grant:2022bla}). As long as the memory is observed in LISA for a period of at least~$10^3\, \mathrm{s}\approx 15 \, \mathrm{min}$ after merger, its SNR is approximately independent of the observation time (c.f. Fig.~\ref{fig:SNRratiovsInc}).

In this work we focus on quasi-circular and non-spinning BBHs. For the dependence of the memory on the spins, mass ratio and eccentricity, we refer the reader to Refs.~\cite{Reisswig:2009vc,Pollney:2010hs,Talbot:2018sgr,Zhao:2021hmx,2021PhRvD.103d3005L,Islam:2021old}.

\section{Fisher Matrix and Parameter Estimation}\label{sec:Parestimation}
 
We use a Fisher matrix analysis (commonly used in GW astronomy~\cite{Finn:1992wt,Cutler:1994ys,Poisson:1995ef,Cutler:1997ta,Vallisneri:2007ev, Berti:2004bd}) to quantify the impact of GW memory on the parameter estimation of the BBH source. We only consider events with merger occurring during the mission lifetime of LISA, since the memory is mainly generated during this phase of the binary evolution. The total signal~$h=h_0+\delta h$ is generated in time-domain and is given by the sum of the (primary) surrogate \texttt{NRHybSur3dq8} waveform~${h}_0$, and the memory~$\delta{h}$ computed through the \texttt{GWMemory} package~\cite{Talbot:2018sgr} from the primary waveform. In this work we focus on non-spinning binaries, so the total number of parameters is reduced compared to the general spinning case. The intrinsic parameters, defined as those that do not depend on the relative position and orientation of the orbital plane and the detector, are simply the redshifted mass~$M_z$ and the mass ratio~$q\equiv M_1/M_2$ (in our convention~$M_1>M_2$). On the other hand, the extrinsic parameters are the luminosity distance~$d_\mathrm{L}$, the inclination angle~$\iota$, and the coalescence phase~$\varphi_\mathrm{c}$. Thus, the input parameters for our waveforms are~$\Theta=\{\ln{M_z},q,\ln{d_\mathrm{L}},\iota,\varphi_\mathrm{c}\}$. 
We only investigate the dependence on these parameters, meaning that we ignore the particular sky position of the source and the LISA response function. Accounting for them should not affect our results, since the LISA orbital motion is a year time scale, whereas the longest inspirals where we see the effect of the memory are of the order of hours.

In the strong-signal limit the probability distribution of the parameters is a multivariate Gaussian distribution centred on the true values~$\Theta=\Bar{\Theta}$, with the covariance matrix~$\Sigma_{ij}$ described by the inverse of the Fisher information matrix $\Gamma_{ij}$, up to corrections of order of the inverse signal-to-noise ratio,
\begin{align}
    \Sigma_{ij}&=(\Gamma^{-1})_{ij}[1+ \mathcal{O}(\mathrm{SNR}^{-1})].\label{eq:Covmatrix}
\end{align}
The SNR~$\rho$ and the Fisher matrix are computed as
\begin{align}
    \rho^2&=(\tilde{h}|\tilde{h}),\qquad
    \Gamma_{ij}\equiv\Big(\frac{\partial \tilde{h}}{\partial \Theta_i}\Big|\frac{\partial  \tilde{h}}{\partial \Theta_j}\Big)\Big |_{\Theta=\Bar{\Theta} },\label{eq:FisherFormula}
\end{align}
where we have used the standard inner product
\begin{equation}
    (\tilde{a}|\tilde{b})\equiv4\mathrm{Re}\int_{f_\mathrm{min}}^{f_\mathrm{max}}\mathrm{d}f\frac{\tilde{a}^*(f)\tilde{b}(f)}{S_n(f)}.
\end{equation}
The $S_n(f)$ is the sky-position- and polarization-averaged, but not inclination-averaged, LISA power spectral density, as found in~\cite{Robson:2018ifk}, with the confusion noise corresponding to $T_{\text{obs}}=4$ years.
Computing the Fisher matrix thus allows us to find the variance of each parameter and correlation of each pair of parameters due to measurement errors,
\begin{equation}
    \sigma^2_i=(\Gamma^{-1})_{ii}, \qquad c_{ij}=\frac{(\Gamma^{-1})_{ij}}{\sigma_i\sigma_j},
\end{equation}
(with no summation implied over repeated indices here).
\begin{figure*}[t!]
    \centering
    \begin{minipage}[b]{0.45\linewidth}
    \includegraphics[width=6.5cm]{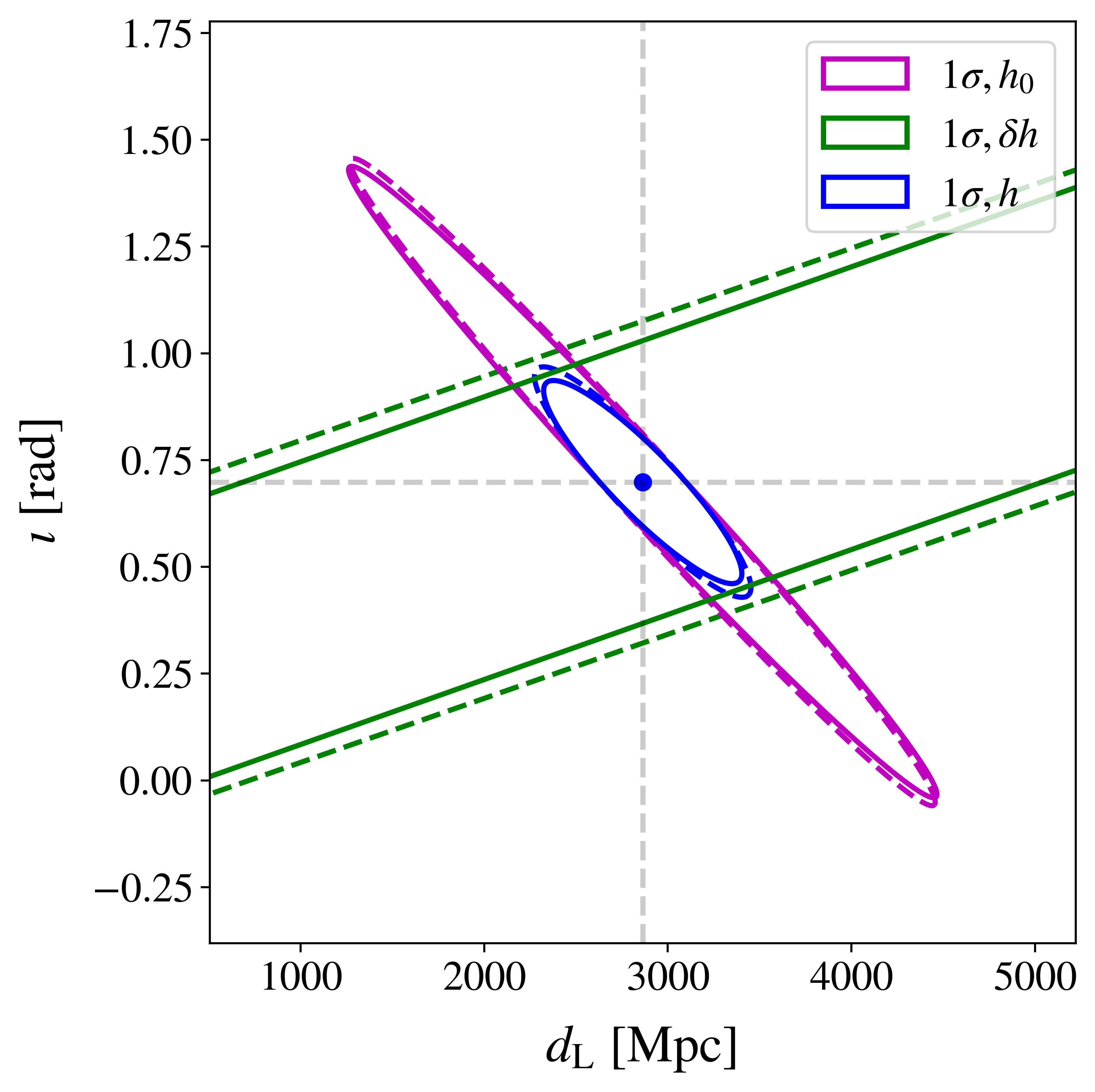}
    \end{minipage}
    \quad
    \begin{minipage}[b]{0.45\linewidth}
    \includegraphics[width=6.5cm]{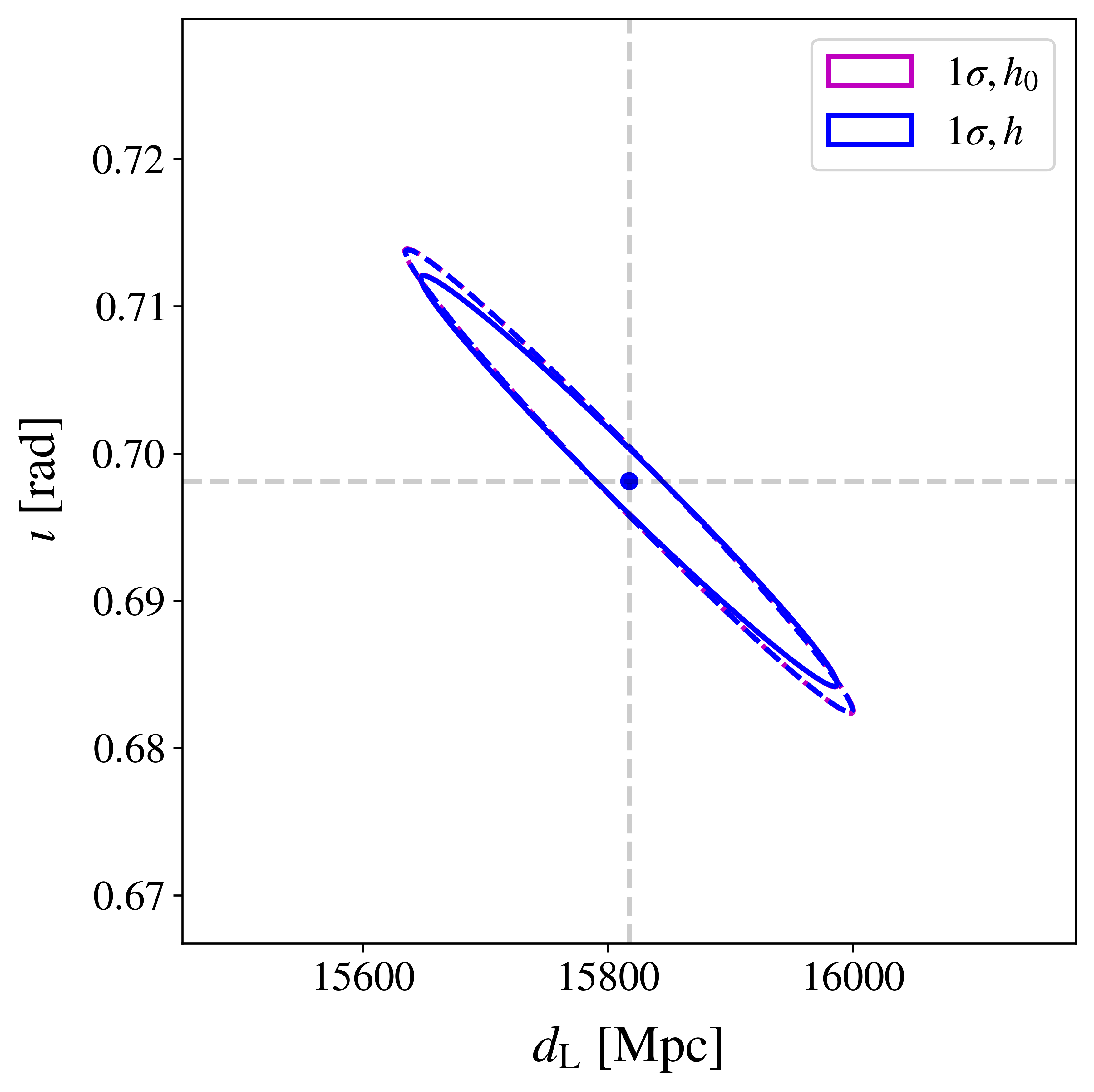}
    \end{minipage}
    \caption{The~$1\sigma$ confidence ellipses computed for the primary waveform without memory (in magenta), with memory (in blue), and with only the memory (in green).
    For each of those we compare the effect of including higher modes (solid lines) and excluding them (dashed lines).
    The left panel shows the result for the ``light'' binary, whereas the right panel shows it for the ``heavy'' binary (c.f. Fig.~\ref{fig:strains}); in both cases we consider the last~$25$ cycles before merger and a line of sight~$\iota=40\, \mathrm{deg}$. For the ``heavy'' binary, the memory has negligible impact, since we cannot distinguish the purple and the blue lines (the green contours of the memory fall beyond the panel).}
    \label{fig:Ellipses}
\end{figure*}

The total frequency-domain signal $\tilde{h}=\tilde{h}_{0}+\delta \tilde{h}$ is given by the sum of the FT of the primary and the memory signals, which we compute numerically via the fast Fourier transform (FFT) implemented in \texttt{NumPy}~\cite{Harris:2020xlr}.
In Appendix~\ref{sec:window} we explain in detail our choices for manipulating the signal, such as the padding and the window function applied, while in Appendix~\ref{NumFIM} we elaborate on the numerical computation of the Fisher matrix.

\begin{figure}
    \centering
    \includegraphics[width=0.9\linewidth]{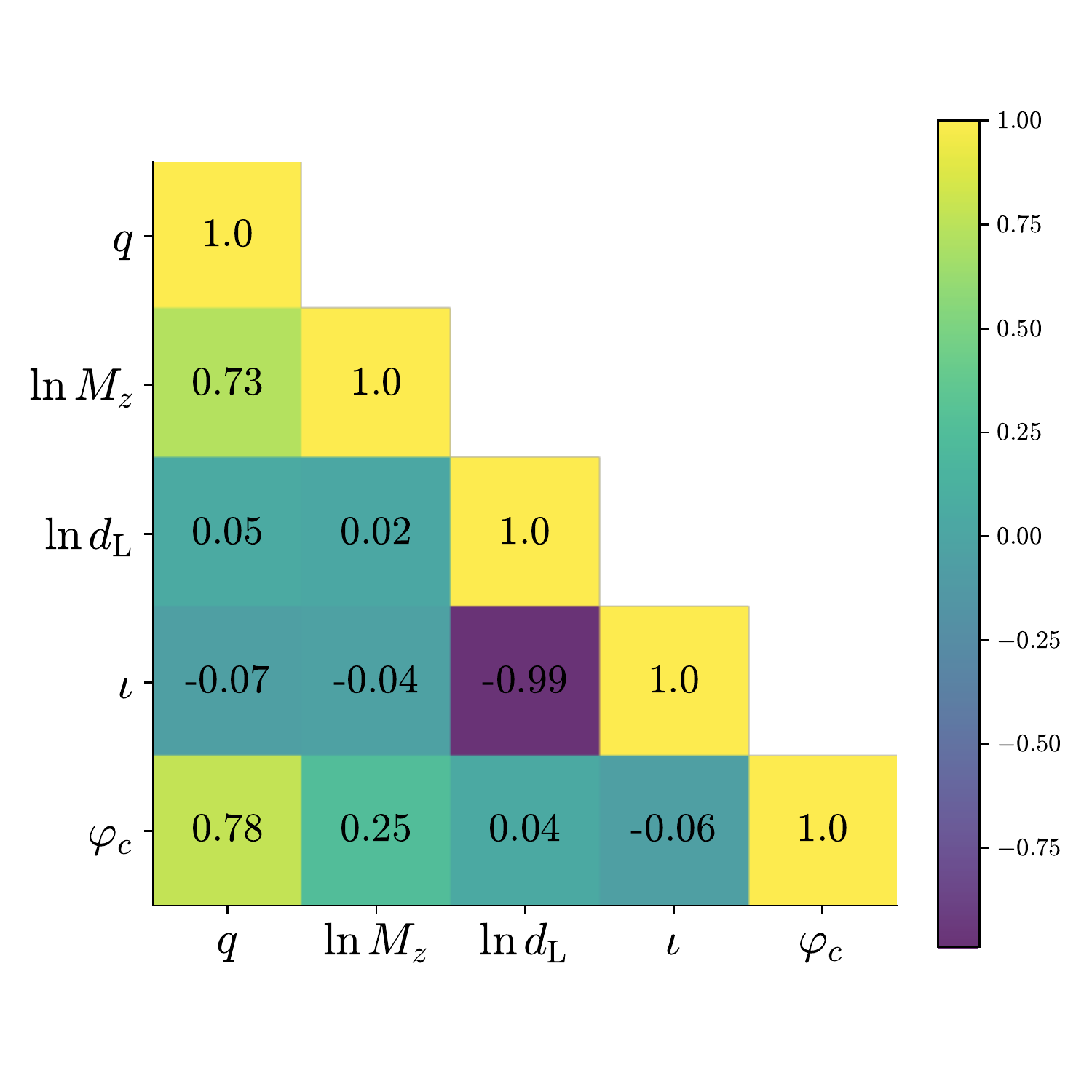}
    \\
    \includegraphics[width=0.9\linewidth]{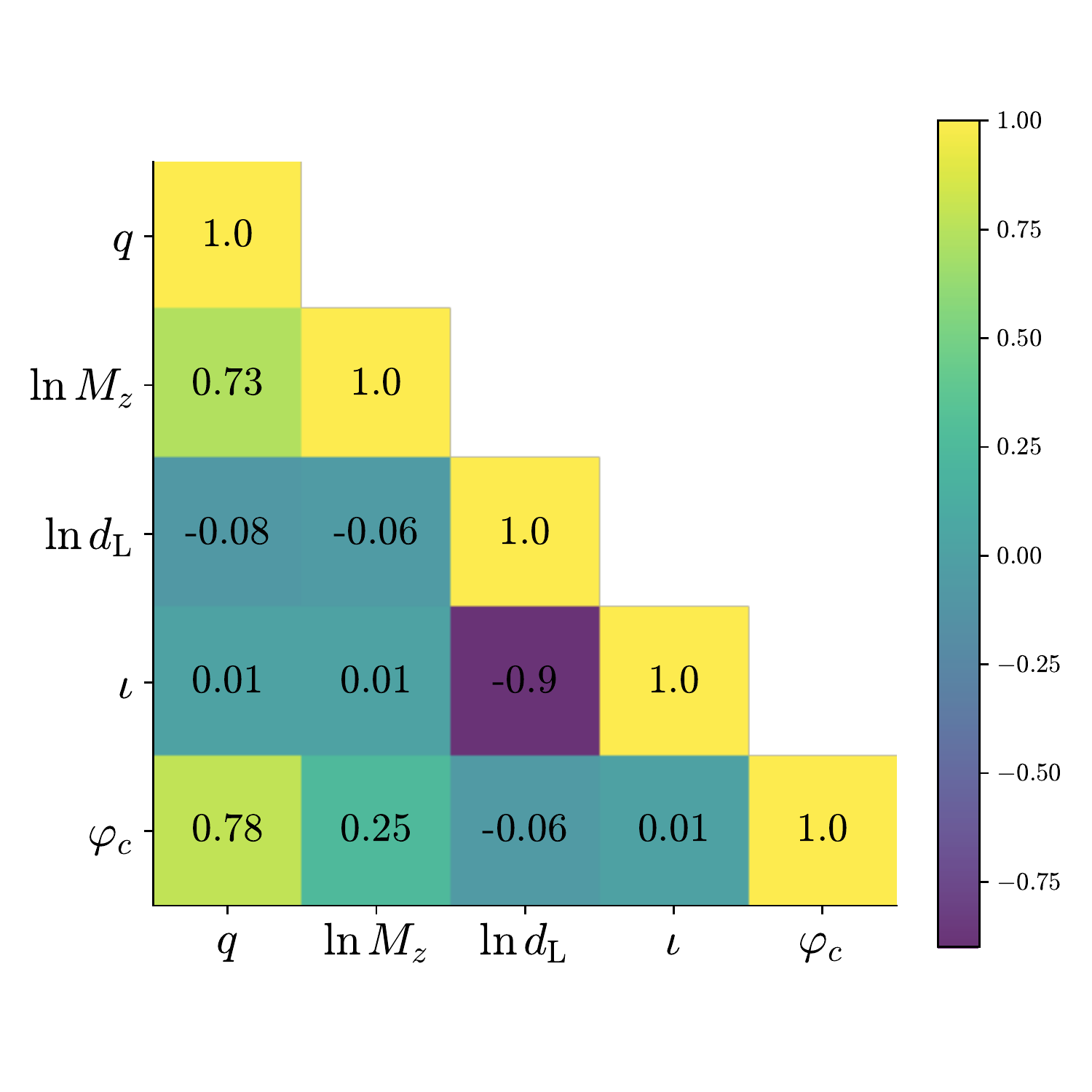}
    \caption{Correlation matrices~$c_{ij}$ for the ``light'' binary of Fig.~\ref{fig:strains}. The upper panel shows the results for the primary signal without memory, while the lower panel is for the total waveform including memory. The major effect of including the memory is decreasing the~$(d_\mathrm{L},\iota)$ correlation.
    }
     \label{fig:FM}
\end{figure}

\section{Results for the distance-inclination estimation}\label{sec:dist-inc}

The goal of this section is to study the impact of the memory on the parameter estimation of intermediate-mass to supermassive BBHs with LISA, with a particular focus on breaking the distance-inclination degeneracy. Our motivation is twofold: (\textit{i}) although nonlinear memory is expected to be observed in single events at LISA, no attention (to our knowledge) has been paid on its impact on the parameter estimation, and (\textit{ii}) the memory signal has an opposite correlation between distance and inclination~$c_{\ln{d_L},\iota}$ with respect to the primary signal, which can make it useful to break the distance-inclination degeneracy, thus improving the accuracy of distance estimations (c.f. Fig.~\ref{fig:Ellipses}).

Firstly, using our Fisher analysis we noted that the primary signal alone already constrains the binary intrinsic parameters quite well, and that the additional information provided by the memory does not constrain them further.
On the other hand, as already anticipated, we found that the dependence of the memory on the extrinsic parameters is complementary to that of the primary signal, and can mitigate the uncertainty on the luminosity distance~$d_\mathrm{L}$ and the inclination~$\iota$ estimations. This is demonstrated in Fig.~\ref{fig:Ellipses} where we explicitly show the constraints coming from the primary signal and the memory, as well as the effect of including all higher modes (HMs) up to~$(\ell,|m|)=(4,4)$. These results are also summarized in Tab.~\ref{tab:HeavyvsLight}.

As it can be seen in Fig.~\ref{fig:Ellipses}, the ellipses computed from the memory signal and the primary waveform are orthogonal, this can be intuitively expected from the opposite monotonic dependence on the inclination of the two components, as explained in Sec.~\ref{sec:phenom}. We analytically derive the opposite correlation for the two signals in Appendix~\ref{sec:AppendixFis}, where we compute the~$2\times2$ Fisher matrix of this pair of parameters $\{\iota,\ln d_\text{L}\}$ for the dominant mode (DM) of the primary waveform and of the memory. This property leads to a lower correlation~$c_{\iota,\ln d_\text{L}}$ of the overall Fisher matrix, as shown in Fig.~\ref{fig:FM}, which can mitigate the error on the luminosity distance estimation. The estimation of the other parameters is less affected, thus justifying our approach to focus on this particular pair of parameters.

\begin{table}[b]
    \centering
    \begin{tabular}{l|cccc}
    \textbf{Light}~~ & ~~$h_{0,\mathrm{DM}}$~~ & ~~$h_{0,\mathrm{HM}}$~~ & ~~$\delta h_{\mathrm{DM}}$~~ & ~~$\delta h_{\mathrm{HM}}$~~ \\ \hline\hline
    SNR & 20.5 & 20.6 & 1.9 & 2.3\\ \hline\hline
    ~ & $h_{0,\mathrm{DM}}$ & $h_{0,\mathrm{HM}}$ & $h_{\mathrm{DM}}$ & $h_{\mathrm{HM}}$ \\ \hline\hline
    $\sigma_{d_\mathrm{L}}/d_\mathrm{L}$ & 0.56 & 0.55 & 0.20 & 0.18 \\
    $\sigma_\iota\, \mathrm{[rad]}$ & 0.75 & 0.73 & 0.27 & 0.23
    \end{tabular}
    
    \vspace{0.3cm}
    
    \begin{tabular}{l|cccc}
    \textbf{Heavy}~~ & ~~$h_{0,\mathrm{DM}}$~~ & ~~$h_{0,\mathrm{HM}}$~~ & ~~$\delta h_{\mathrm{DM}}$~~ & ~~$\delta h_{\mathrm{HM}}$~~ \\ \hline\hline
    SNR & 1001.4 & 1006.3 & 3.5 & 4.3\\ \hline\hline
    ~ & $h_{0,\mathrm{DM}}$ & $h_{0,\mathrm{HM}}$ & $h_{\mathrm{DM}}$ & $h_{\mathrm{HM}}$ \\ \hline\hline
    $\sigma_{d_\mathrm{L}}/d_\mathrm{L}$ & 0.0116 & 0.0108 & 0.0115 & 0.0107 \\
    $\sigma_\iota\, \mathrm{[rad]}$ & 0.0158 & 0.0140 & 0.0157 & 0.0139
    \end{tabular}
    \caption{Summary of the results shown in Fig.~\ref{fig:Ellipses}.
    The first two columns correspond to the results coming from the primary signal alone, whereas the last two correspond to the total (including the memory) waveform. While the memory improves significantly the estimation of the distance-inclination of the ``light'' binary, it does not help much with the ``heavy'' binary parameters.
    }
    \label{tab:HeavyvsLight}
\end{table}

For signals with sufficiently large SNR or whose sources are at sufficiently high redshifts, the uncertainty on the luminosity distance estimation becomes dominated by weak lensing effects (see, e.g., Ref.~\cite{Hirata:2010ba,Tamanini:2016zlh}), and our Fisher analysis (which neglects lensing) ceases to be valid. However, as we show below, the memory is helpful to parameter estimation only for binaries whose primary signal is not very loud, and whose total redshifted mass is in the range~$10^4M_\odot \lesssim M_z \lesssim 10^5M_\odot$. The memory of these sources will only be observed if they are sufficiently close~$z\lesssim1.5$, where lensing effects are not too strong. Therefore, our results indicate that the memory has an impact on the distance estimation only in cases where its intrinsic uncertainty (even after adding the memory) is much larger than the contribution due to lensing.
 
It is natural to compare the effect of the memory with that of HMs, as the latter can also improve the BBH parameter estimation and partially break the aforementioned degeneracy~\cite{Arun:2007qv,Porter:2008kn,Trias:2007fp,Klein:2009gza,CalderonBustillo:2015lrt, Marsat:2020rtl,Graff:2015bba,Payne:2019wmy,Varma:2014jxa,LIGOScientific:2020stg,2018PhRvL.120p1102L}.
Their main contribution to the SNR comes from extending the signal to higher frequencies as compared to the dominant $(2,2)$-mode, thus being more relevant when the merger falls well inside the sensitivity curve of the detector. Importantly, HMs have a different dependence on the inclination angle than the DM, which is again why they may be useful in breaking the distance-inclination degeneracy.  
The memory signal, on the other hand, is always subdominant, but can play a significant role, as we will see, when the information from the primary at lower frequencies (i.e., from the inspiral) is absent/degraded, and the memory becomes the only contribution at those frequencies.
In the following we explain in detail the dependence of memory-assisted parameter estimation on the binary mass, duration of the signal, and line of sight, and we discuss the impact of HMs.

\begin{figure*}[t!]
    \centering
    \includegraphics[width=0.7\textwidth]{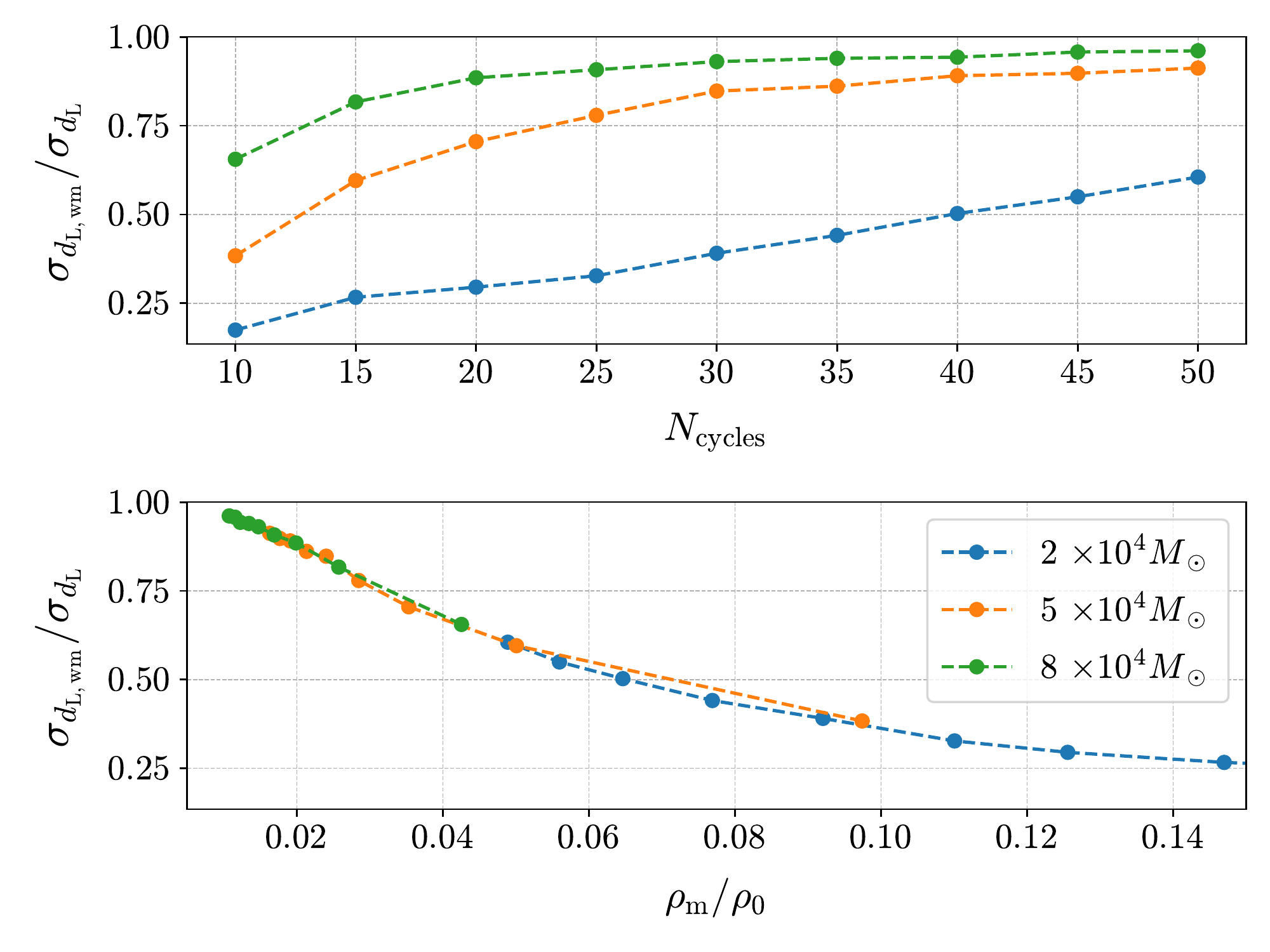}
    \caption{Effect of the memory on the luminosity distance estimation when the primary signal is truncated at increasing number of cycles prior to merger. The memory becomes less important for more massive binaries, for which the merger occurs before or close to the peak of LISA sensitivity and the memory adds negligible contribution to the total SNR. The sources are kept at redshift~$z=0.5$ and line of sight~$\iota=40\, \mathrm{deg}$.}
    \label{fig:Mass&Ndep}
\end{figure*}

\subsection{Dependence on the binary mass} \label{sec:dep_M}

By evaluating the SNR defined in Eq.~\eqref{eq:FisherFormula}, we find that LISA can detect the memory (i.e., can distinguish the memory over the detector noise) when its SNR is~$>1$,\footnote{Indeed, it has been shown that this is a reasonable criteria to ensure the distinguishability of two waveforms that differ by~$\delta h$, over the detector noise~\cite{Hu:2022rjq,Lindblom:2008cm}. It is equivalent to require that, if we parametrize the model waveform as~$h= h_0+\lambda\delta h $ with the fudge parameter~$\lambda\in [0,1]$, the statistical error of the fudge parameter is~$\sigma_\lambda<1$ and, thus, distinguishable from zero.} which happens for binary mergers with total redshifted mass in the range~$[10^4,10^8]\,M_\odot$, thus confirming previous results~\cite{Favata:2009ii,Islo:2019qht,Sun:2022pvh}.
For lower masses, the memory strain is too weak to be detected, whereas for larger masses the turnover frequency at which the memory drops off falls below the LISA band.
Because of the dependence of the memory characteristic strain on the binary mass, LISA will be most sensitive to the low-frequency plateau part of the signal for light binaries, and to the subsequent high-frequency features associated with the merger/ringdown stage for more massive binaries (cf. Fig.~\ref{fig:FTmemHM}).\footnote{See also Fig.~4 of Ref.~\cite{Johnson:2018xly}.}  
We found that nearly all the SNR of the memory  accumulates during the merger and it is maximized for~$M_z\sim 10^6 M_\odot$, in which case the memory can be detectable up to redshift~$z\sim 14$.

For short/degraded primary signals (with fixed number of cycles), we found that the memory is most helpful in parameter estimation for binaries with total redshifted mass~$M_z\lesssim 10^5 M_\odot$, in which case the memory falls in the most sensitive part of the LISA sensitivity band, whereas the merger covers only the high-frequency edge of the spectrum. For this reason, the hierarchy of the SNR between the memory and the primary signals is less severe than for more massive binaries~$M_z\gtrsim10^6 M_\odot$, whose mergers occur in the middle/low-frequency part of the LISA band. 

The difference between ``light''  and ``heavy'' binaries is clear from Fig.~\ref{fig:Ellipses}, which shows the effect of considering short signals (in this case, the last 25 cycles) for two different binary masses. In the left panel, for the ``light'' binary, there is a manifest reduction in the parameter uncertainties, which is due to the intersection of the primary and memory confidence ellipses that results in the shrinking of the total signal's confidence ellipse. Such an effect is not present in the right panel, for the ``heavy'' binary, because, due to the large SNR of the primary signal, its confidence ellipse is already entirely enclosed within the memory's confidence ellipse. The main factor controlling the relative sizes of the confidence ellipses, and thus the impact of the memory in parameter estimation, is the ratio between the memory and the primary signal SNRs (cf. Fig.~\ref{fig:Mass&Ndep}). 

\begin{figure}
\centering
    \includegraphics[width=0.485\textwidth]{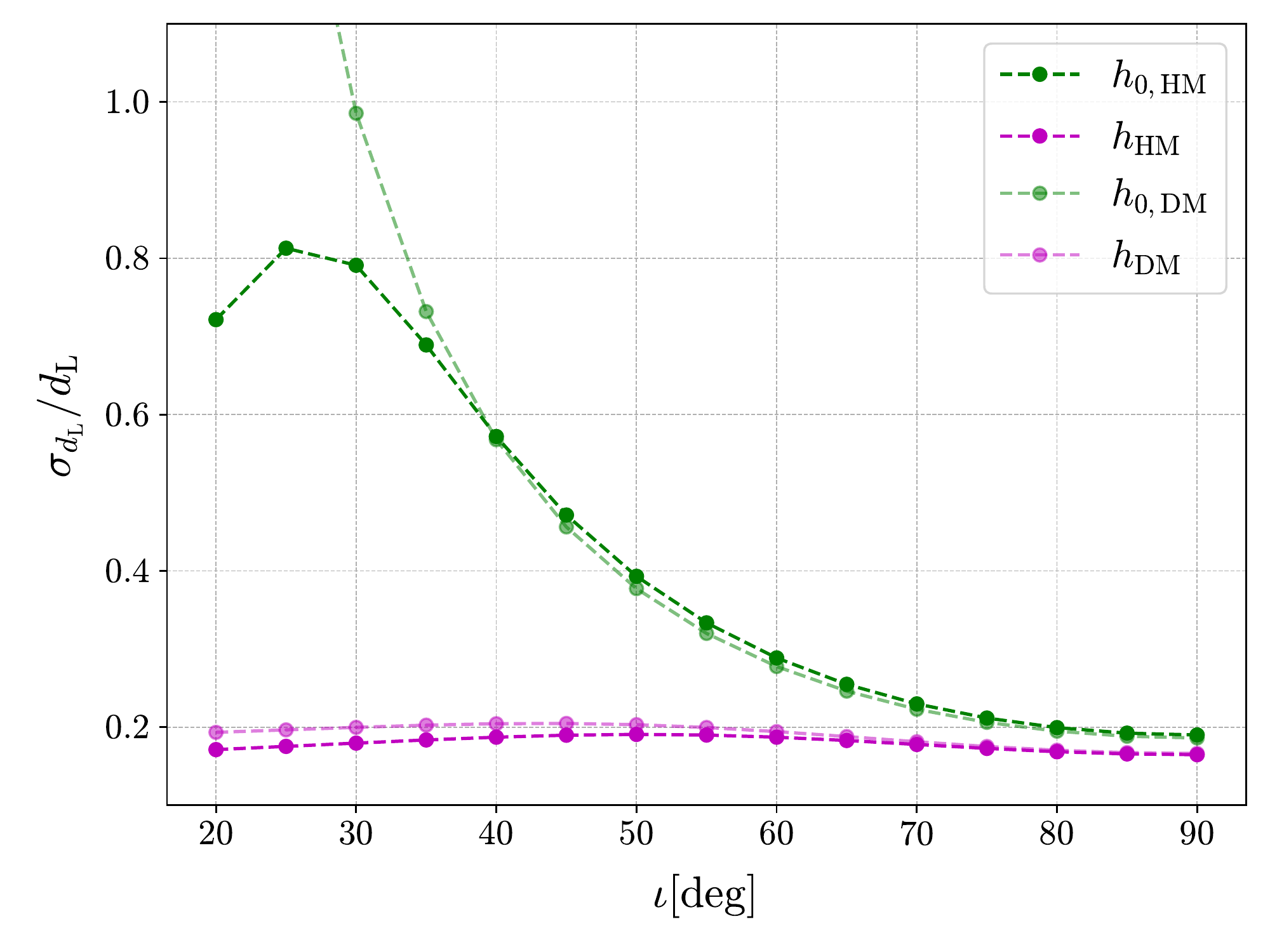}
    \includegraphics[width=0.485\textwidth]{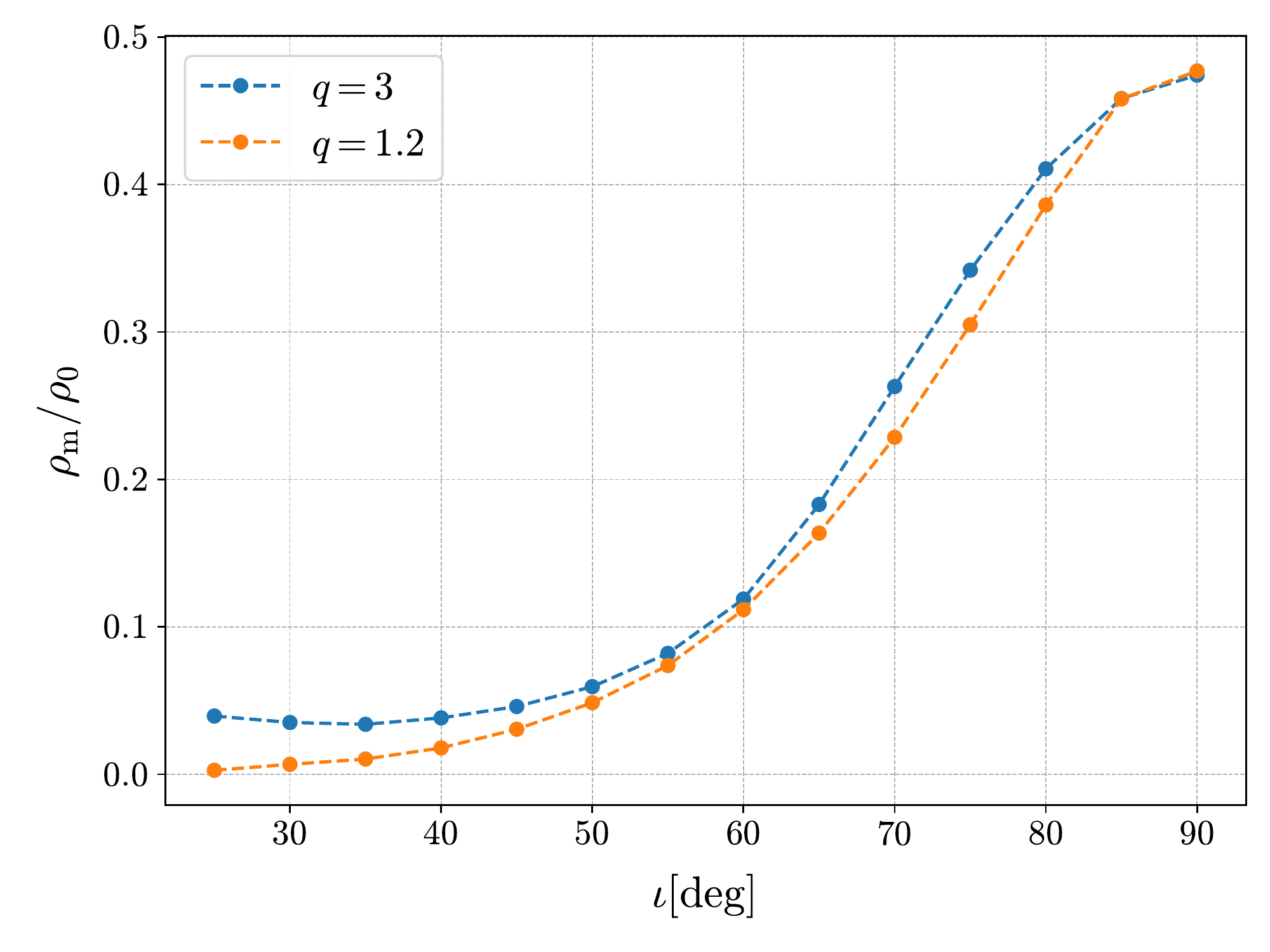}
    \caption{Upper panel: the effect of the memory on the estimation of the luminosity distance as a function of the inclination. The parameters of the source are those of the ``light'' binary of Fig.~\ref{fig:Ellipses} with mass-ratio~$q=1.2$, and the number of cycles before merger is~$N_{\text{cycles}}=25$. In green we show the relative uncertainties~$\sigma_{d_\mathrm{L}}$ computed from the waveform without the memory~$h_0$, with only the dominant mode DM and with higher modes HM. In purple the same but for the total waveform~$h=h_0+\delta h$; in the DM case the memory is computed using only the dominant modes.
    Lower panel: the~$\rho$-ratio required to achieve a 10\% improvement in the luminosity distance estimation (i.e., such that~$\sigma_{d_\mathrm{L},\mathrm{wm}}/\sigma_{d_\mathrm{L}}=0.9$) as a function of inclination for two different values of the mass-ratio~$q=\{1.2,3\}$. }
    \label{fig:SNRratiovsInc}
\end{figure}

\subsection{Dependence on the signal duration}
\label{sec:signal-duration}

As just discussed, our results show that the most reliable indicator for how much the memory contributes to constrain the binary parameters (for a fixed line of sight) is the ratio between the SNR of the memory and of the primary signal,~$\rho_{\mathrm{m}}/\rho_0$.
This ratio depends on the total mass of the binary (cf. Tab.~\ref{tab:HeavyvsLight}), but it is also a function of the inclination and the time duration of the data taken before the merger.
Decreasing the mass and the duration of the signal prior to merger tends to increase this ratio.
We quantify the improvement in parameter estimation in terms of the ratio between the standard deviation of the luminosity distance with memory $\sigma_{d_\mathrm{L},\mathrm{wm}}$ and without $\sigma_{d_\mathrm{L}}$ -- the memory contributes most when the ratio~$\sigma_{d_\mathrm{L},\mathrm{wm}}/\sigma_{d_\mathrm{L}}$ is minimized.

Our results are presented in Fig.~\ref{fig:Mass&Ndep} for different masses in the range~$[10^4,10^5]M_\odot$. In the upper panel, we show the~$\sigma$-ratio as a function of the number of cycles~$N_{\text{cycles}}$ observed prior to merger. We note that for any given number of observed cycles, the memory of more massive binaries leads to a smaller improvement in distance estimation as compared to lighter binaries (as explained in Sec.~\ref{sec:dep_M}). In the lower panel, we show that there exists a monotonic, one-to-one relationship between the~$\sigma$-ratio and the~$\rho$-ratio, for a fixed inclination, which is approximately insensitive to the mass of the binary. This observation will allows us to determine, in the next section, the ``critical'' $\rho$-ratio needed to achieve a given improvement in the distance estimation as a function of the inclination. 

\subsection{Dependence on the inclination}

Fixing the other parameters, the impact of the memory depends strongly on the inclination of the binary, as we show in the upper panel of Fig.~\ref{fig:SNRratiovsInc}. In this plot we present the relative uncertainty in the luminosity distance at different inclination angles, considering the ``light'' binary (discussed previously) with mass-ratio~$q=1.2$ and fixing the observed signal to 25 cycles prior to merger. We restrict the plot to values of $\iota\in[0,\pi/2]$, since its behavior is symmetric with respect to $\iota=\pi/2$.
Note that, in the absence of the memory and the HMs, the uncertainty on the luminosity distance diverges for face-on~$\iota\to0$ (and face-off~$\iota\to\pi$) configurations. 
As we discuss in App.~\ref{sec:AppendixFis}, this is due to the fact that the Fisher matrix is singular at this inclination and, thus, the Fisher analysis for the primary DM ceases to be valid in a neighborhood of~$\iota=0$ (and of~$\iota=\pi$). However, for unequal-mass binaries the HMs regularize the Fisher matrix, and even a slightly asymmetric binary as~$q=1.2$ has a relative uncertainty in the distance smaller than~$\sim0.8$ for all inclination angles.

In the upper panel of Fig.~\ref{fig:SNRratiovsInc} we compare the impact of HMs on the distance estimation with that of the memory for different inclination angles, keeping the mass-ratio $q=1.2$ fixed. However, we also verified that HMs become more relevant the more asymmetric the binary is, whereas the impact of the memory is larger for smaller~$q\sim 1$.\footnote{The influence of the mass-ratio on the impact of the memory can be understood from Eq.~\eqref{eq:OPNres}, $\delta h^{(0\textrm{PN})}\propto q/(1+q)^2$, which results in the memory characteristic strain~$$f \widetilde{\delta h}^{(0\textrm{PN})}\propto q/(1+q)^2,$$ whereas the primary characteristic strain is~\cite{Cutler:1994ys}~$$f \widetilde{h}_0^{(0\textrm{PN})}\propto \sqrt{q}/(1+q).$$} The large effect of the memory observed in the plot -- improving the distance estimation by more than a factor of~$4$ for some inclinations -- is partially related to the short duration of the signal considered (truncated at 6 minutes prior to merger); as discussed in Sec.~\ref{sec:signal-duration}, for longer signals the information in the primary (which accumulates over the inspiral) eventually becomes dominant, with the memory contributing negligibly to parameter estimation (cf. Fig.~\ref{fig:Mass&Ndep}).

In the lower panel of Fig.~\ref{fig:SNRratiovsInc} we show the critical~$\rho$-ratio needed to achieve a~$10\%$ reduction in~$\sigma_{d_\mathrm{L}}$ (i.e., such that~$\sigma_{d_\mathrm{L},\mathrm{wm}}/\sigma_{d_\mathrm{L}}=0.9$) as function of the inclination, for two different mass-ratios~$q=\{1.2,3\}$. From the discussion in Sec.~\ref{sec:signal-duration}, these curves are approximately independent of the binary's total mass. Moreover, we see from this plot that they are also only mildly dependent on the mass-ratio~$q$.
Another important observation is that closer to face-on a smaller~$\rho$-ratio is needed to achieve a given improvement in the distance estimation compared to edge-on.
So, close to face-on the memory can be relevant even if the primary is observed for relatively long periods ($\sim$ few hours). 
For example, if our ``light'' binary (with~$q=1.2$) is seen close to face-on, the inclusion of the memory information leads to a~$10\%$ reduction in~$\sigma_{d_\mathrm{L}}$ if the inspiral is observed over less than 6 hours prior to merger.

\subsection{Impact of higher modes}

Here we discuss the contribution of HMs to the SNR of the primary and memory signals, and its consequent influence on the estimation of the luminosity distance. We consider the ``light'' and ``heavy'' binaries of Fig.~\ref{fig:strains}, both of them with mass ratio~$q=1.2$ and seen at an intermediate inclination angle~$\iota=40\, \mathrm{deg}$.
The outcomes are shown in Fig.~\ref{fig:Ellipses} and Tab.~\ref{tab:HeavyvsLight}.
These results show that, for fixed mass-ratio, the relative increase of the SNR due to the inclusion of HMs is always larger for the memory than for the primary signal. However, the overall SNR of the memory decreases faster with increasing mass-ratio as compared to the SNR of the primary. 
Regarding parameter estimation, Tab.~\ref{tab:HeavyvsLight} shows that HMs have a greater impact for the ``heavy'' binary than for the ``light'' one.
This can be understood from the fact that the HMs add information mostly close to merger (i.e., at high frequencies), which is therefore masked for the ``light'' binary. This is opposed to the memory which, as we have seen, contributes the most to parameter estimation for the ``light'' binary. 
Moreover, we confirmed that, although the overall SNR is suppressed with increasing mass-ratio, the increase in power on HMs for more asymmetric binaries can considerably improve the distance-inclination estimation with the primary waveform, reducing the impact of including the memory component; thus, the equal-mass scenario presented in Tab.~\ref{tab:HeavyvsLight} corresponds to the maximum impact of the memory in parameter estimation over mass-ratio. For instance, for the parameters of the ``light'' binary of Tab.~\ref{tab:HeavyvsLight}, but with~$q=4$, the uncertainty on the luminosity distance is~$\sigma_{d_\text{L}}/d_\text{L}=0.93$ using only the DM of the primary waveform, which is improved to~$\sigma_{d_\text{L}}/d_\text{L}=0.48$ considering the HMs of the primary, and to~$\sigma_{d_\text{L}}/d_\text{L}=0.38$ by including the memory. Interestingly, even though it has a smaller~$\rm{SNR}=12.53$, this binary is better localized in space than the equal-mass one. For the parameters of the ``heavy'' binary, but with $q=4$, the uncertainty ~$\sigma_{d_\text{L}}/d_\text{L}=0.017$ is improved to ~$\sigma_{d_\text{L}}/d_\text{L}=0.008$ with the inclusion of HMs in the primary waveform, and, as for the close to equal-mass case, we do not see further improvements by adding the memory.

Therefore we conclude that, in some cases, the information contained in the memory and in the HMs is complementary for parameter estimation, due to their different dependence on the mass-ratio and total mass.

\section{Population forecasts}
\label{sec:pop}

\begin{figure*}[t!]
    \centering
    \begin{minipage}[b]{0.45\linewidth}
    \includegraphics[scale=0.4]{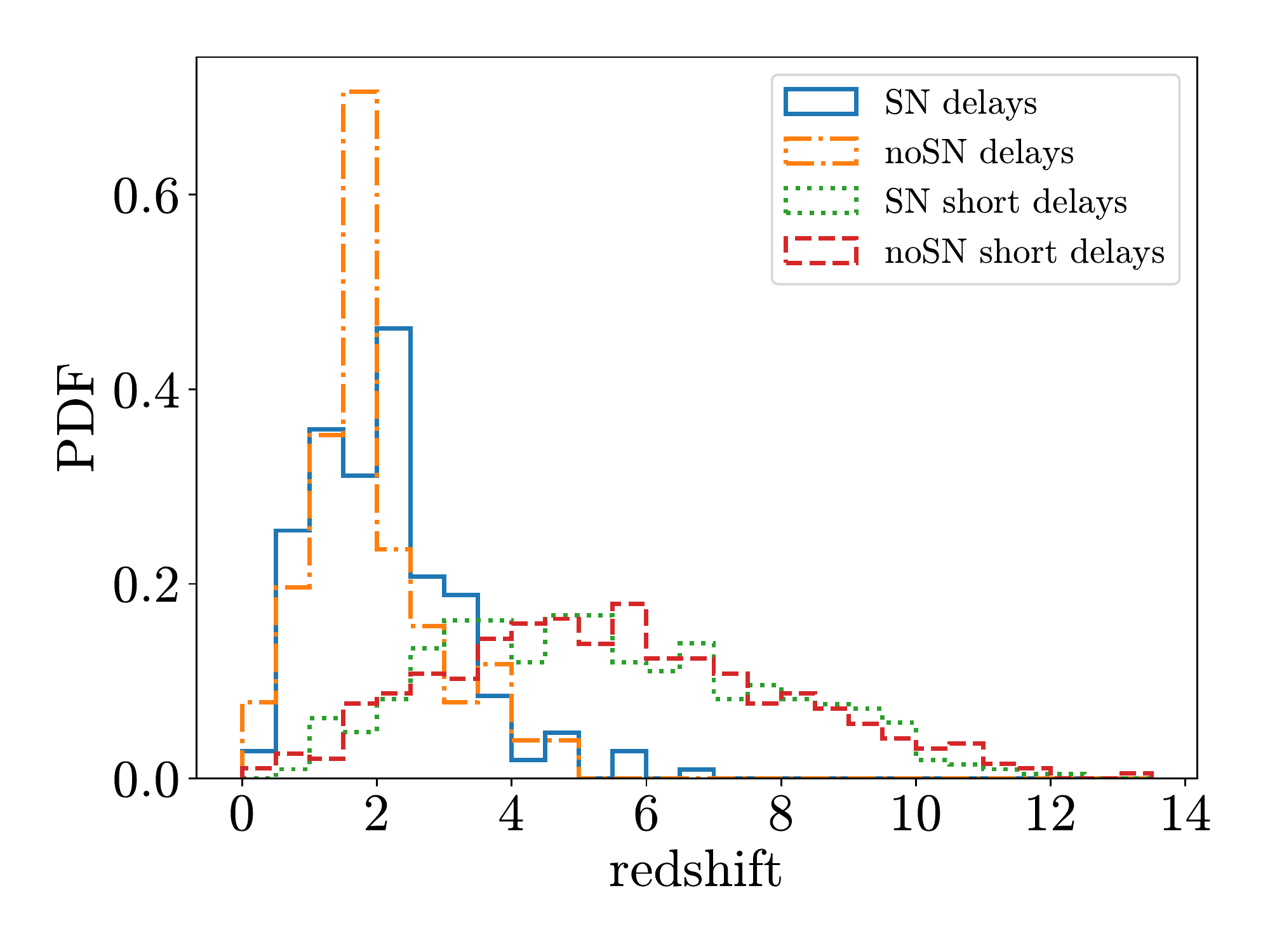}
    \end{minipage}
    \quad
    \begin{minipage}[b]{0.45\linewidth}
    \includegraphics[scale=0.4]{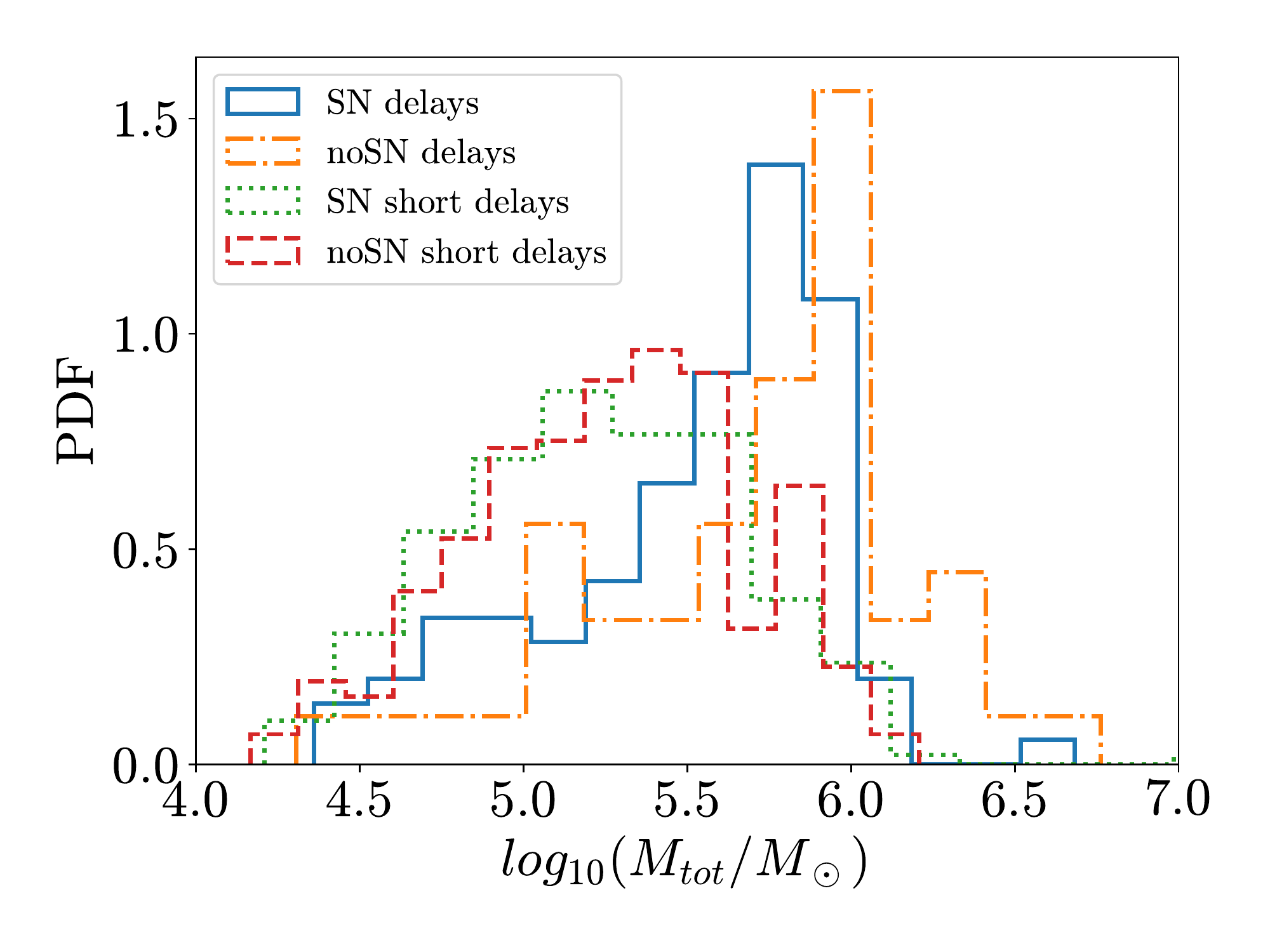}
    \end{minipage}
    \caption{The redshift and total mass distributions of mergers with observable memory~$\rho_\mathrm{m}>1$ for the various HS population models. The models with ``delays'' have mergers typically at lower redshift, explaining the corresponding higher fraction of events with observable memory (cf. Table~\ref{tab:AstroPOP}). The different redshift distribution translates into slightly different peak locations in the total mass distributions.}
    \label{fig:Red&Mass_Dis}
\end{figure*}

\begin{figure*}[t!]
    \centering
    \begin{minipage}[b]{0.45\linewidth}
    \includegraphics[scale=0.4]{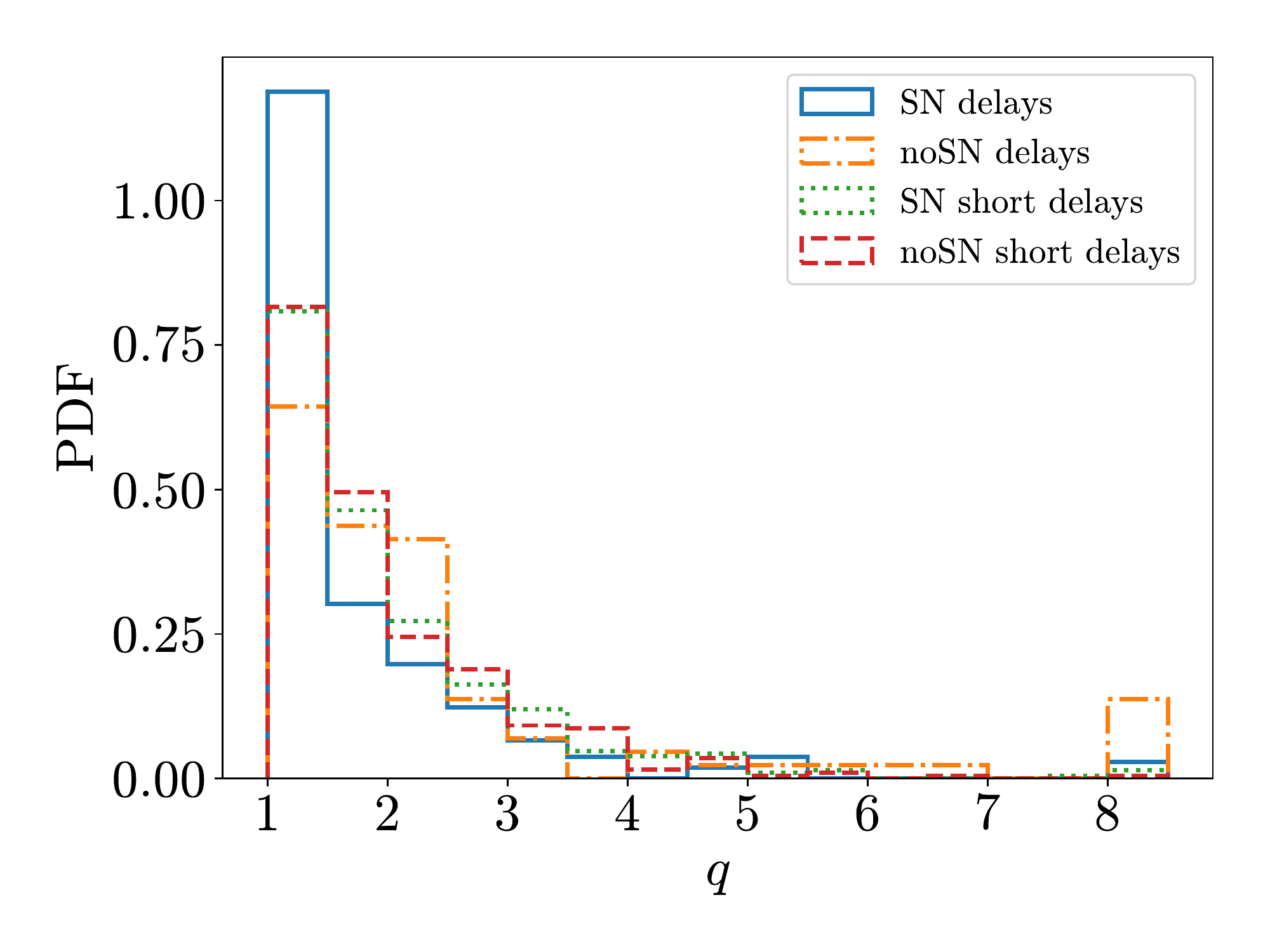}
    \end{minipage}
    \quad
    \begin{minipage}[b]{0.45\linewidth}
    \includegraphics[scale=0.4]{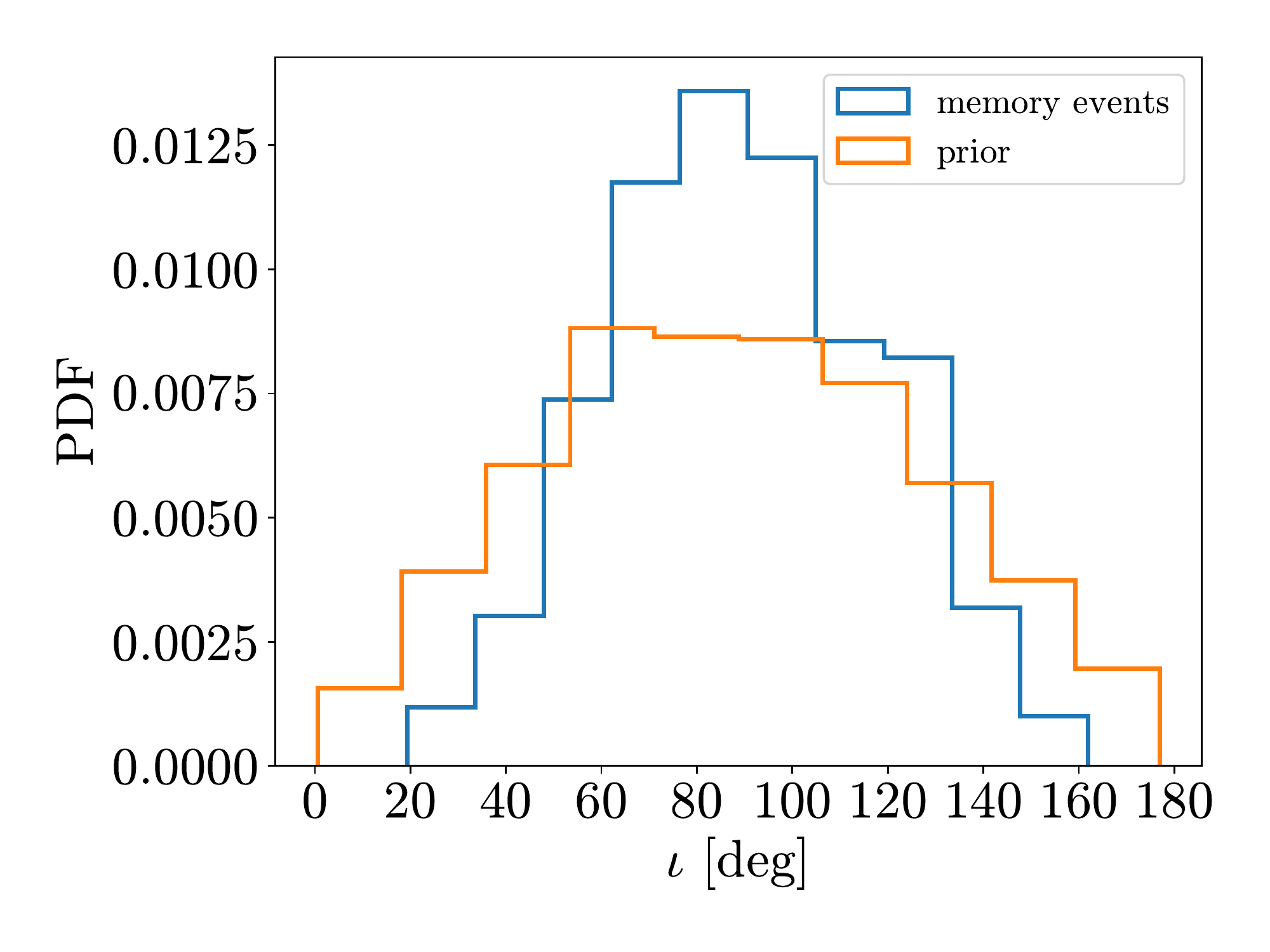}
    \end{minipage}
    \caption{Left panel: the mass-ratio distribution for the events of Fig.~\ref{fig:Red&Mass_Dis}. Right panel: the inclination angle distribution of the events with detectable memory of the ``SN short delays'' model (with $N_{\text{th}}=418$) as compared to the prior isotropic (cosine) distribution.}
    \label{fig:Massratio_Dis}
\end{figure*}

In this section we study the detectability of the nonlinear memory for realistic population models of massive black holes, and assess its potential impact on parameter estimation considering the presence of gaps in the data stream. We update the previous forecasts of Refs.~\cite{Sun:2022pvh,Islo:2019qht} for the measurability of the memory in single events with space-based detectors, by using the  recent population models described in Refs.~\cite{Barausse:2020gbp, Barausse:2020mdt} (and recently used in Ref.~\cite{Dey:2021dem}).

\subsection{Framework}

In our analysis we consider all 8 models described in Refs.~\cite{Barausse:2020gbp,Barausse:2020mdt}, each of those corresponding to different assumptions about some of the main uncertainties in the cosmological evolution of massive BHs. These involve~\cite{Barausse:2020gbp,Barausse:2020mdt}: (\textit{i}) the high-redshift mass function 
of the ``seeds'' of the massive black hole population [``light seeds'' (LS) originating from population III stars, or ``heavy seeds'' (HS) originating from direct collapse of protogalactic gaseous disks], (\textit{ii}) the time delay between the galaxy merger and the corresponding BBH mergers (realistic ``delays'', or ``short delays'' neglecting the contribution from scales of the order of hundreds of pc), and (\textit{iii}) the presence or not of supernova feedback (``SN'' or ``noSN'') on
the accretion disk of massive black holes.
For each of the 8 models we compute the memory random realizations of mergers corresponding to 4 years of LISA mission, and present average results over many such realizations.

We consider only the final~$20$ cycles prior to merger, which is enough to give us a reliable estimate of the SNR of the memory (cf. Sec.~\ref{sec:phenom}).
Due to the limited parameter space covered by the waveforms~\texttt{NRHybSur3dq8}, we restrict the mass-ratio to~$q\leq8$ (i.e., we artificially fix~$q=8$ for every merger with higher values of~$q$). This assumption is stronger for the LS case than for the HS one, since the two cases have quite different mass-ratio distributions, with the latter having a sharper peak close to equal mass (see Fig.~11 of Ref.~\cite{Barausse:2020gbp}). Since the waveforms~\texttt{NRHybSur3dq8} do not cover precessing BBHs, we consider only the spin components orthogonal to the orbital plane and restrict them to~$|\chi_{i,\hat{z}}|\leq0.8$, $i \in \{1,2\}$.\footnote{Using the waveforms \texttt{NrSur7dq2} which allow for precessing BBHs, we found that introducing the spin components along the orbital plane generally leads to an increase of the memory SNR.} 
We removed from the catalogs the sources with~$M_z\geq 10^8\, M_{\odot}$, because evaluating their SNR is computationally very expensive and, as discussed in Sec.~\ref{sec:dep_M}, they fall outside the parameter space of interest for memory observation with LISA. We take the inclination angle and the coalescence phase to be uniformly distributed in~$\cos{\iota}\in [-1,1]$ and~$\varphi_\mathrm{c}\in [0,2\pi]$. 

\begin{table}[b!]
\centering
    \begin{tabular}{p{2cm}|p{3cm}p{3cm}}
    \multicolumn{3}{c}{$\qquad\qquad$Astrophysical Catalogues} \\
    &\textbf{Light seeds}& \textbf{Heavy seeds}\\
    \hline\hline
    SN-delays  & $N_\mathrm{tot}=47$ &   $N_\mathrm{tot}=27.3$\\
    ~ & $N_\mathrm{th}= 0.4\,(0.1)$ & $N_\mathrm{th}= 21.2\,(10)$ \\
    ~ & $\langle \rho\rangle=0.04$& $\langle \rho\rangle=6$\\
    ~  &$\rho_\mathrm{max}=7$ &$\rho_\mathrm{max}=97$\\
    \hline
    noSN-delay & $N_\mathrm{tot}=191$ &   $N_\mathrm{tot}=10$\\
    ~ & $N_\mathrm{th}= 6\,(1)$  & $N_\mathrm{th}= 7.5\,(4)$ \\
    ~ & $\langle \rho\rangle=0.17$ & $\langle \rho\rangle=6.9$\\
    ~ &$\rho_\mathrm{max}=11.64$ &$\rho_\mathrm{max}=68.7$\\
    \hline
    SN-short  & $N_\mathrm{tot}=149$ &   $N_\mathrm{tot}=1245$\\
    Delays & $N_\mathrm{th}= 1\,(1)$ & $N_\mathrm{th}= 418\,(33)$ \\
    ~& $\langle \rho\rangle=0.04$ & $\langle \rho\rangle=1$\\
    ~&$\rho_\mathrm{max}=5.01$ &$\rho_\mathrm{max}=43$\\
    \hline
    noSN-short  & $N_\mathrm{tot}=1203$ &   $N_\mathrm{tot}=1251$\\
    Delays & $N_\mathrm{th}= 12\,(2)$ & $N_\mathrm{th}= 392\,(29)$ \\
    ~& $\langle \rho\rangle=0.06$& $\langle \rho\rangle=1.1$\\
    ~&$\rho_\mathrm{max}=17$ &$\rho_\mathrm{max}=51$
    \end{tabular}
\caption{SNR of the memory for the astrophysical models of Refs.~\cite{Barausse:2020gbp, Barausse:2020mdt}. For each model we consider a random realization of 4 years of events. We denote the total number of events by~$N_\mathrm{tot}$ and the number of those with SNR above the threshold value~$\rho_\mathrm{m}>1\,(5)$ by~$N_\mathrm{th}$. The~$\langle \rho \rangle$ and~$\rho_\mathrm{max}$ are, respectively, the average and the maximum~$\rho_\mathrm{m}$ in the particular realization of events. For the SN-delays models and the NoSN-delay HS model, the reported values are the averages over 10 realizations of 4-year events, since for these models the total number of mergers are much smaller than for the others.}\label{tab:AstroPOP}
\end{table}

\subsection{Detectability of memory in single events}

In the different 4-year realizations, we looked for events with SNR of the memory above the threshold value~$\rho_\mathrm{th}\equiv1$.
Table~\ref{tab:AstroPOP} summarizes our results.
For each population model we denote by~$N_\mathrm{tot}$ the total number of events considered in the 4-year realizations, and by~$N_\mathrm{th}$ the number of those with memory SNR above the threshold~$\rho_\mathrm{th}=1$ (inside parentheses~$\rho_\mathrm{th}=5$).
We also indicate the average SNR of the memory~$\langle \rho \rangle$ and its maximum value~$\rho_\mathrm{max}$ in the particular 4-year realization of events.

The number of events with significant memory SNR depends strongly on the astrophysical population model, with clear differences between the LS and the HS models. Our results are especially promising for the  HS scenario, as it suggests that about~$75\%-78\%$ of events for the ``delays'' model and~$31\%-33\%$ for the ``short-delays'' model will have observable memory with $\rho_{\text{m}}>1$. The numbers inside the parenthesis can be directly compared with the results of Ref.~\cite{Sun:2022pvh}, where the ``Q3d'' and ``Q3nod'' models considered there (and presented in Ref.~\cite{Klein:2015hvg}, based on Refs.~\cite{EB12,Sesana:2014bea,Antonini:2015sza}) can be compared, respectively, with ``delays'' and ``short-delays'' HS models. For those, we found that about~$36\%-40\%$ and~$25\%$, respectively, of the total events have detectable memory with~$\rho_\mathrm{m}>5$, as opposed to~$3.7\%$ and~$1\%$ for the ``Q3d'' and ``Q3nod'' models. The main reason for this large mismatch is that the new population models of Refs.~\cite{Barausse:2020gbp,Barausse:2020mdt} that we used in this work have a different (more realistic) delay model, which shifts the mergers to lower redshift.

Figure~\ref{fig:Red&Mass_Dis} shows the distribution of the redshift and the total mass of the events with~$\rho_\mathrm{m}>1$ for the various HS models. The distribution of models with ``delays'' peaks at lower redshift, while that of ``short-delays'' models extends up to $z\sim 13$, which explains why the former have a bigger fraction of events with memory SNR above the threshold than the latter. Interestingly, we found some events with particularly high SNR ($\rho_\mathrm{m}\gtrsim50$), as can be seen in Tab.~\ref{tab:AstroPOP}; these belong to the low redshift tail of the distributions ($z\lesssim1$). The location of the peak of the total mass distribution changes slightly for the various models, but it is such that the total redshifted mass is about~$\sim 10^6 M_\odot$. 
In Fig.~\ref{fig:Massratio_Dis} we show the mass-ratio distribution for the same events of Fig.~\ref{fig:Red&Mass_Dis}.
Most of the events with detectable memory have a mass-ratio close to unity with a sharp suppression at higher values, so that the restriction to~$q<8$ turns out not to affect our results for the HS models. The ``noSN delays'' model is the only one presenting a mild accumulation of events with~$q=8$ due to this restriction. 

The situation is quite different for the LS models, which have a much broader distribution in the mass-ratio, resulting in a substantial (fictitious) accumulation of events at~$q=8$. So, we repeated our analysis removing directly the binaries with $q>8$ from the catalog. In this more conservative approach, for ``noSN-short delays'' we noted a reduction from 12 to 9 events with $\rho_\mathrm{m}>1$, but no change in the number of events with~$\rho_\mathrm{m}>5$. For ``SN-short delays'' and ``SN delays'' models we found no events with detectable memory, and for ``noSN-short delays'' a reduction from 6 to 4 events with $\rho_\mathrm{m}>1$, and no events at all with $\rho_\mathrm{m}>5$. 
Therefore, for LS population models our results indicate that the prospects of observing the memory with LISA do not seem promising. 

We repeated our analysis of LS population models also for the future generation ground-based detectors Cosmic Explorer (CE)~\cite{Reitze:2019iox} and Einstein Telescope (ET)~\cite{Maggiore:2019uih}, which have better sensitivity at higher frequencies, and thus to lower masses.\footnote{We used the sensitivity curve of the configurations ET-D of Ref.~\cite{Hild:2010id} and CE\_40km\_lf of \url{https://cosmicexplorer.org/sensitivity.html}, which have the best sensitivity at low frequencies.} We have found almost no events with observable memory~$\rho_{\text{m}}>1$ in 4 years of observation (there was just one event with~$\rho_{\mathrm{m}}\simeq2$ for the ``SN short delays'' model with ET), and an average memory SNR within~$10^{-1}-10^{-3}$.

\subsection{Impact of the memory on distance estimation}

Given that HS models predict such a large number of events with observable memory at LISA (which can be almost up to~$80\%$ of the total number of events, in the most optimistic scenario), we consider here the impact of the memory on the luminosity distance estimation for these sources. As we have shown in Sec.~\ref{sec:signal-duration} the impact of including the memory is highly dependent on the ratio of the SNR of the memory and the primary signals, with the memory helping substantially to constrain the distance when the information (or the duration) of the primary signal is somehow limited. Among other possibilities, this could happen due to the presence of gaps in the data stream, which causes a partial loss of  signal.

Two kinds of gaps are expected at LISA: the scheduled ones, related to the regular maintenance of the detector, and the unscheduled ones, due to unexpected problems/events. In Ref.~\cite{Dey:2021dem} it was shown that the scheduled gaps have little or no impact at all on the parameter estimation of massive BBHs, but the unscheduled ones could degrade significantly the parameter estimation. Thus, there is the intriguing possibility that, in the presence of unscheduled gaps, the memory may add useful information to constrain the binary parameters.\footnote{Note that the LISA data analysis will be further complicated by the presence of many overlapping signals~\cite{Littenberg:2023xpl}.}

To quantify this effect we consider the particular gap model used in Ref.~\cite{Dey:2021dem}, which is consistent with a 75\% duty cycle, as expected for LISA~\cite{Seoane:2021kkk}. We simulate the presence of gaps by windowing the signals as in Ref.~\cite{Dey:2021dem}, considering scheduled gaps with a typical duration of 3.5 hours every week and unscheduled gaps with a duration of 3 days. The time interval between two unscheduled gaps is treated as a random variable following an exponential probability distribution~$p(\Delta T)=\lambda\exp{(-\lambda\Delta T)}$ with~$1/\lambda=9$ days. With these choices, we simulate the effective data taking of the mission and we distribute the merger times uniformly over the 4-year mission duration. From our study in Sec.~\ref{sec:dist-inc}, we expect the memory to be helpful in constraining the binary (extrinsic) parameters for a particular chunk of data if the merger happens within the first few hours from the last gap.

For concreteness, let us compute the (average) total number of mergers occurring within 6 hours from the last gap. The number of unscheduled gaps can be estimated by~$N_{\text{gap}}\sim T_{\text{mission}}/T_{\text{gap}}$, where~$T_{\text{gap}}$ is the sum of the average time interval between gaps and the gap duration,~$\langle \Delta T \rangle + 3\simeq 12$ days, thus~$N_{\text{gap}}\sim 120$. We focus on the ``SN-short delays'' HS population model, the most optimistic scenario with the highest number of events with observable memory,~$N_{\text{th}}=418$; note that due to the presence of gaps this number is reduced by~$75\%$. Thus, we can estimate the number of mergers by multiplying the probability of having at least one merger in 6 hours by~$0.75 N_{\text{th}}N_{\text{gap}}$, which gives~$\sim 6.4$ events. We checked this result numerically by simulating 50 times the distribution of mergers over the gap realization and we found consistent results.

To find the number of events for which the inclusion of the memory decreases by more than 5\% the uncertainty on the luminosity distance, we computed the~$\rho-$ratio (i.e., $\rho_\textrm{m}/\rho_0$) for each event occurring within 6 hours from the last gap in  50 numerical realizations, neglecting the information accumulated in the inspiral before the gap. Subsequently, we compared those $\rho$-ratios with the critical values needed to achieve a 5\% improvement on the luminosity distance estimation, which depend on the particular binary inclination (as in the lower panel of Fig.~\ref{fig:SNRratiovsInc}, but for~$\sigma_{d_L \text{wm}}/\sigma_{{d_L}}=0.95$). We found that, on average, only 0.14 events of the 6.4 occurring close after a gap have an improvement of more than 5\% on the distance estimation from including the memory; this corresponds to 0.04\% of the total number of events with observable memory in this population model.

We believe that this low value is due to the fact that most of the BBH mergers with observable memory correspond to configurations relatively close to edge-on (c.f. right panel of Fig.~\ref{fig:Massratio_Dis}), where the critical~$\rho-$ratio is much higher. Another reason is that the majority of the BBHs with observable memory in the population considered have a redshifted total mass $\gtrsim 10^6 M_\odot$, whereas as discussed in sec~\ref{sec:dep_M} the memory is more helpful for lighter binaries~$M_z\lesssim 10^5 M_\odot$. The ``noSN-short delays'' HS population model, which is the second most optimistic in terms of number of events with detectable memory, suffers from these same issues and is, thus, expected to give a similarly small result. 
The other population models have far fewer events with observable memory, thus it is very unlikely that any of these mergers will happen sufficiently close to a gap to have a sufficiently large~$\rho-$ratio.

In summary, applying our Fisher analysis to state-of-the-art synthetic catalogs of massive BBHs indicates that the memory will not help constraining further the binary parameters at LISA, even in the presence of gaps in the data stream. However, there is substantial uncertainty on the assumptions adopted in this analysis, in particular, regarding the population and gap models, and we cannot exclude the possibility that there may exist additional effects leading to a larger degradation of the primary signal than those considered here.

\section{Conclusion}
\label{sec:concl}

In this work we have investigated the prospects of using the nonlinear GW memory to help infer the parameters of merging BBHs.
In particular, we have focused on massive BBHs detections with the future space-based interferometer LISA, as these are the most promising individual sources of memory.
Our motivation is to use the additional source of information provided by the memory signal to break the degeneracy between inclination~$\iota$ and luminosity distance~$d_\mathrm{L}$, which is present in the leading-order GW signal.
This is especially important for attempts to use these BBHs as standard sirens (either via statistical identification of the host galaxy~\cite{Palmese_2020}, or possibly using an electromagnetic counterpart due to the merger taking place in a gas-rich environment~\cite{Tamanini:2016zlh,Hotokezaka:2018dfi,Chen:2018omi}), as the uncertainty on the Hubble constant $H_0$ crucially depends on the uncertainty on $d_\mathrm{L}$.

We find that the memory can indeed play a significant role in breaking this inclination--luminosity distance degeneracy.
This occurs in cases where the redshifted total mass is relatively small ($\lesssim10^5\,M_\odot$), the binary is seen not very close to edge-on, and the observation time is limited to a few hours prior to merger.
The limitation on the observation time could occur due to, e.g., gaps in the data stream caused by interferometer downtime, or confusion noise from the presence of many other simultaneous signals in the LISA frequency band.

In order to understand the relevance of these results for the LISA mission, we started by performing a population study using new synthetic catalogs of massive BBHs to forecast the number of BBH events with observable memory ($\rho_{\text{m}}>1$).  
While there are currently large theoretical uncertainties on the astrophysical processes leading to these mergers, we find a substantially larger number of events with significant memory as compared to previous forecasts. The prospects are particularly bright for the heavy seed model with ``short delays''~\cite{Barausse:2020gbp,Barausse:2020mdt}, which presents about 400 memory events for a 4-year mission time. On the other hand, most of the mergers coming from light seed models~\cite{Barausse:2020gbp,Barausse:2020mdt} are undetectable by LISA (and so is their memory).

Finally, we considered a commonly used gap model, which includes both the scheduled and unscheduled types, to quantify the benefit of the memory in the estimation of the luminosity distance. For the most optimistic ``short delays'' heavy seed models~\cite{Barausse:2020gbp,Barausse:2020mdt}, we found that, out of the $\sim 0.75\times400$ observable memory events in a 4-year mission time, just 0.14 events will produce a larger than 5\% decrease in $\sigma_{d_L}$. Thus, our analysis indicates that the information in the memory signal will not help constraining further the binary parameters at LISA, even in the presence of gaps in the data stream. This is due to the fact that most of the events with observable memory are seen close to edge-on, in which case the luminosity distance and inclination are only slightly correlated in the primary signal and, thus, the information added by the memory is negligible for parameter estimation.

Our study, based on a Fisher matrix analysis, could be further improved by performing a full Bayesian analysis and by investigating the effect of priors on the luminosity distance estimation, which is especially important when the parameters are not well constrained. Another interesting extension of our work would be to consider the impact of the memory on parameter estimation for binaries with precession. However, we expect that our key finding --- that the memory signal can only play an important role in BBH parameter estimation when there is limited information from the inspiral --- holds generically, due to the different orders of magnitude of the primary and memory signal characteristic strains. We also leave open the possibility that some currently unforeseen effects may lead to a much larger degradation of the primary signal than the one due to the presence of gaps, which could make the memory information more relevant to parameter estimation.

As a final remark, we note that even if the information in the memory turns out not to be very useful in constraining the binary parameters, the amount of events with detectable memory we found for LISA (c.f. Table~\ref{tab:AstroPOP}) suggests that it may still play a significant role as a test of GR in the strong-gravity nonlinear regime, since most of the memory is generated close to merger. We leave these questions for future work.

\begin{acknowledgments}

The authors would like to thank Juan Calderon Bustillo, Xisco Jimenez Forteza, Giada Caneva and Marc Andrés for their technical help in the first stage of this project. 
We are also grateful to Neil Cornish for his valuable comments on a draft of the paper.
In this study we used the software packages \texttt{matplotlib}~\cite{4160265}, \texttt{numpy}~\cite{Harris:2020xlr}, \texttt{scipy}~\cite{2020SciPy-NMeth}, \texttt{LISA Sensitivity}~\cite{Robson:2018ifk}, \texttt{gwmemory}~\cite{Lasky:2016knh}, \texttt{GWsurrogate}~\cite{Field:2013cfa}, \texttt{surfinBH}~\cite{vijay}, and \texttt{qnm}~\cite{Stein:2019mop}.
RV is supported by grant no. FJC2021-046551-I funded by MCIN/AEI/10.13039/501100011033 and by the European Union NextGenerationEU/PRTR. RV also acknowledges support by grant no. CERN/FIS-PAR/0023/2019. 
DB is supported by a `Ayuda Beatriz Galindo Senior' from the Spanish `Ministerio de Universidades', grant BG20/00228. 
The research leading to these results has received funding from the Spanish Ministry of Science and Innovation (PID2020-115845GB-I00/AEI/10.13039/501100011033).
IFAE is partially funded by the CERCA program of the Generalitat de Catalunya. This work was partly enabled by the UCL Cosmoparticle Initiative.
EB acknowledges support from the European Union's H2020 ERC Consolidator Grant ``GRavity from Astrophysical to Microscopic Scales'' (Grant No.  GRAMS-815673) and the EU Horizon 2020 Research and Innovation Programme under the Marie Sklodowska-Curie Grant Agreement No. 101007855.
\end{acknowledgments}

\appendix
\section{Signal processing}
\label{sec:window}
In this section we provide more details about our choices in manipulating the BBH waveforms. We first generate the primary signal with a sampling time~$\Delta t=1/4\,\mathrm{s}$, and we subsequently generate its memory via the \texttt{GWmemory} package~\cite{Talbot:2018sgr}.
As explained in Sec.~\ref{sec:Parestimation}, we compute the total signal in frequency domain, summing the individual FFTs of the primary waveform and of the memory.
However, we find that a spurious contribution of the primary waveform at frequencies~$f<f_{\text{in}}$ generates cross-terms between the primary and the memory signal of order~$\mathcal{O}(h_\mathrm{c})$ which affect the computation of the SNR and the Fisher matrix. To prevent these artifacts from affecting our results, we removed the contribution from~$f<f_{\text{in}}$ of the primary signal before summing the individual FFTs. 

We follow a standard procedure to manipulate the primary waveform, namely,   applying a window function, padding the signal, and taking the FFT. We apply the following window function to the primary signal:
 \begin{equation}
        z(t)=\frac{1}{4}\Big[1+\tanh \big( \tfrac{t-t_0}{\sigma_0/4}\big)\Big]\Big[1-\tanh \big(\tfrac{t-t_\mathrm{h}}{\sigma_\mathrm{h}/4} \big)\Big], 
    \end{equation}
with~$M t_0=150$ and~$M t_\mathrm{h}=110$, respectively, at the beginning and at the end of the time series. The duration of the windowing is set by~$M \sigma_0=50$ and~$M \sigma_\mathrm{h}=20$. 

\begin{figure}[t!]
    \centering
    \includegraphics[width=0.5\textwidth]{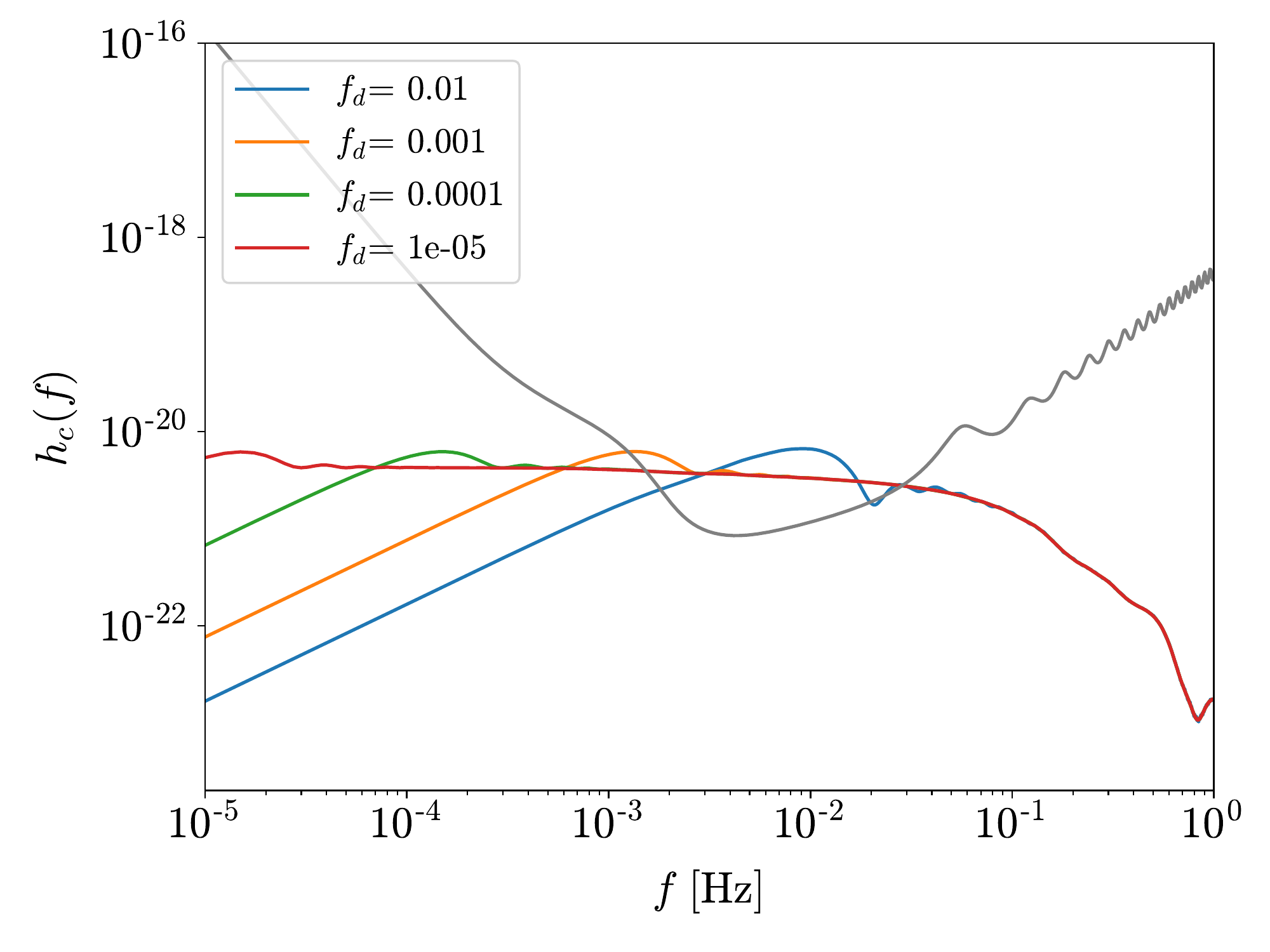}
    \caption{Characteristic strain~$h_\mathrm{c}(f,\iota,\Phi)$ of the memory for different choices of decay frequency of the window function~\eqref{eq:window_mem}. The parameters of the binary are the same as those of the ``light'' binary in Fig.~\ref{fig:strains}, but with an inclination~$\iota=90\, \mathrm{deg}$.}
    \label{fig:windows}
\end{figure}

For the memory signal we follow a different procedure.
We first extend the generated~$\delta h(t)$ (evaluated numerically for~$t_0\leq t \leq t_\mathrm{f}$) at the beginning and the end with the constant values~$\delta h(t_{0})$ and~$\delta h(t_{\mathrm{f}})$, respectively, using the same padding length as for the primary signal.
Subsequently, we apply the following window (as in Ref.~\cite{Richardson:2021lib}):
\begin{equation}\label{eq:window_mem}
        w(t)= \begin{cases} 1, \qquad & t-t_\mathrm{d}<0\\
        \frac{1}{2}\big(1+\cos [2\pi f_\mathrm{d}(t-t_\mathrm{d})]\big), \qquad & t-t_d\geq0 \\ 0, \qquad & t-t_\mathrm{d}\geq\frac{1}{2f_\mathrm{d}}
        \end{cases}
\end{equation}
and choose $M t_\mathrm{d}=140$. The choice of the decay frequency of the window function~$f_\mathrm{d}$ greatly impacts the spectral shape of the memory at low frequencies, as can be seen in Fig.~\ref{fig:windows}, where we show the characteristic strain~$h_c$ of the memory for different values of $f_\mathrm{d}\in\{10^{-2},10^{-3},10^{-4},10^{-5}\}\, \mathrm{Hz}$. Note that higher values of~$f_\mathrm{d}$ inject spurious power at the frequencies for which LISA is most sensitive, thus leading to an artificial increase of the respective memory SNR $\rho_\mathrm{m}=\{7.37,5.35,5.13,5.12\}$. Thus, while taking a lower~$f_d$ is more reliable, in the sense that it does not overestimate the memory SNR, it implies a corresponding longer observation time of the memory, which can become inconsistent with the maximum observation time considered for the primary signal. However, this does not pose a real problem since most of our analysis applies to cases where the primary signal is observed for a few hours, whereas the SNR of the memory does not greatly change as long as the memory is observed for more than 15 minutes ($f_d\lesssim10^{-3}\,\mathrm{Hz}$). 

In computing the SNR and the Fisher matrix we take~$f_{\text{min}}=1/T$, where $T$ is the total length of the signal, and $f_\mathrm{max}=\mathrm{min}\{1\, \mathrm{Hz}, f_{440}\}$, since we find that the QNM~$f_{440}\equiv \omega_{440}/2\pi M_z$ is a good measure of the maximum frequency present in the signal (note that our waveforms include modes up to~$\ell_\mathrm{max}=4$). For total masses between~$[10^4,10^5]M_\odot$ we find that fixing the minimum frequency at~$f_{\text{min}}=10^{-4}\, \mathrm{Hz}$ does not change our SNR and Fisher forecasts.  

\section{Analytic considerations on the distance-inclination Fisher matrix}\label{sec:AppendixFis}

Here we review the Fisher matrix derivation of the $(d_\mathrm{L},\iota)$ degeneracy by computing the relative $2\times 2$ matrix analytically. Including other parameters have little effect close to the degenerate points, since the main source of error comes from this submatrix. Subsequently, we show that the results represented in Fig.~\ref{fig:Ellipses} can be understood by taken into account simply the main angular dependence of the primary and the memory signals. In order to estimate the effect of the memory on this degeneracy it is enough to focus on the part of the Fisher matrix regarding the extrinsic parameters $\{d_\mathrm{L},\cos(\iota),\phi,\varphi_\mathrm{c}\}$ since, at linear order in $\mathrm{SNR}^{-1}$, it is decoupled from the one associated to the intrinsic parameters $\{M_z,q,t_\mathrm{c},\psi,\mathrm{Spins}\}$~\cite{Cutler:1994ys}, where~$\psi$ is the polarization angle. Indeed, while the intrinsic parameters are mainly extracted from the phase evolution of the waveform, the extrinsic parameters depend on the amplitudes~$h_{+}$ and~$h_{\times}$~\cite{Usman:2018imj}.
Here we ignore the dependence on the coalescence phase~$\varphi_\mathrm{c}$, since (at leading order) it does not affect the memory. At Newtonian (0 PN) order the primary waveform is
\begin{align}
h_{+,0}&=2\frac{\eta M_z}{d_\mathrm{L}}[M \omega(t)]^{\frac{2}{3}} (1+\cos^2\iota) \cos[2 \varphi(t)], \label{eq:0PNosc_+}\\
h_{\times,0}&=4\frac{\eta M_z}{d_\mathrm{L}}[M \omega(t)]^{\frac{2}{3}}\cos\iota \sin[2 \varphi(t)],
\end{align}
using the polarization conventions of Ref.~\cite{Kidder:2007rt}.
The GW amplitude in the detector can be written in the frequency domain as~\cite{Robson:2018ifk}
\begin{equation}
\tilde{h}(f)=F^+(f) \tilde{h}_{+}(f)+F^\times(f) \tilde{h}_{\times}(f),
\end{equation}
where~$F^{+,\times}(\iota,\phi,\psi,f)$ are the frequency-dependent detector response functions, which also depend on the source sky-location and polarization angle.
Substituting the primary waveform FTs in the last expression we find
\begin{equation}
    \tilde{h}_0= \frac{\kappa_0}{d_\mathrm{L}}\Big\{F^+(1+\cos^2\iota)-2i F^\times \cos \iota\Big\},
\end{equation}
where~$\kappa_0$ is independent of both the luminosity distance (for fixed~$M_z$) and the inclination angle. 
The sky- and polarization-averaged Fisher matrix~\eqref{eq:FisherFormula} has then the form\footnote{Where we used~$\langle F^+(f)F^{\times*}(f) \rangle=0$ for the sky- and polarization-averaging of the cross terms~\cite{Robson:2018ifk}.}
\begin{equation}
    \Gamma^0_{i j}=\left(\frac{\rho_{\kappa_0}}{d_\mathrm{L}}\right)^2 \hat{\Gamma}^0_{i j},
\end{equation}
with~$i,j\in \{\log{d_\mathrm{L}},\iota\}$, where~$\rho_{\kappa_0}^2=(\kappa_0|\kappa_0)$ and the matrix
\begin{equation*}
    \hat{\Gamma}^0=
    \begin{pmatrix}
    (1+\cos^2\iota)^2+4\cos^2\iota & (3+\cos^2\iota)\sin(2\iota)\\
    (3+\cos^2\iota)\sin(2\iota)& 4(1+\cos^2\iota)\sin^2\iota
    \end{pmatrix}\,,
\end{equation*}
depends only on the inclination.
This matrix is clearly singular for face-on/off binaries, and it is diagonal for edge-on ones (implying that the two parameters are uncorrelated),
\begin{align*}
    &\hat{\Gamma}^0(\iota\in \{0,\pi\})= \begin{pmatrix}
    1&0\\0&0 \end{pmatrix},\qquad \hat{\Gamma}^0(\iota=\tfrac{\pi}{2} )=\begin{pmatrix}
    1&0\\0&1
    \end{pmatrix}
    .
\end{align*}
It is easy to see that for inclination angles~$0<\iota<\frac{\pi}{2}$ ($\frac{\pi}{2}<\iota< \pi$) the distance and the inclination are \emph{negatively} (\emph{positively}) correlated.

This result shows that the degeneracy we focused in  in this work is driven by the dependence of the (leading) quadrupole waveform on the inclination, and the particular combination of plus and cross populations measured by the detector, which, in particular, lead to a singular Fisher matrix for face-on/off configurations. This is a well-known issue in the literature and special care must be taken close to the singular points, where one should use a beyond-Gaussian analysis~\cite{Vallisneri:2007ev,Cutler:1997ta}.
However, for these face-on/off configurations the memory is almost vanishing, so that in this work our focus is on intermediate inclination angles that are not too close to the singular points. Despite that, in the main text our analysis includes higher modes, which break the complete degeneracy (``regularizing'' the Fisher matrix) (c.f. Fig.~\ref{fig:SNRratiovsInc}). 

Now we repeat the above computation, but for the memory signal. Using the 0 PN waveform in Eq.~\eqref{eq:OPNres} we find that (in the frequency domain) the GW memory at the detector is
\begin{equation}
    \widetilde{\delta h}= \frac{\kappa_\mathrm{m} F^+}{d_\mathrm{L}}\sin^2\iota (17+\cos^2\iota),
\end{equation}
with a factor~$\kappa_ \mathrm{m}$ independent of both the distance and the inclination, and such that~$\kappa_0/\kappa_ \mathrm{m}\sim \mathcal{O}(100)$. The sky- and polarization-averaged Fisher matrix of the memory is
\begin{equation}
    \Gamma^{\mathrm{m}}_{i j}=\left(\frac{\rho_{\kappa_\mathrm{m}}}{d_\mathrm{L}}\right)^2 \hat{\Gamma}^\mathrm{m}_{i j},
\end{equation}
where~$\rho_{\kappa_\mathrm{m}}^2=(\kappa_\mathrm{m}|\kappa_\mathrm{m})$ and the matrix elements
\begin{align*}\label{eq:Fishermem}
    &\hat{\Gamma}^{\mathrm{m}}_{\log{d_\mathrm{L}},\log{d_\mathrm{L}}}= \frac{\sin^4\iota}{2}(17+\cos^2\iota)^2,\\
    &\hat{\Gamma}^{\mathrm{m}}_{\log{d_\mathrm{L}},\iota}= -\sin(2\iota)\sin^2\iota\,(8+\cos^2\iota)(17+\cos^2\iota),\\
    &\hat{\Gamma}^{\mathrm{m}}_{\iota,\iota}=2\sin^2(2\iota)(8+\cos^2\iota)^2, 
\end{align*}
depend only on the inclination.
Because of the simple structure of the memory signal (at leading order), its Fisher matrix is singular for all inclination angles. This is not an issue, since this singularity is cured through the inclusion of (subleading) higher modes of the memory. Contrarily to what happens with the primary signal, here the distance and the inclination are \emph{positively} (\emph{negatively}) correlated for inclination angles~$0<\iota<\frac{\pi}{2}$ ($\frac{\pi}{2}<\iota< \pi$). This opposite behavior is nicely illustrated by the orthogonality of the two confidence ellipses in Fig.~\ref{fig:Ellipses}. Note that although these results were derived for the 0 PN waveforms, this picture still holds generically, since it relies mostly on the leading dependence on the inclination.  

The Fisher matrix for the total (primary + memory) waveform $\Gamma^{\mathrm{tot}}$ includes additional cross-terms,
\begin{equation}
    \Gamma^{\mathrm{tot}}_{i j}=\Gamma^0_{i j}+\Gamma^{\mathrm{m}}_{i j}+(\partial_i\widetilde{\delta h}|\partial_j\tilde{h}_ 0)+(\partial_i\tilde{h}_ 0|\partial_j\widetilde{\delta h}).
\end{equation}
Above we focused on~$\Gamma^0_{i j}$ and~$\Gamma^{\mathrm{m}}_{i j}$. This is because we verified that, due to the rapid oscillations of the integrands in the cross-terms, the individual Fisher matrices dominate with respect to those.

\section{Numerical Fisher matrix}
\label{NumFIM}
To calculate the Fisher matrix elements we need to numerically compute derivatives of the waveform. We do so using a second-order finite differences,
\begin{equation}
    \frac{\partial\tilde{h}}{\partial \Theta_i}\approx \frac{\partial\tilde{h}(\Theta_i+\Delta\Theta_i)-\partial\tilde{h}(\Theta_i-\Delta\Theta_i)}{2\Delta\Theta_i},
\end{equation}
except for the luminosity distance, for which we have the exact result
\begin{equation}
    \frac{\partial\tilde{h}}{\partial \log d_\mathrm{L}}=-\tilde{h},
\end{equation}
since~$\tilde{h} \propto 1/d_\mathrm{L}$ (keeping~$M_z$ fixed).
We checked that our Fisher matrices are numerically stable in our region of interest in the parameter space for the increments: $\Delta\Theta_{M_z}= 10^{-3}M_\odot$, $\Delta\Theta_q= 10^{-4}$, $\Delta\Theta_{\iota,\varphi_\mathrm{c}}= 10^{-6}\, \mathrm{rad}$. Varying by an order of magnitude the finite increments gives just a few per cent change in the final matrix elements. We also checked that the Fisher matrices are stable by computing the derivatives with a higher-order finite differences method.   
We verified the reliability of our Fisher matrix inversion by confirming that in all cases,
\begin{equation}
    \mathrm{max}(|\Gamma_{ij}\Gamma^{-1}_{ij}-I_{ij}|)<10^{-6},
\end{equation}
where $I_{ij}$ is the identity matrix.

\bibliography{literature}

\begin{thebibliography}{129}%
\makeatletter
\providecommand \@ifxundefined [1]{%
 \@ifx{#1\undefined}
}%
\providecommand \@ifnum [1]{%
 \ifnum #1\expandafter \@firstoftwo
 \else \expandafter \@secondoftwo
 \fi
}%
\providecommand \@ifx [1]{%
 \ifx #1\expandafter \@firstoftwo
 \else \expandafter \@secondoftwo
 \fi
}%
\providecommand \natexlab [1]{#1}%
\providecommand \enquote  [1]{``#1''}%
\providecommand \bibnamefont  [1]{#1}%
\providecommand \bibfnamefont [1]{#1}%
\providecommand \citenamefont [1]{#1}%
\providecommand \href@noop [0]{\@secondoftwo}%
\providecommand \href [0]{\begingroup \@sanitize@url \@href}%
\providecommand \@href[1]{\@@startlink{#1}\@@href}%
\providecommand \@@href[1]{\endgroup#1\@@endlink}%
\providecommand \@sanitize@url [0]{\catcode `\\12\catcode `\$12\catcode
  `\&12\catcode `\#12\catcode `\^12\catcode `\_12\catcode `\%12\relax}%
\providecommand \@@startlink[1]{}%
\providecommand \@@endlink[0]{}%
\providecommand \url  [0]{\begingroup\@sanitize@url \@url }%
\providecommand \@url [1]{\endgroup\@href {#1}{\urlprefix }}%
\providecommand \urlprefix  [0]{URL }%
\providecommand \Eprint [0]{\href }%
\providecommand \doibase [0]{https://doi.org/}%
\providecommand \selectlanguage [0]{\@gobble}%
\providecommand \bibinfo  [0]{\@secondoftwo}%
\providecommand \bibfield  [0]{\@secondoftwo}%
\providecommand \translation [1]{[#1]}%
\providecommand \BibitemOpen [0]{}%
\providecommand \bibitemStop [0]{}%
\providecommand \bibitemNoStop [0]{.\EOS\space}%
\providecommand \EOS [0]{\spacefactor3000\relax}%
\providecommand \BibitemShut  [1]{\csname bibitem#1\endcsname}%
\let\auto@bib@innerbib\@empty
\bibitem [{\citenamefont {{Einstein}}(1916)}]{Einstein1916}%
  \BibitemOpen
  \bibfield  {author} {\bibinfo {author} {\bibfnamefont {A.}~\bibnamefont
  {{Einstein}}},\ }\bibfield  {title} {\bibinfo {title} {{N{\"a}herungsweise
  Integration der Feldgleichungen der Gravitation}},\ }\href@noop {} {\bibfield
   {journal} {\bibinfo  {journal} {Sitzber. Preuss. Akad. Wiss.}\ ,\ \bibinfo
  {pages} {688}} (\bibinfo {year} {1916})}\BibitemShut {NoStop}%
\bibitem [{\citenamefont {Abbott}\ \emph
  {et~al.}(2019{\natexlab{a}})\citenamefont {Abbott} \emph
  {et~al.}}]{LIGOScientific:2019fpa}%
  \BibitemOpen
  \bibfield  {author} {\bibinfo {author} {\bibfnamefont {B.~P.}\ \bibnamefont
  {Abbott}} \emph {et~al.} (\bibinfo {collaboration} {LIGO Scientific,
  Virgo}),\ }\bibfield  {title} {\bibinfo {title} {{Tests of General Relativity
  with the Binary Black Hole Signals from the LIGO-Virgo Catalog GWTC-1}},\
  }\href {https://doi.org/10.1103/PhysRevD.100.104036} {\bibfield  {journal}
  {\bibinfo  {journal} {Phys. Rev. D}\ }\textbf {\bibinfo {volume} {100}},\
  \bibinfo {pages} {104036} (\bibinfo {year} {2019}{\natexlab{a}})},\ \Eprint
  {https://arxiv.org/abs/1903.04467} {arXiv:1903.04467 [gr-qc]} \BibitemShut
  {NoStop}%
\bibitem [{\citenamefont {Abbott}\ \emph
  {et~al.}(2021{\natexlab{a}})\citenamefont {Abbott} \emph
  {et~al.}}]{LIGOScientific:2020tif}%
  \BibitemOpen
  \bibfield  {author} {\bibinfo {author} {\bibfnamefont {R.}~\bibnamefont
  {Abbott}} \emph {et~al.} (\bibinfo {collaboration} {LIGO Scientific,
  Virgo}),\ }\bibfield  {title} {\bibinfo {title} {{Tests of general relativity
  with binary black holes from the second LIGO-Virgo gravitational-wave
  transient catalog}},\ }\href {https://doi.org/10.1103/PhysRevD.103.122002}
  {\bibfield  {journal} {\bibinfo  {journal} {Phys. Rev. D}\ }\textbf {\bibinfo
  {volume} {103}},\ \bibinfo {pages} {122002} (\bibinfo {year}
  {2021}{\natexlab{a}})},\ \Eprint {https://arxiv.org/abs/2010.14529}
  {arXiv:2010.14529 [gr-qc]} \BibitemShut {NoStop}%
\bibitem [{\citenamefont {Abbott}\ \emph
  {et~al.}(2021{\natexlab{b}})\citenamefont {Abbott} \emph
  {et~al.}}]{LIGOScientific:2021sio}%
  \BibitemOpen
  \bibfield  {author} {\bibinfo {author} {\bibfnamefont {R.}~\bibnamefont
  {Abbott}} \emph {et~al.} (\bibinfo {collaboration} {LIGO Scientific, VIRGO,
  KAGRA}),\ }\bibfield  {title} {\bibinfo {title} {{Tests of General Relativity
  with GWTC-3}},\ }\href@noop {} {\bibfield  {journal} {\bibinfo  {journal}
  {arXiv e-prints}\ } (\bibinfo {year} {2021}{\natexlab{b}})},\ \Eprint
  {https://arxiv.org/abs/2112.06861} {arXiv:2112.06861 [gr-qc]} \BibitemShut
  {NoStop}%
\bibitem [{\citenamefont {Abbott}\ \emph
  {et~al.}(2019{\natexlab{b}})\citenamefont {Abbott} \emph
  {et~al.}}]{LIGOScientific:2018mvr}%
  \BibitemOpen
  \bibfield  {author} {\bibinfo {author} {\bibfnamefont {B.~P.}\ \bibnamefont
  {Abbott}} \emph {et~al.} (\bibinfo {collaboration} {LIGO Scientific,
  Virgo}),\ }\bibfield  {title} {\bibinfo {title} {{GWTC-1: A
  Gravitational-Wave Transient Catalog of Compact Binary Mergers Observed by
  LIGO and Virgo during the First and Second Observing Runs}},\ }\href
  {https://doi.org/10.1103/PhysRevX.9.031040} {\bibfield  {journal} {\bibinfo
  {journal} {Phys. Rev. X}\ }\textbf {\bibinfo {volume} {9}},\ \bibinfo {pages}
  {031040} (\bibinfo {year} {2019}{\natexlab{b}})},\ \Eprint
  {https://arxiv.org/abs/1811.12907} {arXiv:1811.12907 [astro-ph.HE]}
  \BibitemShut {NoStop}%
\bibitem [{\citenamefont {Abbott}\ \emph
  {et~al.}(2021{\natexlab{c}})\citenamefont {Abbott} \emph
  {et~al.}}]{LIGOScientific:2020ibl}%
  \BibitemOpen
  \bibfield  {author} {\bibinfo {author} {\bibfnamefont {R.}~\bibnamefont
  {Abbott}} \emph {et~al.} (\bibinfo {collaboration} {LIGO Scientific,
  Virgo}),\ }\bibfield  {title} {\bibinfo {title} {{GWTC-2: Compact Binary
  Coalescences Observed by LIGO and Virgo During the First Half of the Third
  Observing Run}},\ }\href {https://doi.org/10.1103/PhysRevX.11.021053}
  {\bibfield  {journal} {\bibinfo  {journal} {Phys. Rev. X}\ }\textbf {\bibinfo
  {volume} {11}},\ \bibinfo {pages} {021053} (\bibinfo {year}
  {2021}{\natexlab{c}})},\ \Eprint {https://arxiv.org/abs/2010.14527}
  {arXiv:2010.14527 [gr-qc]} \BibitemShut {NoStop}%
\bibitem [{\citenamefont {Abbott}\ \emph
  {et~al.}(2021{\natexlab{d}})\citenamefont {Abbott} \emph
  {et~al.}}]{LIGOScientific:2021djp}%
  \BibitemOpen
  \bibfield  {author} {\bibinfo {author} {\bibfnamefont {R.}~\bibnamefont
  {Abbott}} \emph {et~al.} (\bibinfo {collaboration} {LIGO Scientific, VIRGO,
  KAGRA}),\ }\bibfield  {title} {\bibinfo {title} {{GWTC-3: Compact Binary
  Coalescences Observed by LIGO and Virgo During the Second Part of the Third
  Observing Run}},\ }\href@noop {} {\bibfield  {journal} {\bibinfo  {journal}
  {arXiv e-prints}\ } (\bibinfo {year} {2021}{\natexlab{d}})},\ \Eprint
  {https://arxiv.org/abs/2111.03606} {arXiv:2111.03606 [gr-qc]} \BibitemShut
  {NoStop}%
\bibitem [{\citenamefont {Berti}\ \emph {et~al.}(2015)\citenamefont {Berti}
  \emph {et~al.}}]{Berti:2015itd}%
  \BibitemOpen
  \bibfield  {author} {\bibinfo {author} {\bibfnamefont {E.}~\bibnamefont
  {Berti}} \emph {et~al.},\ }\bibfield  {title} {\bibinfo {title} {{Testing
  General Relativity with Present and Future Astrophysical Observations}},\
  }\href {https://doi.org/10.1088/0264-9381/32/24/243001} {\bibfield  {journal}
  {\bibinfo  {journal} {Class. Quant. Grav.}\ }\textbf {\bibinfo {volume}
  {32}},\ \bibinfo {pages} {243001} (\bibinfo {year} {2015})},\ \Eprint
  {https://arxiv.org/abs/1501.07274} {arXiv:1501.07274 [gr-qc]} \BibitemShut
  {NoStop}%
\bibitem [{\citenamefont {Barack}\ \emph {et~al.}(2019)\citenamefont {Barack}
  \emph {et~al.}}]{Barack:2018yly}%
  \BibitemOpen
  \bibfield  {author} {\bibinfo {author} {\bibfnamefont {L.}~\bibnamefont
  {Barack}} \emph {et~al.},\ }\bibfield  {title} {\bibinfo {title} {{Black
  holes, gravitational waves and fundamental physics: a roadmap}},\ }\href
  {https://doi.org/10.1088/1361-6382/ab0587} {\bibfield  {journal} {\bibinfo
  {journal} {Class. Quant. Grav.}\ }\textbf {\bibinfo {volume} {36}},\ \bibinfo
  {pages} {143001} (\bibinfo {year} {2019})},\ \Eprint
  {https://arxiv.org/abs/1806.05195} {arXiv:1806.05195 [gr-qc]} \BibitemShut
  {NoStop}%
\bibitem [{\citenamefont {Barausse}\ \emph
  {et~al.}(2020{\natexlab{a}})\citenamefont {Barausse} \emph
  {et~al.}}]{manifesto}%
  \BibitemOpen
  \bibfield  {author} {\bibinfo {author} {\bibfnamefont {E.}~\bibnamefont
  {Barausse}} \emph {et~al.},\ }\bibfield  {title} {\bibinfo {title}
  {{Prospects for Fundamental Physics with LISA}},\ }\href
  {https://doi.org/10.1007/s10714-020-02691-1} {\bibfield  {journal} {\bibinfo
  {journal} {Gen. Rel. Grav.}\ }\textbf {\bibinfo {volume} {52}},\ \bibinfo
  {pages} {81} (\bibinfo {year} {2020}{\natexlab{a}})},\ \Eprint
  {https://arxiv.org/abs/2001.09793} {arXiv:2001.09793 [gr-qc]} \BibitemShut
  {NoStop}%
\bibitem [{\citenamefont {{Amaro-Seoane}}\ \emph {et~al.}(2017)\citenamefont
  {{Amaro-Seoane}} \emph {et~al.}}]{Amaro-Seoane2017}%
  \BibitemOpen
  \bibfield  {author} {\bibinfo {author} {\bibfnamefont {P.}~\bibnamefont
  {{Amaro-Seoane}}} \emph {et~al.},\ }\bibfield  {title} {\bibinfo {title}
  {{Laser Interferometer Space Antenna}},\ }\href@noop {} {\bibfield  {journal}
  {\bibinfo  {journal} {arXiv e-prints}\ } (\bibinfo {year} {2017})},\ \Eprint
  {https://arxiv.org/abs/1702.00786} {arXiv:1702.00786 [astro-ph.IM]}
  \BibitemShut {NoStop}%
\bibitem [{\citenamefont {Seoane}\ \emph {et~al.}(2013)\citenamefont {Seoane}
  \emph {et~al.}}]{eLISA:2013xep}%
  \BibitemOpen
  \bibfield  {author} {\bibinfo {author} {\bibfnamefont {P.~A.}\ \bibnamefont
  {Seoane}} \emph {et~al.} (\bibinfo {collaboration} {eLISA}),\ }\bibfield
  {title} {\bibinfo {title} {{The Gravitational Universe}},\ }\href@noop {}
  {\bibfield  {journal} {\bibinfo  {journal} {arXiv e-prints}\ } (\bibinfo
  {year} {2013})},\ \Eprint {https://arxiv.org/abs/1305.5720} {arXiv:1305.5720
  [astro-ph.CO]} \BibitemShut {NoStop}%
\bibitem [{\citenamefont {Baibhav}\ \emph {et~al.}(2021)\citenamefont {Baibhav}
  \emph {et~al.}}]{Baibhav:2019rsa}%
  \BibitemOpen
  \bibfield  {author} {\bibinfo {author} {\bibfnamefont {V.}~\bibnamefont
  {Baibhav}} \emph {et~al.},\ }\bibfield  {title} {\bibinfo {title} {{Probing
  the nature of black holes: Deep in the mHz gravitational-wave sky}},\ }\href
  {https://doi.org/10.1007/s10686-021-09741-9} {\bibfield  {journal} {\bibinfo
  {journal} {Exper. Astron.}\ }\textbf {\bibinfo {volume} {51}},\ \bibinfo
  {pages} {1385} (\bibinfo {year} {2021})},\ \Eprint
  {https://arxiv.org/abs/1908.11390} {arXiv:1908.11390 [astro-ph.HE]}
  \BibitemShut {NoStop}%
\bibitem [{\citenamefont {Arun}\ \emph {et~al.}(2022)\citenamefont {Arun} \emph
  {et~al.}}]{LISA:2022kgy}%
  \BibitemOpen
  \bibfield  {author} {\bibinfo {author} {\bibfnamefont {K.~G.}\ \bibnamefont
  {Arun}} \emph {et~al.} (\bibinfo {collaboration} {LISA}),\ }\bibfield
  {title} {\bibinfo {title} {{New horizons for fundamental physics with
  LISA}},\ }\href {https://doi.org/10.1007/s41114-022-00036-9} {\bibfield
  {journal} {\bibinfo  {journal} {Living Rev. Rel.}\ }\textbf {\bibinfo
  {volume} {25}},\ \bibinfo {pages} {4} (\bibinfo {year} {2022})},\ \Eprint
  {https://arxiv.org/abs/2205.01597} {arXiv:2205.01597 [gr-qc]} \BibitemShut
  {NoStop}%
\bibitem [{\citenamefont {Cutler}\ and\ \citenamefont
  {Flanagan}(1994)}]{Cutler:1994ys}%
  \BibitemOpen
  \bibfield  {author} {\bibinfo {author} {\bibfnamefont {C.}~\bibnamefont
  {Cutler}}\ and\ \bibinfo {author} {\bibfnamefont {E.~E.}\ \bibnamefont
  {Flanagan}},\ }\bibfield  {title} {\bibinfo {title} {{Gravitational waves
  from merging compact binaries: How accurately can one extract the binary's
  parameters from the inspiral wave form?}},\ }\href
  {https://doi.org/10.1103/PhysRevD.49.2658} {\bibfield  {journal} {\bibinfo
  {journal} {Phys. Rev. D}\ }\textbf {\bibinfo {volume} {49}},\ \bibinfo
  {pages} {2658} (\bibinfo {year} {1994})},\ \Eprint
  {https://arxiv.org/abs/gr-qc/9402014} {arXiv:gr-qc/9402014} \BibitemShut
  {NoStop}%
\bibitem [{\citenamefont {Usman}\ \emph {et~al.}(2019)\citenamefont {Usman},
  \citenamefont {Mills},\ and\ \citenamefont {Fairhurst}}]{Usman:2018imj}%
  \BibitemOpen
  \bibfield  {author} {\bibinfo {author} {\bibfnamefont {S.~A.}\ \bibnamefont
  {Usman}}, \bibinfo {author} {\bibfnamefont {J.~C.}\ \bibnamefont {Mills}},\
  and\ \bibinfo {author} {\bibfnamefont {S.}~\bibnamefont {Fairhurst}},\
  }\bibfield  {title} {\bibinfo {title} {{Constraining the Inclinations of
  Binary Mergers from Gravitational-wave Observations}},\ }\href
  {https://doi.org/10.3847/1538-4357/ab0b3e} {\bibfield  {journal} {\bibinfo
  {journal} {Astrophys. J.}\ }\textbf {\bibinfo {volume} {877}},\ \bibinfo
  {pages} {82} (\bibinfo {year} {2019})},\ \Eprint
  {https://arxiv.org/abs/1809.10727} {arXiv:1809.10727 [gr-qc]} \BibitemShut
  {NoStop}%
\bibitem [{\citenamefont {Chassande-Mottin}\ \emph {et~al.}(2019)\citenamefont
  {Chassande-Mottin}, \citenamefont {Leyde}, \citenamefont {Mastrogiovanni},\
  and\ \citenamefont {Steer}}]{Chassande-Mottin:2019nnz}%
  \BibitemOpen
  \bibfield  {author} {\bibinfo {author} {\bibfnamefont {E.}~\bibnamefont
  {Chassande-Mottin}}, \bibinfo {author} {\bibfnamefont {K.}~\bibnamefont
  {Leyde}}, \bibinfo {author} {\bibfnamefont {S.}~\bibnamefont
  {Mastrogiovanni}},\ and\ \bibinfo {author} {\bibfnamefont {D.~A.}\
  \bibnamefont {Steer}},\ }\bibfield  {title} {\bibinfo {title} {{Gravitational
  wave observations, distance measurement uncertainties, and cosmology}},\
  }\href {https://doi.org/10.1103/PhysRevD.100.083514} {\bibfield  {journal}
  {\bibinfo  {journal} {Phys. Rev. D}\ }\textbf {\bibinfo {volume} {100}},\
  \bibinfo {pages} {083514} (\bibinfo {year} {2019})},\ \Eprint
  {https://arxiv.org/abs/1906.02670} {arXiv:1906.02670 [astro-ph.CO]}
  \BibitemShut {NoStop}%
\bibitem [{\citenamefont {Schutz}(1986)}]{Schutz:1986gp}%
  \BibitemOpen
  \bibfield  {author} {\bibinfo {author} {\bibfnamefont {B.~F.}\ \bibnamefont
  {Schutz}},\ }\bibfield  {title} {\bibinfo {title} {{Determining the Hubble
  Constant from Gravitational Wave Observations}},\ }\href
  {https://doi.org/10.1038/323310a0} {\bibfield  {journal} {\bibinfo  {journal}
  {Nature}\ }\textbf {\bibinfo {volume} {323}},\ \bibinfo {pages} {310}
  (\bibinfo {year} {1986})}\BibitemShut {NoStop}%
\bibitem [{\citenamefont {Holz}\ and\ \citenamefont
  {Hughes}(2005)}]{Holz:2005df}%
  \BibitemOpen
  \bibfield  {author} {\bibinfo {author} {\bibfnamefont {D.~E.}\ \bibnamefont
  {Holz}}\ and\ \bibinfo {author} {\bibfnamefont {S.~A.}\ \bibnamefont
  {Hughes}},\ }\bibfield  {title} {\bibinfo {title} {{Using gravitational-wave
  standard sirens}},\ }\href {https://doi.org/10.1086/431341} {\bibfield
  {journal} {\bibinfo  {journal} {Astrophys. J.}\ }\textbf {\bibinfo {volume}
  {629}},\ \bibinfo {pages} {15} (\bibinfo {year} {2005})},\ \Eprint
  {https://arxiv.org/abs/astro-ph/0504616} {arXiv:astro-ph/0504616}
  \BibitemShut {NoStop}%
\bibitem [{\citenamefont {Nissanke}\ \emph {et~al.}(2010)\citenamefont
  {Nissanke}, \citenamefont {Holz}, \citenamefont {Hughes}, \citenamefont
  {Dalal},\ and\ \citenamefont {Sievers}}]{Nissanke:2009kt}%
  \BibitemOpen
  \bibfield  {author} {\bibinfo {author} {\bibfnamefont {S.}~\bibnamefont
  {Nissanke}}, \bibinfo {author} {\bibfnamefont {D.~E.}\ \bibnamefont {Holz}},
  \bibinfo {author} {\bibfnamefont {S.~A.}\ \bibnamefont {Hughes}}, \bibinfo
  {author} {\bibfnamefont {N.}~\bibnamefont {Dalal}},\ and\ \bibinfo {author}
  {\bibfnamefont {J.~L.}\ \bibnamefont {Sievers}},\ }\bibfield  {title}
  {\bibinfo {title} {{Exploring short gamma-ray bursts as gravitational-wave
  standard sirens}},\ }\href {https://doi.org/10.1088/0004-637X/725/1/496}
  {\bibfield  {journal} {\bibinfo  {journal} {Astrophys. J.}\ }\textbf
  {\bibinfo {volume} {725}},\ \bibinfo {pages} {496} (\bibinfo {year}
  {2010})},\ \Eprint {https://arxiv.org/abs/0904.1017} {arXiv:0904.1017
  [astro-ph.CO]} \BibitemShut {NoStop}%
\bibitem [{\citenamefont {Tamanini}\ \emph {et~al.}(2016)\citenamefont
  {Tamanini}, \citenamefont {Caprini}, \citenamefont {Barausse}, \citenamefont
  {Sesana}, \citenamefont {Klein},\ and\ \citenamefont
  {Petiteau}}]{Tamanini:2016zlh}%
  \BibitemOpen
  \bibfield  {author} {\bibinfo {author} {\bibfnamefont {N.}~\bibnamefont
  {Tamanini}}, \bibinfo {author} {\bibfnamefont {C.}~\bibnamefont {Caprini}},
  \bibinfo {author} {\bibfnamefont {E.}~\bibnamefont {Barausse}}, \bibinfo
  {author} {\bibfnamefont {A.}~\bibnamefont {Sesana}}, \bibinfo {author}
  {\bibfnamefont {A.}~\bibnamefont {Klein}},\ and\ \bibinfo {author}
  {\bibfnamefont {A.}~\bibnamefont {Petiteau}},\ }\bibfield  {title} {\bibinfo
  {title} {{Science with the space-based interferometer eLISA. III: Probing the
  expansion of the Universe using gravitational wave standard sirens}},\ }\href
  {https://doi.org/10.1088/1475-7516/2016/04/002} {\bibfield  {journal}
  {\bibinfo  {journal} {JCAP}\ }\textbf {\bibinfo {volume} {04}},\ \bibinfo
  {pages} {002}},\ \Eprint {https://arxiv.org/abs/1601.07112} {arXiv:1601.07112
  [astro-ph.CO]} \BibitemShut {NoStop}%
\bibitem [{\citenamefont {Abbott}\ \emph {et~al.}(2017)\citenamefont {Abbott}
  \emph {et~al.}}]{LIGOScientific:2017adf}%
  \BibitemOpen
  \bibfield  {author} {\bibinfo {author} {\bibfnamefont {B.~P.}\ \bibnamefont
  {Abbott}} \emph {et~al.} (\bibinfo {collaboration} {LIGO Scientific, Virgo,
  1M2H, Dark Energy Camera GW-E, DES, DLT40, Las Cumbres Observatory, VINROUGE,
  MASTER}),\ }\bibfield  {title} {\bibinfo {title} {{A gravitational-wave
  standard siren measurement of the Hubble constant}},\ }\href
  {https://doi.org/10.1038/nature24471} {\bibfield  {journal} {\bibinfo
  {journal} {Nature}\ }\textbf {\bibinfo {volume} {551}},\ \bibinfo {pages}
  {85} (\bibinfo {year} {2017})},\ \Eprint {https://arxiv.org/abs/1710.05835}
  {arXiv:1710.05835 [astro-ph.CO]} \BibitemShut {NoStop}%
\bibitem [{\citenamefont {Poisson}\ and\ \citenamefont
  {Will}(2014)}]{poisson_will_2014}%
  \BibitemOpen
  \bibfield  {author} {\bibinfo {author} {\bibfnamefont {E.}~\bibnamefont
  {Poisson}}\ and\ \bibinfo {author} {\bibfnamefont {C.~M.}\ \bibnamefont
  {Will}},\ }\bibinfo {title} {{Gravity: Newtonian, Post-Newtonian,
  Relativistic}}\ (\bibinfo  {publisher} {{Cambridge University Press}},\
  \bibinfo {year} {2014})\BibitemShut {NoStop}%
\bibitem [{\citenamefont {Hotokezaka}\ \emph {et~al.}(2019)\citenamefont
  {Hotokezaka}, \citenamefont {Nakar}, \citenamefont {Gottlieb}, \citenamefont
  {Nissanke}, \citenamefont {Masuda}, \citenamefont {Hallinan}, \citenamefont
  {Mooley},\ and\ \citenamefont {Deller}}]{Hotokezaka:2018dfi}%
  \BibitemOpen
  \bibfield  {author} {\bibinfo {author} {\bibfnamefont {K.}~\bibnamefont
  {Hotokezaka}}, \bibinfo {author} {\bibfnamefont {E.}~\bibnamefont {Nakar}},
  \bibinfo {author} {\bibfnamefont {O.}~\bibnamefont {Gottlieb}}, \bibinfo
  {author} {\bibfnamefont {S.}~\bibnamefont {Nissanke}}, \bibinfo {author}
  {\bibfnamefont {K.}~\bibnamefont {Masuda}}, \bibinfo {author} {\bibfnamefont
  {G.}~\bibnamefont {Hallinan}}, \bibinfo {author} {\bibfnamefont {K.~P.}\
  \bibnamefont {Mooley}},\ and\ \bibinfo {author} {\bibfnamefont {A.~T.}\
  \bibnamefont {Deller}},\ }\bibfield  {title} {\bibinfo {title} {{A Hubble
  constant measurement from superluminal motion of the jet in GW170817}},\
  }\href {https://doi.org/10.1038/s41550-019-0820-1} {\bibfield  {journal}
  {\bibinfo  {journal} {Nature Astron.}\ }\textbf {\bibinfo {volume} {3}},\
  \bibinfo {pages} {940} (\bibinfo {year} {2019})},\ \Eprint
  {https://arxiv.org/abs/1806.10596} {arXiv:1806.10596 [astro-ph.CO]}
  \BibitemShut {NoStop}%
\bibitem [{\citenamefont {Chen}\ \emph {et~al.}(2019)\citenamefont {Chen},
  \citenamefont {Vitale},\ and\ \citenamefont {Narayan}}]{Chen:2018omi}%
  \BibitemOpen
  \bibfield  {author} {\bibinfo {author} {\bibfnamefont {H.-Y.}\ \bibnamefont
  {Chen}}, \bibinfo {author} {\bibfnamefont {S.}~\bibnamefont {Vitale}},\ and\
  \bibinfo {author} {\bibfnamefont {R.}~\bibnamefont {Narayan}},\ }\bibfield
  {title} {\bibinfo {title} {{Viewing angle of binary neutron star mergers}},\
  }\href {https://doi.org/10.1103/PhysRevX.9.031028} {\bibfield  {journal}
  {\bibinfo  {journal} {Phys. Rev. X}\ }\textbf {\bibinfo {volume} {9}},\
  \bibinfo {pages} {031028} (\bibinfo {year} {2019})},\ \Eprint
  {https://arxiv.org/abs/1807.05226} {arXiv:1807.05226 [astro-ph.HE]}
  \BibitemShut {NoStop}%
\bibitem [{\citenamefont {Lang}\ and\ \citenamefont
  {Hughes}(2006)}]{Lang:2006bsg}%
  \BibitemOpen
  \bibfield  {author} {\bibinfo {author} {\bibfnamefont {R.~N.}\ \bibnamefont
  {Lang}}\ and\ \bibinfo {author} {\bibfnamefont {S.~A.}\ \bibnamefont
  {Hughes}},\ }\bibfield  {title} {\bibinfo {title} {{Measuring coalescing
  massive binary black holes with gravitational waves: The Impact of
  spin-induced precession}},\ }\href
  {https://doi.org/10.1103/PhysRevD.75.089902} {\bibfield  {journal} {\bibinfo
  {journal} {Phys. Rev. D}\ }\textbf {\bibinfo {volume} {74}},\ \bibinfo
  {pages} {122001} (\bibinfo {year} {2006})},\ \bibinfo {note} {[Erratum:
  Phys.Rev.D 75, 089902 (2007), Erratum: Phys.Rev.D 77, 109901 (2008)]},\
  \Eprint {https://arxiv.org/abs/gr-qc/0608062} {arXiv:gr-qc/0608062}
  \BibitemShut {NoStop}%
\bibitem [{\citenamefont {Apostolatos}\ \emph {et~al.}(1994)\citenamefont
  {Apostolatos}, \citenamefont {Cutler}, \citenamefont {Sussman},\ and\
  \citenamefont {Thorne}}]{Apostolatos:1994mx}%
  \BibitemOpen
  \bibfield  {author} {\bibinfo {author} {\bibfnamefont {T.~A.}\ \bibnamefont
  {Apostolatos}}, \bibinfo {author} {\bibfnamefont {C.}~\bibnamefont {Cutler}},
  \bibinfo {author} {\bibfnamefont {G.~J.}\ \bibnamefont {Sussman}},\ and\
  \bibinfo {author} {\bibfnamefont {K.~S.}\ \bibnamefont {Thorne}},\ }\bibfield
   {title} {\bibinfo {title} {{Spin induced orbital precession and its
  modulation of the gravitational wave forms from merging binaries}},\ }\href
  {https://doi.org/10.1103/PhysRevD.49.6274} {\bibfield  {journal} {\bibinfo
  {journal} {Phys. Rev. D}\ }\textbf {\bibinfo {volume} {49}},\ \bibinfo
  {pages} {6274} (\bibinfo {year} {1994})}\BibitemShut {NoStop}%
\bibitem [{\citenamefont {Lang}\ \emph {et~al.}(2011)\citenamefont {Lang},
  \citenamefont {Hughes},\ and\ \citenamefont {Cornish}}]{Lang:2011je}%
  \BibitemOpen
  \bibfield  {author} {\bibinfo {author} {\bibfnamefont {R.~N.}\ \bibnamefont
  {Lang}}, \bibinfo {author} {\bibfnamefont {S.~A.}\ \bibnamefont {Hughes}},\
  and\ \bibinfo {author} {\bibfnamefont {N.~J.}\ \bibnamefont {Cornish}},\
  }\bibfield  {title} {\bibinfo {title} {{Measuring parameters of massive black
  hole binaries with partially aligned spins}},\ }\href
  {https://doi.org/10.1103/PhysRevD.84.022002} {\bibfield  {journal} {\bibinfo
  {journal} {Phys. Rev. D}\ }\textbf {\bibinfo {volume} {84}},\ \bibinfo
  {pages} {022002} (\bibinfo {year} {2011})},\ \Eprint
  {https://arxiv.org/abs/1101.3591} {arXiv:1101.3591 [gr-qc]} \BibitemShut
  {NoStop}%
\bibitem [{\citenamefont {Arun}\ \emph {et~al.}(2007)\citenamefont {Arun},
  \citenamefont {Iyer}, \citenamefont {Sathyaprakash},\ and\ \citenamefont
  {Sinha}}]{Arun:2007qv}%
  \BibitemOpen
  \bibfield  {author} {\bibinfo {author} {\bibfnamefont {K.~G.}\ \bibnamefont
  {Arun}}, \bibinfo {author} {\bibfnamefont {B.~R.}\ \bibnamefont {Iyer}},
  \bibinfo {author} {\bibfnamefont {B.~S.}\ \bibnamefont {Sathyaprakash}},\
  and\ \bibinfo {author} {\bibfnamefont {S.}~\bibnamefont {Sinha}},\ }\bibfield
   {title} {\bibinfo {title} {{Higher harmonics increase LISA's mass reach for
  supermassive black holes}},\ }\href
  {https://doi.org/10.1103/PhysRevD.75.124002} {\bibfield  {journal} {\bibinfo
  {journal} {Phys. Rev. D}\ }\textbf {\bibinfo {volume} {75}},\ \bibinfo
  {pages} {124002} (\bibinfo {year} {2007})},\ \Eprint
  {https://arxiv.org/abs/0704.1086} {arXiv:0704.1086 [gr-qc]} \BibitemShut
  {NoStop}%
\bibitem [{\citenamefont {Porter}\ and\ \citenamefont
  {Cornish}(2008)}]{Porter:2008kn}%
  \BibitemOpen
  \bibfield  {author} {\bibinfo {author} {\bibfnamefont {E.~K.}\ \bibnamefont
  {Porter}}\ and\ \bibinfo {author} {\bibfnamefont {N.~J.}\ \bibnamefont
  {Cornish}},\ }\bibfield  {title} {\bibinfo {title} {{The Effect of Higher
  Harmonic Corrections on the Detection of massive black hole binaries with
  LISA}},\ }\href {https://doi.org/10.1103/PhysRevD.78.064005} {\bibfield
  {journal} {\bibinfo  {journal} {Phys. Rev. D}\ }\textbf {\bibinfo {volume}
  {78}},\ \bibinfo {pages} {064005} (\bibinfo {year} {2008})},\ \Eprint
  {https://arxiv.org/abs/0804.0332} {arXiv:0804.0332 [gr-qc]} \BibitemShut
  {NoStop}%
\bibitem [{\citenamefont {Trias}\ and\ \citenamefont
  {Sintes}(2008)}]{Trias:2007fp}%
  \BibitemOpen
  \bibfield  {author} {\bibinfo {author} {\bibfnamefont {M.}~\bibnamefont
  {Trias}}\ and\ \bibinfo {author} {\bibfnamefont {A.~M.}\ \bibnamefont
  {Sintes}},\ }\bibfield  {title} {\bibinfo {title} {{LISA observations of
  supermassive black holes: Parameter estimation using full post-Newtonian
  inspiral waveforms}},\ }\href {https://doi.org/10.1103/PhysRevD.77.024030}
  {\bibfield  {journal} {\bibinfo  {journal} {Phys. Rev. D}\ }\textbf {\bibinfo
  {volume} {77}},\ \bibinfo {pages} {024030} (\bibinfo {year} {2008})},\
  \Eprint {https://arxiv.org/abs/0707.4434} {arXiv:0707.4434 [gr-qc]}
  \BibitemShut {NoStop}%
\bibitem [{\citenamefont {Klein}\ \emph {et~al.}(2009)\citenamefont {Klein},
  \citenamefont {Jetzer},\ and\ \citenamefont {Sereno}}]{Klein:2009gza}%
  \BibitemOpen
  \bibfield  {author} {\bibinfo {author} {\bibfnamefont {A.}~\bibnamefont
  {Klein}}, \bibinfo {author} {\bibfnamefont {P.}~\bibnamefont {Jetzer}},\ and\
  \bibinfo {author} {\bibfnamefont {M.}~\bibnamefont {Sereno}},\ }\bibfield
  {title} {\bibinfo {title} {{Parameter estimation for coalescing massive
  binary black holes with LISA using the full 2PN gravitational waveform and
  spin-orbit precession}},\ }\href {https://doi.org/10.1103/PhysRevD.80.064027}
  {\bibfield  {journal} {\bibinfo  {journal} {Phys. Rev. D}\ }\textbf {\bibinfo
  {volume} {80}},\ \bibinfo {pages} {064027} (\bibinfo {year} {2009})},\
  \Eprint {https://arxiv.org/abs/0907.3318} {arXiv:0907.3318 [astro-ph.CO]}
  \BibitemShut {NoStop}%
\bibitem [{\citenamefont {London}\ \emph {et~al.}(2018)\citenamefont {London},
  \citenamefont {Khan}, \citenamefont {Fauchon-Jones}, \citenamefont
  {Garc\'\i{}a}, \citenamefont {Hannam}, \citenamefont {Husa}, \citenamefont
  {Jim\'enez-Forteza}, \citenamefont {Kalaghatgi}, \citenamefont {Ohme},\ and\
  \citenamefont {Pannarale}}]{London:2017bcn}%
  \BibitemOpen
  \bibfield  {author} {\bibinfo {author} {\bibfnamefont {L.}~\bibnamefont
  {London}}, \bibinfo {author} {\bibfnamefont {S.}~\bibnamefont {Khan}},
  \bibinfo {author} {\bibfnamefont {E.}~\bibnamefont {Fauchon-Jones}}, \bibinfo
  {author} {\bibfnamefont {C.}~\bibnamefont {Garc\'\i{}a}}, \bibinfo {author}
  {\bibfnamefont {M.}~\bibnamefont {Hannam}}, \bibinfo {author} {\bibfnamefont
  {S.}~\bibnamefont {Husa}}, \bibinfo {author} {\bibfnamefont {X.}~\bibnamefont
  {Jim\'enez-Forteza}}, \bibinfo {author} {\bibfnamefont {C.}~\bibnamefont
  {Kalaghatgi}}, \bibinfo {author} {\bibfnamefont {F.}~\bibnamefont {Ohme}},\
  and\ \bibinfo {author} {\bibfnamefont {F.}~\bibnamefont {Pannarale}},\
  }\bibfield  {title} {\bibinfo {title} {{First higher-multipole model of
  gravitational waves from spinning and coalescing black-hole binaries}},\
  }\href {https://doi.org/10.1103/PhysRevLett.120.161102} {\bibfield  {journal}
  {\bibinfo  {journal} {Phys. Rev. Lett.}\ }\textbf {\bibinfo {volume} {120}},\
  \bibinfo {pages} {161102} (\bibinfo {year} {2018})},\ \Eprint
  {https://arxiv.org/abs/1708.00404} {arXiv:1708.00404 [gr-qc]} \BibitemShut
  {NoStop}%
\bibitem [{\citenamefont {Borhanian}\ \emph {et~al.}(2020)\citenamefont
  {Borhanian}, \citenamefont {Dhani}, \citenamefont {Gupta}, \citenamefont
  {Arun},\ and\ \citenamefont {Sathyaprakash}}]{Borhanian:2020vyr}%
  \BibitemOpen
  \bibfield  {author} {\bibinfo {author} {\bibfnamefont {S.}~\bibnamefont
  {Borhanian}}, \bibinfo {author} {\bibfnamefont {A.}~\bibnamefont {Dhani}},
  \bibinfo {author} {\bibfnamefont {A.}~\bibnamefont {Gupta}}, \bibinfo
  {author} {\bibfnamefont {K.~G.}\ \bibnamefont {Arun}},\ and\ \bibinfo
  {author} {\bibfnamefont {B.~S.}\ \bibnamefont {Sathyaprakash}},\ }\bibfield
  {title} {\bibinfo {title} {{Dark Sirens to Resolve the
  Hubble\textendash{}Lema\^\i{}tre Tension}},\ }\href
  {https://doi.org/10.3847/2041-8213/abcaf5} {\bibfield  {journal} {\bibinfo
  {journal} {Astrophys. J. Lett.}\ }\textbf {\bibinfo {volume} {905}},\
  \bibinfo {pages} {L28} (\bibinfo {year} {2020})},\ \Eprint
  {https://arxiv.org/abs/2007.02883} {arXiv:2007.02883 [astro-ph.CO]}
  \BibitemShut {NoStop}%
\bibitem [{\citenamefont {Yang}\ \emph
  {et~al.}(2022{\natexlab{a}})\citenamefont {Yang}, \citenamefont {Cai},
  \citenamefont {Cao},\ and\ \citenamefont {Lee}}]{Yang:2022tig}%
  \BibitemOpen
  \bibfield  {author} {\bibinfo {author} {\bibfnamefont {T.}~\bibnamefont
  {Yang}}, \bibinfo {author} {\bibfnamefont {R.-G.}\ \bibnamefont {Cai}},
  \bibinfo {author} {\bibfnamefont {Z.}~\bibnamefont {Cao}},\ and\ \bibinfo
  {author} {\bibfnamefont {H.~M.}\ \bibnamefont {Lee}},\ }\bibfield  {title}
  {\bibinfo {title} {{Eccentricity of Long Inspiraling Compact Binaries Sheds
  Light on Dark Sirens}},\ }\href
  {https://doi.org/10.1103/PhysRevLett.129.191102} {\bibfield  {journal}
  {\bibinfo  {journal} {Phys. Rev. Lett.}\ }\textbf {\bibinfo {volume} {129}},\
  \bibinfo {pages} {191102} (\bibinfo {year} {2022}{\natexlab{a}})},\ \Eprint
  {https://arxiv.org/abs/2202.08608} {arXiv:2202.08608 [gr-qc]} \BibitemShut
  {NoStop}%
\bibitem [{\citenamefont {Yang}\ \emph
  {et~al.}(2022{\natexlab{b}})\citenamefont {Yang}, \citenamefont {Cai},\ and\
  \citenamefont {Lee}}]{Yang:2022iwn}%
  \BibitemOpen
  \bibfield  {author} {\bibinfo {author} {\bibfnamefont {T.}~\bibnamefont
  {Yang}}, \bibinfo {author} {\bibfnamefont {R.-G.}\ \bibnamefont {Cai}},\ and\
  \bibinfo {author} {\bibfnamefont {H.~M.}\ \bibnamefont {Lee}},\ }\bibfield
  {title} {\bibinfo {title} {{Space-borne atom interferometric gravitational
  wave detections. Part III. Eccentricity on dark sirens}},\ }\href
  {https://doi.org/10.1088/1475-7516/2022/10/061} {\bibfield  {journal}
  {\bibinfo  {journal} {JCAP}\ }\textbf {\bibinfo {volume} {10}},\ \bibinfo
  {pages} {061}},\ \Eprint {https://arxiv.org/abs/2208.10998} {arXiv:2208.10998
  [gr-qc]} \BibitemShut {NoStop}%
\bibitem [{\citenamefont {Yang}\ \emph
  {et~al.}(2022{\natexlab{c}})\citenamefont {Yang}, \citenamefont {Cai},
  \citenamefont {Cao},\ and\ \citenamefont {Lee}}]{Yang:2022fgp}%
  \BibitemOpen
  \bibfield  {author} {\bibinfo {author} {\bibfnamefont {T.}~\bibnamefont
  {Yang}}, \bibinfo {author} {\bibfnamefont {R.-G.}\ \bibnamefont {Cai}},
  \bibinfo {author} {\bibfnamefont {Z.}~\bibnamefont {Cao}},\ and\ \bibinfo
  {author} {\bibfnamefont {H.~M.}\ \bibnamefont {Lee}},\ }\bibfield  {title}
  {\bibinfo {title} {{Parameter Estimation of Eccentric Gravitational Waves
  with Decihertz Observatory and Its Cosmological Implications}},\ }\href@noop
  {} {\  (\bibinfo {year} {2022}{\natexlab{c}})},\ \Eprint
  {https://arxiv.org/abs/2212.11131} {arXiv:2212.11131 [gr-qc]} \BibitemShut
  {NoStop}%
\bibitem [{\citenamefont {Xie}\ \emph {et~al.}(2022)\citenamefont {Xie},
  \citenamefont {Chatterjee}, \citenamefont {Holder}, \citenamefont {Holz},
  \citenamefont {Perkins}, \citenamefont {Yagi},\ and\ \citenamefont
  {Yunes}}]{Xie:2022brn}%
  \BibitemOpen
  \bibfield  {author} {\bibinfo {author} {\bibfnamefont {Y.}~\bibnamefont
  {Xie}}, \bibinfo {author} {\bibfnamefont {D.}~\bibnamefont {Chatterjee}},
  \bibinfo {author} {\bibfnamefont {G.}~\bibnamefont {Holder}}, \bibinfo
  {author} {\bibfnamefont {D.~E.}\ \bibnamefont {Holz}}, \bibinfo {author}
  {\bibfnamefont {S.}~\bibnamefont {Perkins}}, \bibinfo {author} {\bibfnamefont
  {K.}~\bibnamefont {Yagi}},\ and\ \bibinfo {author} {\bibfnamefont
  {N.}~\bibnamefont {Yunes}},\ }\bibfield  {title} {\bibinfo {title} {{Breaking
  Bad Degeneracies with Love: Improving gravitational-wave measurements through
  universal relations}},\ }\href@noop {} {\  (\bibinfo {year} {2022})},\
  \Eprint {https://arxiv.org/abs/2210.09386} {arXiv:2210.09386 [gr-qc]}
  \BibitemShut {NoStop}%
\bibitem [{\citenamefont {Hoy}\ \emph {et~al.}(2022)\citenamefont {Hoy},
  \citenamefont {Mills},\ and\ \citenamefont {Fairhurst}}]{Hoy:2021dqg}%
  \BibitemOpen
  \bibfield  {author} {\bibinfo {author} {\bibfnamefont {C.}~\bibnamefont
  {Hoy}}, \bibinfo {author} {\bibfnamefont {C.}~\bibnamefont {Mills}},\ and\
  \bibinfo {author} {\bibfnamefont {S.}~\bibnamefont {Fairhurst}},\ }\bibfield
  {title} {\bibinfo {title} {{Evidence for subdominant multipole moments and
  precession in merging black-hole-binaries from GWTC-2.1}},\ }\href
  {https://doi.org/10.1103/PhysRevD.106.023019} {\bibfield  {journal} {\bibinfo
   {journal} {Phys. Rev. D}\ }\textbf {\bibinfo {volume} {106}},\ \bibinfo
  {pages} {023019} (\bibinfo {year} {2022})},\ \Eprint
  {https://arxiv.org/abs/2111.10455} {arXiv:2111.10455 [gr-qc]} \BibitemShut
  {NoStop}%
\bibitem [{\citenamefont {Hannam}\ \emph {et~al.}(2021)\citenamefont {Hannam},
  \citenamefont {Hoy}, \citenamefont {Thompson}, \citenamefont {Fairhurst},\
  and\ \citenamefont {Raymond}}]{Hannam:2021pit}%
  \BibitemOpen
  \bibfield  {author} {\bibinfo {author} {\bibfnamefont {M.}~\bibnamefont
  {Hannam}}, \bibinfo {author} {\bibfnamefont {C.}~\bibnamefont {Hoy}},
  \bibinfo {author} {\bibfnamefont {J.~E.}\ \bibnamefont {Thompson}}, \bibinfo
  {author} {\bibfnamefont {S.}~\bibnamefont {Fairhurst}},\ and\ \bibinfo
  {author} {\bibfnamefont {V.}~\bibnamefont {Raymond}} (\bibinfo
  {collaboration} {LIGO/Virgo}),\ }\bibfield  {title} {\bibinfo {title}
  {{Measurement of general-relativistic precession in a black-hole binary}},\
  }\href@noop {} {\  (\bibinfo {year} {2021})},\ \Eprint
  {https://arxiv.org/abs/2112.11300} {arXiv:2112.11300 [gr-qc]} \BibitemShut
  {NoStop}%
\bibitem [{\citenamefont {Payne}\ \emph {et~al.}(2022)\citenamefont {Payne},
  \citenamefont {Hourihane}, \citenamefont {Golomb}, \citenamefont {Udall},
  \citenamefont {Davis},\ and\ \citenamefont {Chatziioannou}}]{Payne:2022spz}%
  \BibitemOpen
  \bibfield  {author} {\bibinfo {author} {\bibfnamefont {E.}~\bibnamefont
  {Payne}}, \bibinfo {author} {\bibfnamefont {S.}~\bibnamefont {Hourihane}},
  \bibinfo {author} {\bibfnamefont {J.}~\bibnamefont {Golomb}}, \bibinfo
  {author} {\bibfnamefont {R.}~\bibnamefont {Udall}}, \bibinfo {author}
  {\bibfnamefont {D.}~\bibnamefont {Davis}},\ and\ \bibinfo {author}
  {\bibfnamefont {K.}~\bibnamefont {Chatziioannou}},\ }\bibfield  {title}
  {\bibinfo {title} {{The curious case of GW200129: interplay between
  spin-precession inference and data-quality issues}},\ }\href@noop {} {\
  (\bibinfo {year} {2022})},\ \Eprint {https://arxiv.org/abs/2206.11932}
  {arXiv:2206.11932 [gr-qc]} \BibitemShut {NoStop}%
\bibitem [{\citenamefont {Ng}\ \emph {et~al.}(2022)\citenamefont {Ng} \emph
  {et~al.}}]{Ng:2022vbz}%
  \BibitemOpen
  \bibfield  {author} {\bibinfo {author} {\bibfnamefont {K.~K.~Y.}\
  \bibnamefont {Ng}} \emph {et~al.},\ }\bibfield  {title} {\bibinfo {title}
  {{Measuring properties of primordial black hole mergers at cosmological
  distances: effect of higher order modes in gravitational waves}},\
  }\href@noop {} {\  (\bibinfo {year} {2022})},\ \Eprint
  {https://arxiv.org/abs/2210.03132} {arXiv:2210.03132 [astro-ph.CO]}
  \BibitemShut {NoStop}%
\bibitem [{\citenamefont {Christodoulou}(1991)}]{Christodoulou:1991cr}%
  \BibitemOpen
  \bibfield  {author} {\bibinfo {author} {\bibfnamefont {D.}~\bibnamefont
  {Christodoulou}},\ }\bibfield  {title} {\bibinfo {title} {{Nonlinear nature
  of gravitation and gravitational wave experiments}},\ }\href
  {https://doi.org/10.1103/PhysRevLett.67.1486} {\bibfield  {journal} {\bibinfo
   {journal} {Phys. Rev. Lett.}\ }\textbf {\bibinfo {volume} {67}},\ \bibinfo
  {pages} {1486} (\bibinfo {year} {1991})}\BibitemShut {NoStop}%
\bibitem [{\citenamefont {Payne}(1983)}]{Payne:1983rrr}%
  \BibitemOpen
  \bibfield  {author} {\bibinfo {author} {\bibfnamefont {P.~N.}\ \bibnamefont
  {Payne}},\ }\bibfield  {title} {\bibinfo {title} {{SMARR'S ZERO FREQUENCY
  LIMIT CALCULATION}},\ }\href {https://doi.org/10.1103/PhysRevD.28.1894}
  {\bibfield  {journal} {\bibinfo  {journal} {Phys. Rev. D}\ }\textbf {\bibinfo
  {volume} {28}},\ \bibinfo {pages} {1894} (\bibinfo {year}
  {1983})}\BibitemShut {NoStop}%
\bibitem [{\citenamefont {Wiseman}\ and\ \citenamefont
  {Will}(1991)}]{Wiseman:1991ss}%
  \BibitemOpen
  \bibfield  {author} {\bibinfo {author} {\bibfnamefont {A.~G.}\ \bibnamefont
  {Wiseman}}\ and\ \bibinfo {author} {\bibfnamefont {C.~M.}\ \bibnamefont
  {Will}},\ }\bibfield  {title} {\bibinfo {title} {{Christodoulou's nonlinear
  gravitational wave memory: Evaluation in the quadrupole approximation}},\
  }\href {https://doi.org/10.1103/PhysRevD.44.R2945} {\bibfield  {journal}
  {\bibinfo  {journal} {Phys. Rev. D}\ }\textbf {\bibinfo {volume} {44}},\
  \bibinfo {pages} {R2945} (\bibinfo {year} {1991})}\BibitemShut {NoStop}%
\bibitem [{\citenamefont {Blanchet}\ and\ \citenamefont
  {Damour}(1992)}]{Blanchet:1992br}%
  \BibitemOpen
  \bibfield  {author} {\bibinfo {author} {\bibfnamefont {L.}~\bibnamefont
  {Blanchet}}\ and\ \bibinfo {author} {\bibfnamefont {T.}~\bibnamefont
  {Damour}},\ }\bibfield  {title} {\bibinfo {title} {{Hereditary effects in
  gravitational radiation}},\ }\href {https://doi.org/10.1103/PhysRevD.46.4304}
  {\bibfield  {journal} {\bibinfo  {journal} {Phys. Rev. D}\ }\textbf {\bibinfo
  {volume} {46}},\ \bibinfo {pages} {4304} (\bibinfo {year}
  {1992})}\BibitemShut {NoStop}%
\bibitem [{\citenamefont {Favata}(2009{\natexlab{a}})}]{Favata:2008yd}%
  \BibitemOpen
  \bibfield  {author} {\bibinfo {author} {\bibfnamefont {M.}~\bibnamefont
  {Favata}},\ }\bibfield  {title} {\bibinfo {title} {{Post-Newtonian
  corrections to the gravitational-wave memory for quasi-circular, inspiralling
  compact binaries}},\ }\href {https://doi.org/10.1103/PhysRevD.80.024002}
  {\bibfield  {journal} {\bibinfo  {journal} {Phys. Rev. D}\ }\textbf {\bibinfo
  {volume} {80}},\ \bibinfo {pages} {024002} (\bibinfo {year}
  {2009}{\natexlab{a}})},\ \Eprint {https://arxiv.org/abs/0812.0069}
  {arXiv:0812.0069 [gr-qc]} \BibitemShut {NoStop}%
\bibitem [{\citenamefont {Blanchet}\ \emph {et~al.}(2008)\citenamefont
  {Blanchet}, \citenamefont {Faye}, \citenamefont {Iyer},\ and\ \citenamefont
  {Sinha}}]{Blanchet:2008je}%
  \BibitemOpen
  \bibfield  {author} {\bibinfo {author} {\bibfnamefont {L.}~\bibnamefont
  {Blanchet}}, \bibinfo {author} {\bibfnamefont {G.}~\bibnamefont {Faye}},
  \bibinfo {author} {\bibfnamefont {B.~R.}\ \bibnamefont {Iyer}},\ and\
  \bibinfo {author} {\bibfnamefont {S.}~\bibnamefont {Sinha}},\ }\bibfield
  {title} {\bibinfo {title} {{The third post-Newtonian gravitational wave
  polarisations and associated spherical harmonic modes for inspiralling
  compact binaries in quasi-circular orbits}},\ }\href
  {https://doi.org/10.1088/0264-9381/25/16/165003} {\bibfield  {journal}
  {\bibinfo  {journal} {Class. Quant. Grav.}\ }\textbf {\bibinfo {volume}
  {25}},\ \bibinfo {pages} {165003} (\bibinfo {year} {2008})},\ \bibinfo {note}
  {[Erratum: Class.Quant.Grav. 29, 239501 (2012)]},\ \Eprint
  {https://arxiv.org/abs/0802.1249} {arXiv:0802.1249 [gr-qc]} \BibitemShut
  {NoStop}%
\bibitem [{\citenamefont {Favata}(2010)}]{Favata:2010zu}%
  \BibitemOpen
  \bibfield  {author} {\bibinfo {author} {\bibfnamefont {M.}~\bibnamefont
  {Favata}},\ }\bibfield  {title} {\bibinfo {title} {{The gravitational-wave
  memory effect}},\ }\href {https://doi.org/10.1088/0264-9381/27/8/084036}
  {\bibfield  {journal} {\bibinfo  {journal} {Class. Quantum Gravity}\ }\textbf
  {\bibinfo {volume} {27}},\ \bibinfo {pages} {084036} (\bibinfo {year}
  {2010})},\ \Eprint {https://arxiv.org/abs/1003.3486} {arXiv:1003.3486
  [gr-qc]} \BibitemShut {NoStop}%
\bibitem [{\citenamefont {Zel'dovich}\ and\ \citenamefont
  {Polnarev}(1974)}]{Zeldovich:1974gvh}%
  \BibitemOpen
  \bibfield  {author} {\bibinfo {author} {\bibfnamefont {Y.~B.}\ \bibnamefont
  {Zel'dovich}}\ and\ \bibinfo {author} {\bibfnamefont {A.~G.}\ \bibnamefont
  {Polnarev}},\ }\bibfield  {title} {\bibinfo {title} {{Radiation of
  gravitational waves by a cluster of superdense stars}},\ }\href@noop {}
  {\bibfield  {journal} {\bibinfo  {journal} {Sov. Astron.}\ }\textbf {\bibinfo
  {volume} {18}},\ \bibinfo {pages} {17} (\bibinfo {year} {1974})}\BibitemShut
  {NoStop}%
\bibitem [{\citenamefont {Braginsky}\ and\ \citenamefont
  {Grishchuk}(1985)}]{Braginsky:1985vlg}%
  \BibitemOpen
  \bibfield  {author} {\bibinfo {author} {\bibfnamefont {V.~B.}\ \bibnamefont
  {Braginsky}}\ and\ \bibinfo {author} {\bibfnamefont {L.~P.}\ \bibnamefont
  {Grishchuk}},\ }\bibfield  {title} {\bibinfo {title} {{Kinematic Resonance
  and Memory Effect in Free Mass Gravitational Antennas}},\ }\href@noop {}
  {\bibfield  {journal} {\bibinfo  {journal} {Sov. Phys. JETP}\ }\textbf
  {\bibinfo {volume} {62}},\ \bibinfo {pages} {427} (\bibinfo {year}
  {1985})}\BibitemShut {NoStop}%
\bibitem [{\citenamefont {{Braginsky}}\ and\ \citenamefont
  {{Thorne}}(1987)}]{Braginsky:1987}%
  \BibitemOpen
  \bibfield  {author} {\bibinfo {author} {\bibfnamefont {V.~B.}\ \bibnamefont
  {{Braginsky}}}\ and\ \bibinfo {author} {\bibfnamefont {K.~S.}\ \bibnamefont
  {{Thorne}}},\ }\bibfield  {title} {\bibinfo {title} {{Gravitational-wave
  bursts with memory and experimental prospects}},\ }\href
  {https://doi.org/10.1038/327123a0} {\bibfield  {journal} {\bibinfo  {journal}
  {\nat}\ }\textbf {\bibinfo {volume} {327}},\ \bibinfo {pages} {123} (\bibinfo
  {year} {1987})}\BibitemShut {NoStop}%
\bibitem [{\citenamefont {Aurrekoetxea}\ \emph {et~al.}(2020)\citenamefont
  {Aurrekoetxea}, \citenamefont {Helfer},\ and\ \citenamefont
  {Lim}}]{Aurrekoetxea:2020tuw}%
  \BibitemOpen
  \bibfield  {author} {\bibinfo {author} {\bibfnamefont {J.~C.}\ \bibnamefont
  {Aurrekoetxea}}, \bibinfo {author} {\bibfnamefont {T.}~\bibnamefont
  {Helfer}},\ and\ \bibinfo {author} {\bibfnamefont {E.~A.}\ \bibnamefont
  {Lim}},\ }\bibfield  {title} {\bibinfo {title} {{Coherent Gravitational
  Waveforms and Memory from Cosmic String Loops}},\ }\href
  {https://doi.org/10.1088/1361-6382/aba28b} {\bibfield  {journal} {\bibinfo
  {journal} {Class. Quant. Grav.}\ }\textbf {\bibinfo {volume} {37}},\ \bibinfo
  {pages} {204001} (\bibinfo {year} {2020})},\ \Eprint
  {https://arxiv.org/abs/2002.05177} {arXiv:2002.05177 [gr-qc]} \BibitemShut
  {NoStop}%
\bibitem [{\citenamefont {Jenkins}\ and\ \citenamefont
  {Sakellariadou}(2021)}]{Jenkins:2021kcj}%
  \BibitemOpen
  \bibfield  {author} {\bibinfo {author} {\bibfnamefont {A.~C.}\ \bibnamefont
  {Jenkins}}\ and\ \bibinfo {author} {\bibfnamefont {M.}~\bibnamefont
  {Sakellariadou}},\ }\bibfield  {title} {\bibinfo {title} {{Nonlinear
  gravitational-wave memory from cusps and kinks on cosmic strings}},\ }\href
  {https://doi.org/10.1088/1361-6382/ac1084} {\bibfield  {journal} {\bibinfo
  {journal} {Class. Quant. Grav.}\ }\textbf {\bibinfo {volume} {38}},\ \bibinfo
  {pages} {165004} (\bibinfo {year} {2021})},\ \Eprint
  {https://arxiv.org/abs/2102.12487} {arXiv:2102.12487 [gr-qc]} \BibitemShut
  {NoStop}%
\bibitem [{\citenamefont {Barausse}\ \emph {et~al.}(2012)\citenamefont
  {Barausse}, \citenamefont {Morozova},\ and\ \citenamefont
  {Rezzolla}}]{Barausse:2012qz}%
  \BibitemOpen
  \bibfield  {author} {\bibinfo {author} {\bibfnamefont {E.}~\bibnamefont
  {Barausse}}, \bibinfo {author} {\bibfnamefont {V.}~\bibnamefont {Morozova}},\
  and\ \bibinfo {author} {\bibfnamefont {L.}~\bibnamefont {Rezzolla}},\
  }\bibfield  {title} {\bibinfo {title} {{On the mass radiated by coalescing
  black-hole binaries}},\ }\href {https://doi.org/10.1088/0004-637X/758/1/63}
  {\bibfield  {journal} {\bibinfo  {journal} {Astrophys. J.}\ }\textbf
  {\bibinfo {volume} {758}},\ \bibinfo {pages} {63} (\bibinfo {year} {2012})},\
  \bibinfo {note} {[Erratum: Astrophys.J. 786, 76 (2014)]},\ \Eprint
  {https://arxiv.org/abs/1206.3803} {arXiv:1206.3803 [gr-qc]} \BibitemShut
  {NoStop}%
\bibitem [{\citenamefont {Favata}(2011)}]{Favata:2011qi}%
  \BibitemOpen
  \bibfield  {author} {\bibinfo {author} {\bibfnamefont {M.}~\bibnamefont
  {Favata}},\ }\bibfield  {title} {\bibinfo {title} {{The Gravitational-wave
  memory from eccentric binaries}},\ }\href
  {https://doi.org/10.1103/PhysRevD.84.124013} {\bibfield  {journal} {\bibinfo
  {journal} {Phys. Rev. D}\ }\textbf {\bibinfo {volume} {84}},\ \bibinfo
  {pages} {124013} (\bibinfo {year} {2011})},\ \Eprint
  {https://arxiv.org/abs/1108.3121} {arXiv:1108.3121 [gr-qc]} \BibitemShut
  {NoStop}%
\bibitem [{\citenamefont {Favata}(2009{\natexlab{b}})}]{Favata:2009ii}%
  \BibitemOpen
  \bibfield  {author} {\bibinfo {author} {\bibfnamefont {M.}~\bibnamefont
  {Favata}},\ }\bibfield  {title} {\bibinfo {title} {{Nonlinear
  gravitational-wave memory from binary black hole mergers}},\ }\href
  {https://doi.org/10.1088/0004-637X/696/2/L159} {\bibfield  {journal}
  {\bibinfo  {journal} {Astrophys. J. Lett.}\ }\textbf {\bibinfo {volume}
  {696}},\ \bibinfo {pages} {L159} (\bibinfo {year} {2009}{\natexlab{b}})},\
  \Eprint {https://arxiv.org/abs/0902.3660} {arXiv:0902.3660 [astro-ph.SR]}
  \BibitemShut {NoStop}%
\bibitem [{\citenamefont {H\"ubner}\ \emph {et~al.}(2020)\citenamefont
  {H\"ubner}, \citenamefont {Talbot}, \citenamefont {Lasky},\ and\
  \citenamefont {Thrane}}]{Hubner:2019sly}%
  \BibitemOpen
  \bibfield  {author} {\bibinfo {author} {\bibfnamefont {M.}~\bibnamefont
  {H\"ubner}}, \bibinfo {author} {\bibfnamefont {C.}~\bibnamefont {Talbot}},
  \bibinfo {author} {\bibfnamefont {P.~D.}\ \bibnamefont {Lasky}},\ and\
  \bibinfo {author} {\bibfnamefont {E.}~\bibnamefont {Thrane}},\ }\bibfield
  {title} {\bibinfo {title} {{Measuring gravitational-wave memory in the first
  LIGO/Virgo gravitational-wave transient catalog}},\ }\href
  {https://doi.org/10.1103/PhysRevD.101.023011} {\bibfield  {journal} {\bibinfo
   {journal} {Phys. Rev. D}\ }\textbf {\bibinfo {volume} {101}},\ \bibinfo
  {pages} {023011} (\bibinfo {year} {2020})},\ \Eprint
  {https://arxiv.org/abs/1911.12496} {arXiv:1911.12496 [astro-ph.HE]}
  \BibitemShut {NoStop}%
\bibitem [{\citenamefont {Ebersold}\ and\ \citenamefont
  {Tiwari}(2020)}]{Ebersold:2020zah}%
  \BibitemOpen
  \bibfield  {author} {\bibinfo {author} {\bibfnamefont {M.}~\bibnamefont
  {Ebersold}}\ and\ \bibinfo {author} {\bibfnamefont {S.}~\bibnamefont
  {Tiwari}},\ }\bibfield  {title} {\bibinfo {title} {{Search for nonlinear
  memory from subsolar mass compact binary mergers}},\ }\href
  {https://doi.org/10.1103/PhysRevD.101.104041} {\bibfield  {journal} {\bibinfo
   {journal} {Phys. Rev. D}\ }\textbf {\bibinfo {volume} {101}},\ \bibinfo
  {pages} {104041} (\bibinfo {year} {2020})},\ \Eprint
  {https://arxiv.org/abs/2005.03306} {arXiv:2005.03306 [gr-qc]} \BibitemShut
  {NoStop}%
\bibitem [{\citenamefont {Zhao}\ \emph {et~al.}(2021)\citenamefont {Zhao},
  \citenamefont {Liu}, \citenamefont {Cao},\ and\ \citenamefont
  {He}}]{Zhao:2021hmx}%
  \BibitemOpen
  \bibfield  {author} {\bibinfo {author} {\bibfnamefont {Z.-C.}\ \bibnamefont
  {Zhao}}, \bibinfo {author} {\bibfnamefont {X.}~\bibnamefont {Liu}}, \bibinfo
  {author} {\bibfnamefont {Z.}~\bibnamefont {Cao}},\ and\ \bibinfo {author}
  {\bibfnamefont {X.}~\bibnamefont {He}},\ }\bibfield  {title} {\bibinfo
  {title} {{Gravitational wave memory of the binary black hole events in
  GWTC-2}},\ }\href {https://doi.org/10.1103/PhysRevD.104.064056} {\bibfield
  {journal} {\bibinfo  {journal} {Phys. Rev. D}\ }\textbf {\bibinfo {volume}
  {104}},\ \bibinfo {pages} {064056} (\bibinfo {year} {2021})},\ \Eprint
  {https://arxiv.org/abs/2111.13882} {arXiv:2111.13882 [gr-qc]} \BibitemShut
  {NoStop}%
\bibitem [{\citenamefont {H\"ubner}\ \emph {et~al.}(2021)\citenamefont
  {H\"ubner}, \citenamefont {Lasky},\ and\ \citenamefont
  {Thrane}}]{Hubner:2021amk}%
  \BibitemOpen
  \bibfield  {author} {\bibinfo {author} {\bibfnamefont {M.}~\bibnamefont
  {H\"ubner}}, \bibinfo {author} {\bibfnamefont {P.}~\bibnamefont {Lasky}},\
  and\ \bibinfo {author} {\bibfnamefont {E.}~\bibnamefont {Thrane}},\
  }\bibfield  {title} {\bibinfo {title} {{Memory remains undetected: Updates
  from the second LIGO/Virgo gravitational-wave transient catalog}},\ }\href
  {https://doi.org/10.1103/PhysRevD.104.023004} {\bibfield  {journal} {\bibinfo
   {journal} {Phys. Rev. D}\ }\textbf {\bibinfo {volume} {104}},\ \bibinfo
  {pages} {023004} (\bibinfo {year} {2021})},\ \Eprint
  {https://arxiv.org/abs/2105.02879} {arXiv:2105.02879 [gr-qc]} \BibitemShut
  {NoStop}%
\bibitem [{\citenamefont {Lasky}\ \emph {et~al.}(2016)\citenamefont {Lasky},
  \citenamefont {Thrane}, \citenamefont {Levin}, \citenamefont {Blackman},\
  and\ \citenamefont {Chen}}]{Lasky:2016knh}%
  \BibitemOpen
  \bibfield  {author} {\bibinfo {author} {\bibfnamefont {P.~D.}\ \bibnamefont
  {Lasky}}, \bibinfo {author} {\bibfnamefont {E.}~\bibnamefont {Thrane}},
  \bibinfo {author} {\bibfnamefont {Y.}~\bibnamefont {Levin}}, \bibinfo
  {author} {\bibfnamefont {J.}~\bibnamefont {Blackman}},\ and\ \bibinfo
  {author} {\bibfnamefont {Y.}~\bibnamefont {Chen}},\ }\bibfield  {title}
  {\bibinfo {title} {{Detecting gravitational-wave memory with LIGO:
  implications of GW150914}},\ }\href
  {https://doi.org/10.1103/PhysRevLett.117.061102} {\bibfield  {journal}
  {\bibinfo  {journal} {Phys. Rev. Lett.}\ }\textbf {\bibinfo {volume} {117}},\
  \bibinfo {pages} {061102} (\bibinfo {year} {2016})},\ \Eprint
  {https://arxiv.org/abs/1605.01415} {arXiv:1605.01415 [astro-ph.HE]}
  \BibitemShut {NoStop}%
\bibitem [{\citenamefont {Yang}\ and\ \citenamefont
  {Martynov}(2018)}]{Yang:2018ceq}%
  \BibitemOpen
  \bibfield  {author} {\bibinfo {author} {\bibfnamefont {H.}~\bibnamefont
  {Yang}}\ and\ \bibinfo {author} {\bibfnamefont {D.}~\bibnamefont
  {Martynov}},\ }\bibfield  {title} {\bibinfo {title} {{Testing Gravitational
  Memory Generation with Compact Binary Mergers}},\ }\href
  {https://doi.org/10.1103/PhysRevLett.121.071102} {\bibfield  {journal}
  {\bibinfo  {journal} {Phys. Rev. Lett.}\ }\textbf {\bibinfo {volume} {121}},\
  \bibinfo {pages} {071102} (\bibinfo {year} {2018})},\ \Eprint
  {https://arxiv.org/abs/1803.02429} {arXiv:1803.02429 [gr-qc]} \BibitemShut
  {NoStop}%
\bibitem [{\citenamefont {Boersma}\ \emph {et~al.}(2020)\citenamefont
  {Boersma}, \citenamefont {Nichols},\ and\ \citenamefont
  {Schmidt}}]{Boersma:2020gxx}%
  \BibitemOpen
  \bibfield  {author} {\bibinfo {author} {\bibfnamefont {O.~M.}\ \bibnamefont
  {Boersma}}, \bibinfo {author} {\bibfnamefont {D.~A.}\ \bibnamefont
  {Nichols}},\ and\ \bibinfo {author} {\bibfnamefont {P.}~\bibnamefont
  {Schmidt}},\ }\bibfield  {title} {\bibinfo {title} {{Forecasts for detecting
  the gravitational-wave memory effect with Advanced LIGO and Virgo}},\ }\href
  {https://doi.org/10.1103/PhysRevD.101.083026} {\bibfield  {journal} {\bibinfo
   {journal} {Phys. Rev. D}\ }\textbf {\bibinfo {volume} {101}},\ \bibinfo
  {pages} {083026} (\bibinfo {year} {2020})},\ \Eprint
  {https://arxiv.org/abs/2002.01821} {arXiv:2002.01821 [astro-ph.HE]}
  \BibitemShut {NoStop}%
\bibitem [{\citenamefont {Grant}\ and\ \citenamefont
  {Nichols}(2022)}]{Grant:2022bla}%
  \BibitemOpen
  \bibfield  {author} {\bibinfo {author} {\bibfnamefont {A.~M.}\ \bibnamefont
  {Grant}}\ and\ \bibinfo {author} {\bibfnamefont {D.~A.}\ \bibnamefont
  {Nichols}},\ }\bibfield  {title} {\bibinfo {title} {{Outlook for detecting
  the gravitational wave displacement and spin memory effects with current and
  future gravitational wave detectors}},\ }\href@noop {} {\  (\bibinfo {year}
  {2022})},\ \Eprint {https://arxiv.org/abs/2210.16266} {arXiv:2210.16266
  [gr-qc]} \BibitemShut {NoStop}%
\bibitem [{\citenamefont {{Buikema}}\ \emph {et~al.}(2020)\citenamefont
  {{Buikema}} \emph {et~al.}}]{Buikema:2020}%
  \BibitemOpen
  \bibfield  {author} {\bibinfo {author} {\bibfnamefont {A.}~\bibnamefont
  {{Buikema}}} \emph {et~al.},\ }\bibfield  {title} {\bibinfo {title}
  {{Sensitivity and performance of the Advanced LIGO detectors in the third
  observing run}},\ }\href {https://doi.org/10.1103/PhysRevD.102.062003}
  {\bibfield  {journal} {\bibinfo  {journal} {\prd}\ }\textbf {\bibinfo
  {volume} {102}},\ \bibinfo {eid} {062003} (\bibinfo {year} {2020})},\ \Eprint
  {https://arxiv.org/abs/2008.01301} {arXiv:2008.01301 [astro-ph.IM]}
  \BibitemShut {NoStop}%
\bibitem [{\citenamefont {Punturo}\ \emph {et~al.}(2010)\citenamefont {Punturo}
  \emph {et~al.}}]{Punturo:2010zz}%
  \BibitemOpen
  \bibfield  {author} {\bibinfo {author} {\bibfnamefont {M.}~\bibnamefont
  {Punturo}} \emph {et~al.},\ }\bibfield  {title} {\bibinfo {title} {{The
  Einstein Telescope: A third-generation gravitational wave observatory}},\
  }\href {https://doi.org/10.1088/0264-9381/27/19/194002} {\bibfield  {journal}
  {\bibinfo  {journal} {Class. Quant. Grav.}\ }\textbf {\bibinfo {volume}
  {27}},\ \bibinfo {pages} {194002} (\bibinfo {year} {2010})}\BibitemShut
  {NoStop}%
\bibitem [{\citenamefont {Reitze}\ \emph {et~al.}(2019)\citenamefont {Reitze}
  \emph {et~al.}}]{Reitze:2019iox}%
  \BibitemOpen
  \bibfield  {author} {\bibinfo {author} {\bibfnamefont {D.}~\bibnamefont
  {Reitze}} \emph {et~al.},\ }\bibfield  {title} {\bibinfo {title} {{Cosmic
  Explorer: The U.S. Contribution to Gravitational-Wave Astronomy beyond
  LIGO}},\ }\href@noop {} {\bibfield  {journal} {\bibinfo  {journal} {Bull. Am.
  Astron. Soc.}\ }\textbf {\bibinfo {volume} {51}},\ \bibinfo {pages} {035}
  (\bibinfo {year} {2019})},\ \Eprint {https://arxiv.org/abs/1907.04833}
  {arXiv:1907.04833 [astro-ph.IM]} \BibitemShut {NoStop}%
\bibitem [{\citenamefont {Johnson}\ \emph {et~al.}(2019)\citenamefont
  {Johnson}, \citenamefont {Kapadia}, \citenamefont {Osborne}, \citenamefont
  {Hixon},\ and\ \citenamefont {Kennefick}}]{Johnson:2018xly}%
  \BibitemOpen
  \bibfield  {author} {\bibinfo {author} {\bibfnamefont {A.~D.}\ \bibnamefont
  {Johnson}}, \bibinfo {author} {\bibfnamefont {S.~J.}\ \bibnamefont
  {Kapadia}}, \bibinfo {author} {\bibfnamefont {A.}~\bibnamefont {Osborne}},
  \bibinfo {author} {\bibfnamefont {A.}~\bibnamefont {Hixon}},\ and\ \bibinfo
  {author} {\bibfnamefont {D.}~\bibnamefont {Kennefick}},\ }\bibfield  {title}
  {\bibinfo {title} {{Prospects of detecting the nonlinear gravitational wave
  memory}},\ }\href {https://doi.org/10.1103/PhysRevD.99.044045} {\bibfield
  {journal} {\bibinfo  {journal} {Phys. Rev. D}\ }\textbf {\bibinfo {volume}
  {99}},\ \bibinfo {pages} {044045} (\bibinfo {year} {2019})},\ \Eprint
  {https://arxiv.org/abs/1810.09563} {arXiv:1810.09563 [gr-qc]} \BibitemShut
  {NoStop}%
\bibitem [{\citenamefont {Islo}\ \emph {et~al.}(2019)\citenamefont {Islo},
  \citenamefont {Simon}, \citenamefont {Burke-Spolaor},\ and\ \citenamefont
  {Siemens}}]{Islo:2019qht}%
  \BibitemOpen
  \bibfield  {author} {\bibinfo {author} {\bibfnamefont {K.}~\bibnamefont
  {Islo}}, \bibinfo {author} {\bibfnamefont {J.}~\bibnamefont {Simon}},
  \bibinfo {author} {\bibfnamefont {S.}~\bibnamefont {Burke-Spolaor}},\ and\
  \bibinfo {author} {\bibfnamefont {X.}~\bibnamefont {Siemens}},\ }\bibfield
  {title} {\bibinfo {title} {{Prospects for Memory Detection with Low-Frequency
  Gravitational Wave Detectors}},\ }\href@noop {} {\  (\bibinfo {year}
  {2019})},\ \Eprint {https://arxiv.org/abs/1906.11936} {arXiv:1906.11936
  [astro-ph.HE]} \BibitemShut {NoStop}%
\bibitem [{\citenamefont {Islam}\ \emph {et~al.}(2021)\citenamefont {Islam},
  \citenamefont {Field}, \citenamefont {Khanna},\ and\ \citenamefont
  {Warburton}}]{Islam:2021old}%
  \BibitemOpen
  \bibfield  {author} {\bibinfo {author} {\bibfnamefont {T.}~\bibnamefont
  {Islam}}, \bibinfo {author} {\bibfnamefont {S.~E.}\ \bibnamefont {Field}},
  \bibinfo {author} {\bibfnamefont {G.}~\bibnamefont {Khanna}},\ and\ \bibinfo
  {author} {\bibfnamefont {N.}~\bibnamefont {Warburton}},\ }\bibfield  {title}
  {\bibinfo {title} {{Survey of gravitational wave memory in intermediate mass
  ratio binaries}},\ }\href@noop {} {\  (\bibinfo {year} {2021})},\ \Eprint
  {https://arxiv.org/abs/2109.00754} {arXiv:2109.00754 [gr-qc]} \BibitemShut
  {NoStop}%
\bibitem [{\citenamefont {Sun}\ \emph {et~al.}(2022)\citenamefont {Sun},
  \citenamefont {Shi}, \citenamefont {Zhang},\ and\ \citenamefont
  {Mei}}]{Sun:2022pvh}%
  \BibitemOpen
  \bibfield  {author} {\bibinfo {author} {\bibfnamefont {S.}~\bibnamefont
  {Sun}}, \bibinfo {author} {\bibfnamefont {C.}~\bibnamefont {Shi}}, \bibinfo
  {author} {\bibfnamefont {J.-d.}\ \bibnamefont {Zhang}},\ and\ \bibinfo
  {author} {\bibfnamefont {J.}~\bibnamefont {Mei}},\ }\bibfield  {title}
  {\bibinfo {title} {{Detecting the gravitational wave memory effect with
  TianQin}},\ }\href@noop {} {\  (\bibinfo {year} {2022})},\ \Eprint
  {https://arxiv.org/abs/2207.13009} {arXiv:2207.13009 [gr-qc]} \BibitemShut
  {NoStop}%
\bibitem [{\citenamefont {Seto}(2009)}]{Seto:2009nv}%
  \BibitemOpen
  \bibfield  {author} {\bibinfo {author} {\bibfnamefont {N.}~\bibnamefont
  {Seto}},\ }\bibfield  {title} {\bibinfo {title} {{Search for Memory and
  Inspiral Gravitational Waves from Super-Massive Binary Black Holes with
  Pulsar Timing Arrays}},\ }\href
  {https://doi.org/10.1111/j.1745-3933.2009.00758.x} {\bibfield  {journal}
  {\bibinfo  {journal} {Mon. Not. Roy. Astron. Soc.}\ }\textbf {\bibinfo
  {volume} {400}},\ \bibinfo {pages} {L38} (\bibinfo {year} {2009})},\ \Eprint
  {https://arxiv.org/abs/0909.1379} {arXiv:0909.1379 [astro-ph.CO]}
  \BibitemShut {NoStop}%
\bibitem [{\citenamefont {Cordes}\ and\ \citenamefont
  {Jenet}(2012)}]{Cordes:2012zz}%
  \BibitemOpen
  \bibfield  {author} {\bibinfo {author} {\bibfnamefont {J.~M.}\ \bibnamefont
  {Cordes}}\ and\ \bibinfo {author} {\bibfnamefont {F.~A.}\ \bibnamefont
  {Jenet}},\ }\bibfield  {title} {\bibinfo {title} {{Detecting gravitational
  wave memory with pulsar timing}},\ }\href
  {https://doi.org/10.1088/0004-637X/752/1/54} {\bibfield  {journal} {\bibinfo
  {journal} {Astrophys. J.}\ }\textbf {\bibinfo {volume} {752}},\ \bibinfo
  {pages} {54} (\bibinfo {year} {2012})}\BibitemShut {NoStop}%
\bibitem [{\citenamefont {Wang}\ \emph {et~al.}(2015)\citenamefont {Wang} \emph
  {et~al.}}]{Wang:2014zls}%
  \BibitemOpen
  \bibfield  {author} {\bibinfo {author} {\bibfnamefont {J.~B.}\ \bibnamefont
  {Wang}} \emph {et~al.},\ }\bibfield  {title} {\bibinfo {title} {{Searching
  for gravitational wave memory bursts with the Parkes Pulsar Timing Array}},\
  }\href {https://doi.org/10.1093/mnras/stu2137} {\bibfield  {journal}
  {\bibinfo  {journal} {Mon. Not. Roy. Astron. Soc.}\ }\textbf {\bibinfo
  {volume} {446}},\ \bibinfo {pages} {1657} (\bibinfo {year} {2015})},\ \Eprint
  {https://arxiv.org/abs/1410.3323} {arXiv:1410.3323 [astro-ph.GA]}
  \BibitemShut {NoStop}%
\bibitem [{\citenamefont {Aggarwal}\ \emph {et~al.}(2020)\citenamefont
  {Aggarwal} \emph {et~al.}}]{NANOGrav:2019vto}%
  \BibitemOpen
  \bibfield  {author} {\bibinfo {author} {\bibfnamefont {K.}~\bibnamefont
  {Aggarwal}} \emph {et~al.} (\bibinfo {collaboration} {NANOGrav}),\ }\bibfield
   {title} {\bibinfo {title} {{The NANOGrav 11 yr Data Set: Limits on
  Gravitational Wave Memory}},\ }\href
  {https://doi.org/10.3847/1538-4357/ab6083} {\bibfield  {journal} {\bibinfo
  {journal} {Astrophys. J.}\ }\textbf {\bibinfo {volume} {889}},\ \bibinfo
  {pages} {38} (\bibinfo {year} {2020})},\ \Eprint
  {https://arxiv.org/abs/1911.08488} {arXiv:1911.08488 [astro-ph.HE]}
  \BibitemShut {NoStop}%
\bibitem [{\citenamefont {Seoane}\ \emph {et~al.}(2022)\citenamefont {Seoane}
  \emph {et~al.}}]{Seoane:2021kkk}%
  \BibitemOpen
  \bibfield  {author} {\bibinfo {author} {\bibfnamefont {P.~A.}\ \bibnamefont
  {Seoane}} \emph {et~al.},\ }\bibfield  {title} {\bibinfo {title} {{The effect
  of mission duration on LISA science objectives}},\ }\href
  {https://doi.org/10.1007/s10714-021-02889-x} {\bibfield  {journal} {\bibinfo
  {journal} {Gen. Rel. Grav.}\ }\textbf {\bibinfo {volume} {54}},\ \bibinfo
  {pages} {3} (\bibinfo {year} {2022})},\ \Eprint
  {https://arxiv.org/abs/2107.09665} {arXiv:2107.09665 [astro-ph.IM]}
  \BibitemShut {NoStop}%
\bibitem [{\citenamefont {Barausse}\ and\ \citenamefont
  {Lapi}(2020)}]{Barausse:2020gbp}%
  \BibitemOpen
  \bibfield  {author} {\bibinfo {author} {\bibfnamefont {E.}~\bibnamefont
  {Barausse}}\ and\ \bibinfo {author} {\bibfnamefont {A.}~\bibnamefont
  {Lapi}},\ }\bibfield  {title} {\bibinfo {title} {{Massive Black Hole
  Mergers}},\ }\href@noop {} {\  (\bibinfo {year} {2020})},\ \Eprint
  {https://arxiv.org/abs/2011.01994} {arXiv:2011.01994 [astro-ph.GA]}
  \BibitemShut {NoStop}%
\bibitem [{\citenamefont {Barausse}\ \emph
  {et~al.}(2020{\natexlab{b}})\citenamefont {Barausse}, \citenamefont
  {Dvorkin}, \citenamefont {Tremmel}, \citenamefont {Volonteri},\ and\
  \citenamefont {Bonetti}}]{Barausse:2020mdt}%
  \BibitemOpen
  \bibfield  {author} {\bibinfo {author} {\bibfnamefont {E.}~\bibnamefont
  {Barausse}}, \bibinfo {author} {\bibfnamefont {I.}~\bibnamefont {Dvorkin}},
  \bibinfo {author} {\bibfnamefont {M.}~\bibnamefont {Tremmel}}, \bibinfo
  {author} {\bibfnamefont {M.}~\bibnamefont {Volonteri}},\ and\ \bibinfo
  {author} {\bibfnamefont {M.}~\bibnamefont {Bonetti}},\ }\bibfield  {title}
  {\bibinfo {title} {{Massive Black Hole Merger Rates: The Effect of Kiloparsec
  Separation Wandering and Supernova Feedback}},\ }\href
  {https://doi.org/10.3847/1538-4357/abba7f} {\bibfield  {journal} {\bibinfo
  {journal} {Astrophys. J.}\ }\textbf {\bibinfo {volume} {904}},\ \bibinfo
  {pages} {16} (\bibinfo {year} {2020}{\natexlab{b}})},\ \Eprint
  {https://arxiv.org/abs/2006.03065} {arXiv:2006.03065 [astro-ph.GA]}
  \BibitemShut {NoStop}%
\bibitem [{\citenamefont {Barausse}(2012)}]{EB12}%
  \BibitemOpen
  \bibfield  {author} {\bibinfo {author} {\bibfnamefont {E.}~\bibnamefont
  {Barausse}},\ }\bibfield  {title} {\bibinfo {title} {{The evolution of
  massive black holes and their spins in their galactic hosts}},\ }\href
  {https://doi.org/10.1111/j.1365-2966.2012.21057.x} {\bibfield  {journal}
  {\bibinfo  {journal} {Mon. Not. Roy. Astron. Soc.}\ }\textbf {\bibinfo
  {volume} {423}},\ \bibinfo {pages} {2533} (\bibinfo {year} {2012})},\ \Eprint
  {https://arxiv.org/abs/1201.5888} {arXiv:1201.5888 [astro-ph.CO]}
  \BibitemShut {NoStop}%
\bibitem [{\citenamefont {Sesana}\ \emph {et~al.}(2014)\citenamefont {Sesana},
  \citenamefont {Barausse}, \citenamefont {Dotti},\ and\ \citenamefont
  {Rossi}}]{Sesana:2014bea}%
  \BibitemOpen
  \bibfield  {author} {\bibinfo {author} {\bibfnamefont {A.}~\bibnamefont
  {Sesana}}, \bibinfo {author} {\bibfnamefont {E.}~\bibnamefont {Barausse}},
  \bibinfo {author} {\bibfnamefont {M.}~\bibnamefont {Dotti}},\ and\ \bibinfo
  {author} {\bibfnamefont {E.~M.}\ \bibnamefont {Rossi}},\ }\bibfield  {title}
  {\bibinfo {title} {{Linking the spin evolution of massive black holes to
  galaxy kinematics}},\ }\href {https://doi.org/10.1088/0004-637X/794/2/104}
  {\bibfield  {journal} {\bibinfo  {journal} {Astrophys. J.}\ }\textbf
  {\bibinfo {volume} {794}},\ \bibinfo {pages} {104} (\bibinfo {year}
  {2014})},\ \Eprint {https://arxiv.org/abs/1402.7088} {arXiv:1402.7088
  [astro-ph.CO]} \BibitemShut {NoStop}%
\bibitem [{\citenamefont {Antonini}\ \emph {et~al.}(2015)\citenamefont
  {Antonini}, \citenamefont {Barausse},\ and\ \citenamefont
  {Silk}}]{Antonini:2015sza}%
  \BibitemOpen
  \bibfield  {author} {\bibinfo {author} {\bibfnamefont {F.}~\bibnamefont
  {Antonini}}, \bibinfo {author} {\bibfnamefont {E.}~\bibnamefont {Barausse}},\
  and\ \bibinfo {author} {\bibfnamefont {J.}~\bibnamefont {Silk}},\ }\bibfield
  {title} {\bibinfo {title} {{The Coevolution of Nuclear Star Clusters, Massive
  Black Holes, and their Host Galaxies}},\ }\href
  {https://doi.org/10.1088/0004-637X/812/1/72} {\bibfield  {journal} {\bibinfo
  {journal} {Astrophys. J.}\ }\textbf {\bibinfo {volume} {812}},\ \bibinfo
  {pages} {72} (\bibinfo {year} {2015})},\ \Eprint
  {https://arxiv.org/abs/1506.02050} {arXiv:1506.02050 [astro-ph.GA]}
  \BibitemShut {NoStop}%
\bibitem [{\citenamefont {Dey}\ \emph {et~al.}(2021)\citenamefont {Dey},
  \citenamefont {Karnesis}, \citenamefont {Toubiana}, \citenamefont {Barausse},
  \citenamefont {Korsakova}, \citenamefont {Baghi},\ and\ \citenamefont
  {Basak}}]{Dey:2021dem}%
  \BibitemOpen
  \bibfield  {author} {\bibinfo {author} {\bibfnamefont {K.}~\bibnamefont
  {Dey}}, \bibinfo {author} {\bibfnamefont {N.}~\bibnamefont {Karnesis}},
  \bibinfo {author} {\bibfnamefont {A.}~\bibnamefont {Toubiana}}, \bibinfo
  {author} {\bibfnamefont {E.}~\bibnamefont {Barausse}}, \bibinfo {author}
  {\bibfnamefont {N.}~\bibnamefont {Korsakova}}, \bibinfo {author}
  {\bibfnamefont {Q.}~\bibnamefont {Baghi}},\ and\ \bibinfo {author}
  {\bibfnamefont {S.}~\bibnamefont {Basak}},\ }\bibfield  {title} {\bibinfo
  {title} {{Effect of data gaps on the detectability and parameter estimation
  of massive black hole binaries with LISA}},\ }\href
  {https://doi.org/10.1103/PhysRevD.104.044035} {\bibfield  {journal} {\bibinfo
   {journal} {Phys. Rev. D}\ }\textbf {\bibinfo {volume} {104}},\ \bibinfo
  {pages} {044035} (\bibinfo {year} {2021})},\ \Eprint
  {https://arxiv.org/abs/2104.12646} {arXiv:2104.12646 [gr-qc]} \BibitemShut
  {NoStop}%
\bibitem [{\citenamefont {Pollney}\ and\ \citenamefont
  {Reisswig}(2011)}]{Pollney:2010hs}%
  \BibitemOpen
  \bibfield  {author} {\bibinfo {author} {\bibfnamefont {D.}~\bibnamefont
  {Pollney}}\ and\ \bibinfo {author} {\bibfnamefont {C.}~\bibnamefont
  {Reisswig}},\ }\bibfield  {title} {\bibinfo {title} {{Gravitational memory in
  binary black hole mergers}},\ }\href
  {https://doi.org/10.1088/2041-8205/732/1/L13} {\bibfield  {journal} {\bibinfo
   {journal} {Astrophys. J. Lett.}\ }\textbf {\bibinfo {volume} {732}},\
  \bibinfo {pages} {L13} (\bibinfo {year} {2011})},\ \Eprint
  {https://arxiv.org/abs/1004.4209} {arXiv:1004.4209 [gr-qc]} \BibitemShut
  {NoStop}%
\bibitem [{\citenamefont {Mitman}\ \emph {et~al.}(2020)\citenamefont {Mitman},
  \citenamefont {Moxon}, \citenamefont {Scheel}, \citenamefont {Teukolsky},
  \citenamefont {Boyle}, \citenamefont {Deppe}, \citenamefont {Kidder},\ and\
  \citenamefont {Throwe}}]{Mitman:2020pbt}%
  \BibitemOpen
  \bibfield  {author} {\bibinfo {author} {\bibfnamefont {K.}~\bibnamefont
  {Mitman}}, \bibinfo {author} {\bibfnamefont {J.}~\bibnamefont {Moxon}},
  \bibinfo {author} {\bibfnamefont {M.~A.}\ \bibnamefont {Scheel}}, \bibinfo
  {author} {\bibfnamefont {S.~A.}\ \bibnamefont {Teukolsky}}, \bibinfo {author}
  {\bibfnamefont {M.}~\bibnamefont {Boyle}}, \bibinfo {author} {\bibfnamefont
  {N.}~\bibnamefont {Deppe}}, \bibinfo {author} {\bibfnamefont {L.~E.}\
  \bibnamefont {Kidder}},\ and\ \bibinfo {author} {\bibfnamefont
  {W.}~\bibnamefont {Throwe}},\ }\bibfield  {title} {\bibinfo {title}
  {{Computation of displacement and spin gravitational memory in numerical
  relativity}},\ }\href {https://doi.org/10.1103/PhysRevD.102.104007}
  {\bibfield  {journal} {\bibinfo  {journal} {Phys. Rev. D}\ }\textbf {\bibinfo
  {volume} {102}},\ \bibinfo {pages} {104007} (\bibinfo {year} {2020})},\
  \Eprint {https://arxiv.org/abs/2007.11562} {arXiv:2007.11562 [gr-qc]}
  \BibitemShut {NoStop}%
\bibitem [{\citenamefont {Ashtekar}\ \emph {et~al.}(2020)\citenamefont
  {Ashtekar}, \citenamefont {De~Lorenzo},\ and\ \citenamefont
  {Khera}}]{Ashtekar:2019viz}%
  \BibitemOpen
  \bibfield  {author} {\bibinfo {author} {\bibfnamefont {A.}~\bibnamefont
  {Ashtekar}}, \bibinfo {author} {\bibfnamefont {T.}~\bibnamefont
  {De~Lorenzo}},\ and\ \bibinfo {author} {\bibfnamefont {N.}~\bibnamefont
  {Khera}},\ }\bibfield  {title} {\bibinfo {title} {{Compact binary
  coalescences: Constraints on waveforms}},\ }\href
  {https://doi.org/10.1007/s10714-020-02764-1} {\bibfield  {journal} {\bibinfo
  {journal} {Gen. Rel. Grav.}\ }\textbf {\bibinfo {volume} {52}},\ \bibinfo
  {pages} {107} (\bibinfo {year} {2020})},\ \Eprint
  {https://arxiv.org/abs/1906.00913} {arXiv:1906.00913 [gr-qc]} \BibitemShut
  {NoStop}%
\bibitem [{\citenamefont {Comp\`ere}\ \emph {et~al.}(2020)\citenamefont
  {Comp\`ere}, \citenamefont {Oliveri},\ and\ \citenamefont
  {Seraj}}]{Compere:2019gft}%
  \BibitemOpen
  \bibfield  {author} {\bibinfo {author} {\bibfnamefont {G.}~\bibnamefont
  {Comp\`ere}}, \bibinfo {author} {\bibfnamefont {R.}~\bibnamefont {Oliveri}},\
  and\ \bibinfo {author} {\bibfnamefont {A.}~\bibnamefont {Seraj}},\ }\bibfield
   {title} {\bibinfo {title} {{The Poincar\'e and BMS flux-balance laws with
  application to binary systems}},\ }\href
  {https://doi.org/10.1007/JHEP10(2020)116} {\bibfield  {journal} {\bibinfo
  {journal} {JHEP}\ }\textbf {\bibinfo {volume} {10}},\ \bibinfo {pages}
  {116}},\ \Eprint {https://arxiv.org/abs/1912.03164} {arXiv:1912.03164
  [gr-qc]} \BibitemShut {NoStop}%
\bibitem [{\citenamefont {Khera}\ \emph {et~al.}(2021)\citenamefont {Khera},
  \citenamefont {Krishnan}, \citenamefont {Ashtekar},\ and\ \citenamefont
  {De~Lorenzo}}]{Khera:2020mcz}%
  \BibitemOpen
  \bibfield  {author} {\bibinfo {author} {\bibfnamefont {N.}~\bibnamefont
  {Khera}}, \bibinfo {author} {\bibfnamefont {B.}~\bibnamefont {Krishnan}},
  \bibinfo {author} {\bibfnamefont {A.}~\bibnamefont {Ashtekar}},\ and\
  \bibinfo {author} {\bibfnamefont {T.}~\bibnamefont {De~Lorenzo}},\ }\bibfield
   {title} {\bibinfo {title} {{Inferring the gravitational wave memory for
  binary coalescence events}},\ }\href
  {https://doi.org/10.1103/PhysRevD.103.044012} {\bibfield  {journal} {\bibinfo
   {journal} {Phys. Rev. D}\ }\textbf {\bibinfo {volume} {103}},\ \bibinfo
  {pages} {044012} (\bibinfo {year} {2021})},\ \Eprint
  {https://arxiv.org/abs/2009.06351} {arXiv:2009.06351 [gr-qc]} \BibitemShut
  {NoStop}%
\bibitem [{\citenamefont {Mitman}\ \emph {et~al.}(2021)\citenamefont {Mitman}
  \emph {et~al.}}]{Mitman:2020bjf}%
  \BibitemOpen
  \bibfield  {author} {\bibinfo {author} {\bibfnamefont {K.}~\bibnamefont
  {Mitman}} \emph {et~al.},\ }\bibfield  {title} {\bibinfo {title} {{Adding
  gravitational memory to waveform catalogs using BMS balance laws}},\ }\href
  {https://doi.org/10.1103/PhysRevD.103.024031} {\bibfield  {journal} {\bibinfo
   {journal} {Phys. Rev. D}\ }\textbf {\bibinfo {volume} {103}},\ \bibinfo
  {pages} {024031} (\bibinfo {year} {2021})},\ \Eprint
  {https://arxiv.org/abs/2011.01309} {arXiv:2011.01309 [gr-qc]} \BibitemShut
  {NoStop}%
\bibitem [{\citenamefont {Thorne}(1992)}]{Thorne:1992sdb}%
  \BibitemOpen
  \bibfield  {author} {\bibinfo {author} {\bibfnamefont {K.~S.}\ \bibnamefont
  {Thorne}},\ }\bibfield  {title} {\bibinfo {title} {{Gravitational-wave bursts
  with memory: The Christodoulou effect}},\ }\href
  {https://doi.org/10.1103/PhysRevD.45.520} {\bibfield  {journal} {\bibinfo
  {journal} {Phys. Rev. D}\ }\textbf {\bibinfo {volume} {45}},\ \bibinfo
  {pages} {520} (\bibinfo {year} {1992})}\BibitemShut {NoStop}%
\bibitem [{\citenamefont {Kidder}(2008)}]{Kidder:2007rt}%
  \BibitemOpen
  \bibfield  {author} {\bibinfo {author} {\bibfnamefont {L.~E.}\ \bibnamefont
  {Kidder}},\ }\bibfield  {title} {\bibinfo {title} {{Using full information
  when computing modes of post-Newtonian waveforms from inspiralling compact
  binaries in circular orbit}},\ }\href
  {https://doi.org/10.1103/PhysRevD.77.044016} {\bibfield  {journal} {\bibinfo
  {journal} {Phys. Rev. D}\ }\textbf {\bibinfo {volume} {77}},\ \bibinfo
  {pages} {044016} (\bibinfo {year} {2008})},\ \Eprint
  {https://arxiv.org/abs/0710.0614} {arXiv:0710.0614 [gr-qc]} \BibitemShut
  {NoStop}%
\bibitem [{\citenamefont {Varma}\ \emph
  {et~al.}(2019{\natexlab{a}})\citenamefont {Varma}, \citenamefont {Field},
  \citenamefont {Scheel}, \citenamefont {Blackman}, \citenamefont {Kidder},\
  and\ \citenamefont {Pfeiffer}}]{Varma:2018mmi}%
  \BibitemOpen
  \bibfield  {author} {\bibinfo {author} {\bibfnamefont {V.}~\bibnamefont
  {Varma}}, \bibinfo {author} {\bibfnamefont {S.~E.}\ \bibnamefont {Field}},
  \bibinfo {author} {\bibfnamefont {M.~A.}\ \bibnamefont {Scheel}}, \bibinfo
  {author} {\bibfnamefont {J.}~\bibnamefont {Blackman}}, \bibinfo {author}
  {\bibfnamefont {L.~E.}\ \bibnamefont {Kidder}},\ and\ \bibinfo {author}
  {\bibfnamefont {H.~P.}\ \bibnamefont {Pfeiffer}},\ }\bibfield  {title}
  {\bibinfo {title} {{Surrogate model of hybridized numerical relativity binary
  black hole waveforms}},\ }\href {https://doi.org/10.1103/PhysRevD.99.064045}
  {\bibfield  {journal} {\bibinfo  {journal} {Phys. Rev. D}\ }\textbf {\bibinfo
  {volume} {99}},\ \bibinfo {pages} {064045} (\bibinfo {year}
  {2019}{\natexlab{a}})},\ \Eprint {https://arxiv.org/abs/1812.07865}
  {arXiv:1812.07865 [gr-qc]} \BibitemShut {NoStop}%
\bibitem [{\citenamefont {Talbot}\ \emph {et~al.}(2018)\citenamefont {Talbot},
  \citenamefont {Thrane}, \citenamefont {Lasky},\ and\ \citenamefont
  {Lin}}]{Talbot:2018sgr}%
  \BibitemOpen
  \bibfield  {author} {\bibinfo {author} {\bibfnamefont {C.}~\bibnamefont
  {Talbot}}, \bibinfo {author} {\bibfnamefont {E.}~\bibnamefont {Thrane}},
  \bibinfo {author} {\bibfnamefont {P.~D.}\ \bibnamefont {Lasky}},\ and\
  \bibinfo {author} {\bibfnamefont {F.}~\bibnamefont {Lin}},\ }\bibfield
  {title} {\bibinfo {title} {{Gravitational-wave memory: waveforms and
  phenomenology}},\ }\href {https://doi.org/10.1103/PhysRevD.98.064031}
  {\bibfield  {journal} {\bibinfo  {journal} {Phys. Rev. D}\ }\textbf {\bibinfo
  {volume} {98}},\ \bibinfo {pages} {064031} (\bibinfo {year} {2018})},\
  \Eprint {https://arxiv.org/abs/1807.00990} {arXiv:1807.00990 [astro-ph.HE]}
  \BibitemShut {NoStop}%
\bibitem [{\citenamefont {Tolish}\ and\ \citenamefont
  {Wald}(2016)}]{Tolish:2016ggo}%
  \BibitemOpen
  \bibfield  {author} {\bibinfo {author} {\bibfnamefont {A.}~\bibnamefont
  {Tolish}}\ and\ \bibinfo {author} {\bibfnamefont {R.~M.}\ \bibnamefont
  {Wald}},\ }\bibfield  {title} {\bibinfo {title} {{Cosmological memory
  effect}},\ }\href {https://doi.org/10.1103/PhysRevD.94.044009} {\bibfield
  {journal} {\bibinfo  {journal} {Phys. Rev. D}\ }\textbf {\bibinfo {volume}
  {94}},\ \bibinfo {pages} {044009} (\bibinfo {year} {2016})},\ \Eprint
  {https://arxiv.org/abs/1606.04894} {arXiv:1606.04894 [gr-qc]} \BibitemShut
  {NoStop}%
\bibitem [{\citenamefont {Bieri}\ \emph {et~al.}(2017)\citenamefont {Bieri},
  \citenamefont {Garfinkle},\ and\ \citenamefont {Yunes}}]{Bieri:2017vni}%
  \BibitemOpen
  \bibfield  {author} {\bibinfo {author} {\bibfnamefont {L.}~\bibnamefont
  {Bieri}}, \bibinfo {author} {\bibfnamefont {D.}~\bibnamefont {Garfinkle}},\
  and\ \bibinfo {author} {\bibfnamefont {N.}~\bibnamefont {Yunes}},\ }\bibfield
   {title} {\bibinfo {title} {{Gravitational wave memory in $\Lambda$CDM
  cosmology}},\ }\href {https://doi.org/10.1088/1361-6382/aa8b52} {\bibfield
  {journal} {\bibinfo  {journal} {Class. Quant. Grav.}\ }\textbf {\bibinfo
  {volume} {34}},\ \bibinfo {pages} {215002} (\bibinfo {year} {2017})},\
  \Eprint {https://arxiv.org/abs/1706.02009} {arXiv:1706.02009 [gr-qc]}
  \BibitemShut {NoStop}%
\bibitem [{\citenamefont {Jokela}\ \emph {et~al.}(2023)\citenamefont {Jokela},
  \citenamefont {Kajantie},\ and\ \citenamefont {Sarkkinen}}]{Jokela:2023suz}%
  \BibitemOpen
  \bibfield  {author} {\bibinfo {author} {\bibfnamefont {N.}~\bibnamefont
  {Jokela}}, \bibinfo {author} {\bibfnamefont {K.}~\bibnamefont {Kajantie}},\
  and\ \bibinfo {author} {\bibfnamefont {M.}~\bibnamefont {Sarkkinen}},\
  }\bibfield  {title} {\bibinfo {title} {{Gravitational wave memory in
  conformally flat spacetimes}},\ }\href@noop {} {\  (\bibinfo {year}
  {2023})},\ \Eprint {https://arxiv.org/abs/2301.07680} {arXiv:2301.07680
  [gr-qc]} \BibitemShut {NoStop}%
\bibitem [{\citenamefont {Stein}(2019)}]{Stein:2019mop}%
  \BibitemOpen
  \bibfield  {author} {\bibinfo {author} {\bibfnamefont {L.~C.}\ \bibnamefont
  {Stein}},\ }\bibfield  {title} {\bibinfo {title} {{qnm: A Python package for
  calculating Kerr quasinormal modes, separation constants, and
  spherical-spheroidal mixing coefficients}},\ }\href
  {https://doi.org/10.21105/joss.01683} {\bibfield  {journal} {\bibinfo
  {journal} {J. Open Source Softw.}\ }\textbf {\bibinfo {volume} {4}},\
  \bibinfo {pages} {1683} (\bibinfo {year} {2019})},\ \Eprint
  {https://arxiv.org/abs/1908.10377} {arXiv:1908.10377 [gr-qc]} \BibitemShut
  {NoStop}%
\bibitem [{\citenamefont {Varma}\ \emph {et~al.}(2018)\citenamefont {Varma},
  \citenamefont {Stein},\ and\ \citenamefont {Gerosa}}]{vijay}%
  \BibitemOpen
  \bibfield  {author} {\bibinfo {author} {\bibfnamefont {V.}~\bibnamefont
  {Varma}}, \bibinfo {author} {\bibfnamefont {L.~C.}\ \bibnamefont {Stein}},\
  and\ \bibinfo {author} {\bibfnamefont {D.}~\bibnamefont {Gerosa}},\
  }\bibfield  {title} {\bibinfo {title} {vijayvarma392/surfinbh: Surrogate
  final bh properties}\ }\href {https://doi.org/10.5281/zenodo.1435832}
  {10.5281/zenodo.1435832} (\bibinfo {year} {2018})\BibitemShut {NoStop}%
\bibitem [{\citenamefont {Varma}\ \emph
  {et~al.}(2019{\natexlab{b}})\citenamefont {Varma}, \citenamefont {Gerosa},
  \citenamefont {Stein}, \citenamefont {H\'ebert},\ and\ \citenamefont
  {Zhang}}]{Varma:2018aht}%
  \BibitemOpen
  \bibfield  {author} {\bibinfo {author} {\bibfnamefont {V.}~\bibnamefont
  {Varma}}, \bibinfo {author} {\bibfnamefont {D.}~\bibnamefont {Gerosa}},
  \bibinfo {author} {\bibfnamefont {L.~C.}\ \bibnamefont {Stein}}, \bibinfo
  {author} {\bibfnamefont {F.}~\bibnamefont {H\'ebert}},\ and\ \bibinfo
  {author} {\bibfnamefont {H.}~\bibnamefont {Zhang}},\ }\bibfield  {title}
  {\bibinfo {title} {{High-accuracy mass, spin, and recoil predictions of
  generic black-hole merger remnants}},\ }\href
  {https://doi.org/10.1103/PhysRevLett.122.011101} {\bibfield  {journal}
  {\bibinfo  {journal} {Phys. Rev. Lett.}\ }\textbf {\bibinfo {volume} {122}},\
  \bibinfo {pages} {011101} (\bibinfo {year} {2019}{\natexlab{b}})},\ \Eprint
  {https://arxiv.org/abs/1809.09125} {arXiv:1809.09125 [gr-qc]} \BibitemShut
  {NoStop}%
\bibitem [{\citenamefont {Buonanno}\ \emph {et~al.}(2007)\citenamefont
  {Buonanno}, \citenamefont {Cook},\ and\ \citenamefont
  {Pretorius}}]{Buonanno:2006ui}%
  \BibitemOpen
  \bibfield  {author} {\bibinfo {author} {\bibfnamefont {A.}~\bibnamefont
  {Buonanno}}, \bibinfo {author} {\bibfnamefont {G.~B.}\ \bibnamefont {Cook}},\
  and\ \bibinfo {author} {\bibfnamefont {F.}~\bibnamefont {Pretorius}},\
  }\bibfield  {title} {\bibinfo {title} {{Inspiral, merger and ring-down of
  equal-mass black-hole binaries}},\ }\href
  {https://doi.org/10.1103/PhysRevD.75.124018} {\bibfield  {journal} {\bibinfo
  {journal} {Phys. Rev. D}\ }\textbf {\bibinfo {volume} {75}},\ \bibinfo
  {pages} {124018} (\bibinfo {year} {2007})},\ \Eprint
  {https://arxiv.org/abs/gr-qc/0610122} {arXiv:gr-qc/0610122} \BibitemShut
  {NoStop}%
\bibitem [{\citenamefont {Berti}\ \emph {et~al.}(2006)\citenamefont {Berti},
  \citenamefont {Cardoso},\ and\ \citenamefont {Will}}]{Berti:2005ys}%
  \BibitemOpen
  \bibfield  {author} {\bibinfo {author} {\bibfnamefont {E.}~\bibnamefont
  {Berti}}, \bibinfo {author} {\bibfnamefont {V.}~\bibnamefont {Cardoso}},\
  and\ \bibinfo {author} {\bibfnamefont {C.~M.}\ \bibnamefont {Will}},\
  }\bibfield  {title} {\bibinfo {title} {{On gravitational-wave spectroscopy of
  massive black holes with the space interferometer LISA}},\ }\href
  {https://doi.org/10.1103/PhysRevD.73.064030} {\bibfield  {journal} {\bibinfo
  {journal} {Phys. Rev. D}\ }\textbf {\bibinfo {volume} {73}},\ \bibinfo
  {pages} {064030} (\bibinfo {year} {2006})},\ \Eprint
  {https://arxiv.org/abs/gr-qc/0512160} {arXiv:gr-qc/0512160} \BibitemShut
  {NoStop}%
\bibitem [{\citenamefont {Reisswig}\ \emph {et~al.}(2009)\citenamefont
  {Reisswig}, \citenamefont {Husa}, \citenamefont {Rezzolla}, \citenamefont
  {Dorband}, \citenamefont {Pollney},\ and\ \citenamefont
  {Seiler}}]{Reisswig:2009vc}%
  \BibitemOpen
  \bibfield  {author} {\bibinfo {author} {\bibfnamefont {C.}~\bibnamefont
  {Reisswig}}, \bibinfo {author} {\bibfnamefont {S.}~\bibnamefont {Husa}},
  \bibinfo {author} {\bibfnamefont {L.}~\bibnamefont {Rezzolla}}, \bibinfo
  {author} {\bibfnamefont {E.~N.}\ \bibnamefont {Dorband}}, \bibinfo {author}
  {\bibfnamefont {D.}~\bibnamefont {Pollney}},\ and\ \bibinfo {author}
  {\bibfnamefont {J.}~\bibnamefont {Seiler}},\ }\bibfield  {title} {\bibinfo
  {title} {{Gravitational-wave detectability of equal-mass black-hole binaries
  with aligned spins}},\ }\href {https://doi.org/10.1103/PhysRevD.80.124026}
  {\bibfield  {journal} {\bibinfo  {journal} {Phys. Rev. D}\ }\textbf {\bibinfo
  {volume} {80}},\ \bibinfo {pages} {124026} (\bibinfo {year} {2009})},\
  \Eprint {https://arxiv.org/abs/0907.0462} {arXiv:0907.0462 [gr-qc]}
  \BibitemShut {NoStop}%
\bibitem [{\citenamefont {{Liu}}\ \emph {et~al.}(2021)\citenamefont {{Liu}},
  \citenamefont {{He}},\ and\ \citenamefont {{Cao}}}]{2021PhRvD.103d3005L}%
  \BibitemOpen
  \bibfield  {author} {\bibinfo {author} {\bibfnamefont {X.}~\bibnamefont
  {{Liu}}}, \bibinfo {author} {\bibfnamefont {X.}~\bibnamefont {{He}}},\ and\
  \bibinfo {author} {\bibfnamefont {Z.}~\bibnamefont {{Cao}}},\ }\bibfield
  {title} {\bibinfo {title} {{Accurate calculation of gravitational wave
  memory}},\ }\href {https://doi.org/10.1103/PhysRevD.103.043005} {\bibfield
  {journal} {\bibinfo  {journal} {\prd}\ }\textbf {\bibinfo {volume} {103}},\
  \bibinfo {eid} {043005} (\bibinfo {year} {2021})}\BibitemShut {NoStop}%
\bibitem [{\citenamefont {Finn}(1992)}]{Finn:1992wt}%
  \BibitemOpen
  \bibfield  {author} {\bibinfo {author} {\bibfnamefont {L.~S.}\ \bibnamefont
  {Finn}},\ }\bibfield  {title} {\bibinfo {title} {{Detection, measurement and
  gravitational radiation}},\ }\href {https://doi.org/10.1103/PhysRevD.46.5236}
  {\bibfield  {journal} {\bibinfo  {journal} {Phys. Rev. D}\ }\textbf {\bibinfo
  {volume} {46}},\ \bibinfo {pages} {5236} (\bibinfo {year} {1992})},\ \Eprint
  {https://arxiv.org/abs/gr-qc/9209010} {arXiv:gr-qc/9209010} \BibitemShut
  {NoStop}%
\bibitem [{\citenamefont {Poisson}\ and\ \citenamefont
  {Will}(1995)}]{Poisson:1995ef}%
  \BibitemOpen
  \bibfield  {author} {\bibinfo {author} {\bibfnamefont {E.}~\bibnamefont
  {Poisson}}\ and\ \bibinfo {author} {\bibfnamefont {C.~M.}\ \bibnamefont
  {Will}},\ }\bibfield  {title} {\bibinfo {title} {{Gravitational waves from
  inspiraling compact binaries: Parameter estimation using second postNewtonian
  wave forms}},\ }\href {https://doi.org/10.1103/PhysRevD.52.848} {\bibfield
  {journal} {\bibinfo  {journal} {Phys. Rev. D}\ }\textbf {\bibinfo {volume}
  {52}},\ \bibinfo {pages} {848} (\bibinfo {year} {1995})},\ \Eprint
  {https://arxiv.org/abs/gr-qc/9502040} {arXiv:gr-qc/9502040} \BibitemShut
  {NoStop}%
\bibitem [{\citenamefont {Cutler}(1998)}]{Cutler:1997ta}%
  \BibitemOpen
  \bibfield  {author} {\bibinfo {author} {\bibfnamefont {C.}~\bibnamefont
  {Cutler}},\ }\bibfield  {title} {\bibinfo {title} {{Angular resolution of the
  LISA gravitational wave detector}},\ }\href
  {https://doi.org/10.1103/PhysRevD.57.7089} {\bibfield  {journal} {\bibinfo
  {journal} {Phys. Rev. D}\ }\textbf {\bibinfo {volume} {57}},\ \bibinfo
  {pages} {7089} (\bibinfo {year} {1998})},\ \Eprint
  {https://arxiv.org/abs/gr-qc/9703068} {arXiv:gr-qc/9703068} \BibitemShut
  {NoStop}%
\bibitem [{\citenamefont {Vallisneri}(2008)}]{Vallisneri:2007ev}%
  \BibitemOpen
  \bibfield  {author} {\bibinfo {author} {\bibfnamefont {M.}~\bibnamefont
  {Vallisneri}},\ }\bibfield  {title} {\bibinfo {title} {{Use and abuse of the
  Fisher information matrix in the assessment of gravitational-wave
  parameter-estimation prospects}},\ }\href
  {https://doi.org/10.1103/PhysRevD.77.042001} {\bibfield  {journal} {\bibinfo
  {journal} {Phys. Rev. D}\ }\textbf {\bibinfo {volume} {77}},\ \bibinfo
  {pages} {042001} (\bibinfo {year} {2008})},\ \Eprint
  {https://arxiv.org/abs/gr-qc/0703086} {arXiv:gr-qc/0703086} \BibitemShut
  {NoStop}%
\bibitem [{\citenamefont {Berti}\ \emph {et~al.}(2005)\citenamefont {Berti},
  \citenamefont {Buonanno},\ and\ \citenamefont {Will}}]{Berti:2004bd}%
  \BibitemOpen
  \bibfield  {author} {\bibinfo {author} {\bibfnamefont {E.}~\bibnamefont
  {Berti}}, \bibinfo {author} {\bibfnamefont {A.}~\bibnamefont {Buonanno}},\
  and\ \bibinfo {author} {\bibfnamefont {C.~M.}\ \bibnamefont {Will}},\
  }\bibfield  {title} {\bibinfo {title} {{Estimating spinning binary parameters
  and testing alternative theories of gravity with LISA}},\ }\href
  {https://doi.org/10.1103/PhysRevD.71.084025} {\bibfield  {journal} {\bibinfo
  {journal} {Phys. Rev. D}\ }\textbf {\bibinfo {volume} {71}},\ \bibinfo
  {pages} {084025} (\bibinfo {year} {2005})},\ \Eprint
  {https://arxiv.org/abs/gr-qc/0411129} {arXiv:gr-qc/0411129} \BibitemShut
  {NoStop}%
\bibitem [{\citenamefont {Robson}\ \emph {et~al.}(2019)\citenamefont {Robson},
  \citenamefont {Cornish},\ and\ \citenamefont {Liu}}]{Robson:2018ifk}%
  \BibitemOpen
  \bibfield  {author} {\bibinfo {author} {\bibfnamefont {T.}~\bibnamefont
  {Robson}}, \bibinfo {author} {\bibfnamefont {N.~J.}\ \bibnamefont
  {Cornish}},\ and\ \bibinfo {author} {\bibfnamefont {C.}~\bibnamefont {Liu}},\
  }\bibfield  {title} {\bibinfo {title} {{The construction and use of LISA
  sensitivity curves}},\ }\href {https://doi.org/10.1088/1361-6382/ab1101}
  {\bibfield  {journal} {\bibinfo  {journal} {Class. Quant. Grav.}\ }\textbf
  {\bibinfo {volume} {36}},\ \bibinfo {pages} {105011} (\bibinfo {year}
  {2019})},\ \Eprint {https://arxiv.org/abs/1803.01944} {arXiv:1803.01944
  [astro-ph.HE]} \BibitemShut {NoStop}%
\bibitem [{\citenamefont {Harris}\ \emph {et~al.}(2020)\citenamefont {Harris}
  \emph {et~al.}}]{Harris:2020xlr}%
  \BibitemOpen
  \bibfield  {author} {\bibinfo {author} {\bibfnamefont {C.~R.}\ \bibnamefont
  {Harris}} \emph {et~al.},\ }\bibfield  {title} {\bibinfo {title} {{Array
  programming with NumPy}},\ }\href {https://doi.org/10.1038/s41586-020-2649-2}
  {\bibfield  {journal} {\bibinfo  {journal} {Nature}\ }\textbf {\bibinfo
  {volume} {585}},\ \bibinfo {pages} {357} (\bibinfo {year} {2020})},\ \Eprint
  {https://arxiv.org/abs/2006.10256} {arXiv:2006.10256 [cs.MS]} \BibitemShut
  {NoStop}%
\bibitem [{\citenamefont {Hirata}\ \emph {et~al.}(2010)\citenamefont {Hirata},
  \citenamefont {Holz},\ and\ \citenamefont {Cutler}}]{Hirata:2010ba}%
  \BibitemOpen
  \bibfield  {author} {\bibinfo {author} {\bibfnamefont {C.~M.}\ \bibnamefont
  {Hirata}}, \bibinfo {author} {\bibfnamefont {D.~E.}\ \bibnamefont {Holz}},\
  and\ \bibinfo {author} {\bibfnamefont {C.}~\bibnamefont {Cutler}},\
  }\bibfield  {title} {\bibinfo {title} {{Reducing the weak lensing noise for
  the gravitational wave Hubble diagram using the non-Gaussianity of the
  magnification distribution}},\ }\href
  {https://doi.org/10.1103/PhysRevD.81.124046} {\bibfield  {journal} {\bibinfo
  {journal} {Phys. Rev. D}\ }\textbf {\bibinfo {volume} {81}},\ \bibinfo
  {pages} {124046} (\bibinfo {year} {2010})},\ \Eprint
  {https://arxiv.org/abs/1004.3988} {arXiv:1004.3988 [astro-ph.CO]}
  \BibitemShut {NoStop}%
\bibitem [{\citenamefont {Calder\'on~Bustillo}\ \emph
  {et~al.}(2016)\citenamefont {Calder\'on~Bustillo}, \citenamefont {Husa},
  \citenamefont {Sintes},\ and\ \citenamefont
  {P\"urrer}}]{CalderonBustillo:2015lrt}%
  \BibitemOpen
  \bibfield  {author} {\bibinfo {author} {\bibfnamefont {J.}~\bibnamefont
  {Calder\'on~Bustillo}}, \bibinfo {author} {\bibfnamefont {S.}~\bibnamefont
  {Husa}}, \bibinfo {author} {\bibfnamefont {A.~M.}\ \bibnamefont {Sintes}},\
  and\ \bibinfo {author} {\bibfnamefont {M.}~\bibnamefont {P\"urrer}},\
  }\bibfield  {title} {\bibinfo {title} {{Impact of gravitational radiation
  higher order modes on single aligned-spin gravitational wave searches for
  binary black holes}},\ }\href {https://doi.org/10.1103/PhysRevD.93.084019}
  {\bibfield  {journal} {\bibinfo  {journal} {Phys. Rev. D}\ }\textbf {\bibinfo
  {volume} {93}},\ \bibinfo {pages} {084019} (\bibinfo {year} {2016})},\
  \Eprint {https://arxiv.org/abs/1511.02060} {arXiv:1511.02060 [gr-qc]}
  \BibitemShut {NoStop}%
\bibitem [{\citenamefont {Marsat}\ \emph {et~al.}(2021)\citenamefont {Marsat},
  \citenamefont {Baker},\ and\ \citenamefont {Dal~Canton}}]{Marsat:2020rtl}%
  \BibitemOpen
  \bibfield  {author} {\bibinfo {author} {\bibfnamefont {S.}~\bibnamefont
  {Marsat}}, \bibinfo {author} {\bibfnamefont {J.~G.}\ \bibnamefont {Baker}},\
  and\ \bibinfo {author} {\bibfnamefont {T.}~\bibnamefont {Dal~Canton}},\
  }\bibfield  {title} {\bibinfo {title} {{Exploring the Bayesian parameter
  estimation of binary black holes with LISA}},\ }\href
  {https://doi.org/10.1103/PhysRevD.103.083011} {\bibfield  {journal} {\bibinfo
   {journal} {Phys. Rev. D}\ }\textbf {\bibinfo {volume} {103}},\ \bibinfo
  {pages} {083011} (\bibinfo {year} {2021})},\ \Eprint
  {https://arxiv.org/abs/2003.00357} {arXiv:2003.00357 [gr-qc]} \BibitemShut
  {NoStop}%
\bibitem [{\citenamefont {Graff}\ \emph {et~al.}(2015)\citenamefont {Graff},
  \citenamefont {Buonanno},\ and\ \citenamefont
  {Sathyaprakash}}]{Graff:2015bba}%
  \BibitemOpen
  \bibfield  {author} {\bibinfo {author} {\bibfnamefont {P.~B.}\ \bibnamefont
  {Graff}}, \bibinfo {author} {\bibfnamefont {A.}~\bibnamefont {Buonanno}},\
  and\ \bibinfo {author} {\bibfnamefont {B.~S.}\ \bibnamefont
  {Sathyaprakash}},\ }\bibfield  {title} {\bibinfo {title} {{Missing Link:
  Bayesian detection and measurement of intermediate-mass black-hole
  binaries}},\ }\href {https://doi.org/10.1103/PhysRevD.92.022002} {\bibfield
  {journal} {\bibinfo  {journal} {Phys. Rev. D}\ }\textbf {\bibinfo {volume}
  {92}},\ \bibinfo {pages} {022002} (\bibinfo {year} {2015})},\ \Eprint
  {https://arxiv.org/abs/1504.04766} {arXiv:1504.04766 [gr-qc]} \BibitemShut
  {NoStop}%
\bibitem [{\citenamefont {Payne}\ \emph {et~al.}(2019)\citenamefont {Payne},
  \citenamefont {Talbot},\ and\ \citenamefont {Thrane}}]{Payne:2019wmy}%
  \BibitemOpen
  \bibfield  {author} {\bibinfo {author} {\bibfnamefont {E.}~\bibnamefont
  {Payne}}, \bibinfo {author} {\bibfnamefont {C.}~\bibnamefont {Talbot}},\ and\
  \bibinfo {author} {\bibfnamefont {E.}~\bibnamefont {Thrane}},\ }\bibfield
  {title} {\bibinfo {title} {{Higher order gravitational-wave modes with
  likelihood reweighting}},\ }\href
  {https://doi.org/10.1103/PhysRevD.100.123017} {\bibfield  {journal} {\bibinfo
   {journal} {Phys. Rev. D}\ }\textbf {\bibinfo {volume} {100}},\ \bibinfo
  {pages} {123017} (\bibinfo {year} {2019})},\ \Eprint
  {https://arxiv.org/abs/1905.05477} {arXiv:1905.05477 [astro-ph.IM]}
  \BibitemShut {NoStop}%
\bibitem [{\citenamefont {Varma}\ \emph {et~al.}(2014)\citenamefont {Varma},
  \citenamefont {Ajith}, \citenamefont {Husa}, \citenamefont {Bustillo},
  \citenamefont {Hannam},\ and\ \citenamefont {P\"urrer}}]{Varma:2014jxa}%
  \BibitemOpen
  \bibfield  {author} {\bibinfo {author} {\bibfnamefont {V.}~\bibnamefont
  {Varma}}, \bibinfo {author} {\bibfnamefont {P.}~\bibnamefont {Ajith}},
  \bibinfo {author} {\bibfnamefont {S.}~\bibnamefont {Husa}}, \bibinfo {author}
  {\bibfnamefont {J.~C.}\ \bibnamefont {Bustillo}}, \bibinfo {author}
  {\bibfnamefont {M.}~\bibnamefont {Hannam}},\ and\ \bibinfo {author}
  {\bibfnamefont {M.}~\bibnamefont {P\"urrer}},\ }\bibfield  {title} {\bibinfo
  {title} {{Gravitational-wave observations of binary black holes: Effect of
  nonquadrupole modes}},\ }\href {https://doi.org/10.1103/PhysRevD.90.124004}
  {\bibfield  {journal} {\bibinfo  {journal} {Phys. Rev. D}\ }\textbf {\bibinfo
  {volume} {90}},\ \bibinfo {pages} {124004} (\bibinfo {year} {2014})},\
  \Eprint {https://arxiv.org/abs/1409.2349} {arXiv:1409.2349 [gr-qc]}
  \BibitemShut {NoStop}%
\bibitem [{\citenamefont {Abbott}\ \emph {et~al.}(2020)\citenamefont {Abbott}
  \emph {et~al.}}]{LIGOScientific:2020stg}%
  \BibitemOpen
  \bibfield  {author} {\bibinfo {author} {\bibfnamefont {R.}~\bibnamefont
  {Abbott}} \emph {et~al.} (\bibinfo {collaboration} {LIGO Scientific,
  Virgo}),\ }\bibfield  {title} {\bibinfo {title} {{GW190412: Observation of a
  Binary-Black-Hole Coalescence with Asymmetric Masses}},\ }\href
  {https://doi.org/10.1103/PhysRevD.102.043015} {\bibfield  {journal} {\bibinfo
   {journal} {Phys. Rev. D}\ }\textbf {\bibinfo {volume} {102}},\ \bibinfo
  {pages} {043015} (\bibinfo {year} {2020})},\ \Eprint
  {https://arxiv.org/abs/2004.08342} {arXiv:2004.08342 [astro-ph.HE]}
  \BibitemShut {NoStop}%
\bibitem [{\citenamefont {{London}}\ \emph {et~al.}(2018)\citenamefont
  {{London}}, \citenamefont {{Khan}}, \citenamefont {{Fauchon-Jones}},
  \citenamefont {{Garc{\'\i}a}}, \citenamefont {{Hannam}}, \citenamefont
  {{Husa}}, \citenamefont {{Jim{\'e}nez-Forteza}}, \citenamefont
  {{Kalaghatgi}}, \citenamefont {{Ohme}},\ and\ \citenamefont
  {{Pannarale}}}]{2018PhRvL.120p1102L}%
  \BibitemOpen
  \bibfield  {author} {\bibinfo {author} {\bibfnamefont {L.}~\bibnamefont
  {{London}}}, \bibinfo {author} {\bibfnamefont {S.}~\bibnamefont {{Khan}}},
  \bibinfo {author} {\bibfnamefont {E.}~\bibnamefont {{Fauchon-Jones}}},
  \bibinfo {author} {\bibfnamefont {C.}~\bibnamefont {{Garc{\'\i}a}}}, \bibinfo
  {author} {\bibfnamefont {M.}~\bibnamefont {{Hannam}}}, \bibinfo {author}
  {\bibfnamefont {S.}~\bibnamefont {{Husa}}}, \bibinfo {author} {\bibfnamefont
  {X.}~\bibnamefont {{Jim{\'e}nez-Forteza}}}, \bibinfo {author} {\bibfnamefont
  {C.}~\bibnamefont {{Kalaghatgi}}}, \bibinfo {author} {\bibfnamefont
  {F.}~\bibnamefont {{Ohme}}},\ and\ \bibinfo {author} {\bibfnamefont
  {F.}~\bibnamefont {{Pannarale}}},\ }\bibfield  {title} {\bibinfo {title}
  {{First Higher-Multipole Model of Gravitational Waves from Spinning and
  Coalescing Black-Hole Binaries}},\ }\href
  {https://doi.org/10.1103/PhysRevLett.120.161102} {\bibfield  {journal}
  {\bibinfo  {journal} {\prl}\ }\textbf {\bibinfo {volume} {120}},\ \bibinfo
  {eid} {161102} (\bibinfo {year} {2018})},\ \Eprint
  {https://arxiv.org/abs/1708.00404} {arXiv:1708.00404 [gr-qc]} \BibitemShut
  {NoStop}%
\bibitem [{\citenamefont {Hu}\ and\ \citenamefont {Veitch}(2022)}]{Hu:2022rjq}%
  \BibitemOpen
  \bibfield  {author} {\bibinfo {author} {\bibfnamefont {Q.}~\bibnamefont
  {Hu}}\ and\ \bibinfo {author} {\bibfnamefont {J.}~\bibnamefont {Veitch}},\
  }\bibfield  {title} {\bibinfo {title} {{Assessing the model waveform accuracy
  of gravitational waves}},\ }\href
  {https://doi.org/10.1103/PhysRevD.106.044042} {\bibfield  {journal} {\bibinfo
   {journal} {Phys. Rev. D}\ }\textbf {\bibinfo {volume} {106}},\ \bibinfo
  {pages} {044042} (\bibinfo {year} {2022})},\ \Eprint
  {https://arxiv.org/abs/2205.08448} {arXiv:2205.08448 [gr-qc]} \BibitemShut
  {NoStop}%
\bibitem [{\citenamefont {Lindblom}\ \emph {et~al.}(2008)\citenamefont
  {Lindblom}, \citenamefont {Owen},\ and\ \citenamefont
  {Brown}}]{Lindblom:2008cm}%
  \BibitemOpen
  \bibfield  {author} {\bibinfo {author} {\bibfnamefont {L.}~\bibnamefont
  {Lindblom}}, \bibinfo {author} {\bibfnamefont {B.~J.}\ \bibnamefont {Owen}},\
  and\ \bibinfo {author} {\bibfnamefont {D.~A.}\ \bibnamefont {Brown}},\
  }\bibfield  {title} {\bibinfo {title} {{Model Waveform Accuracy Standards for
  Gravitational Wave Data Analysis}},\ }\href
  {https://doi.org/10.1103/PhysRevD.78.124020} {\bibfield  {journal} {\bibinfo
  {journal} {Phys. Rev. D}\ }\textbf {\bibinfo {volume} {78}},\ \bibinfo
  {pages} {124020} (\bibinfo {year} {2008})},\ \Eprint
  {https://arxiv.org/abs/0809.3844} {arXiv:0809.3844 [gr-qc]} \BibitemShut
  {NoStop}%
\bibitem [{\citenamefont {Klein}\ \emph {et~al.}(2016)\citenamefont {Klein}
  \emph {et~al.}}]{Klein:2015hvg}%
  \BibitemOpen
  \bibfield  {author} {\bibinfo {author} {\bibfnamefont {A.}~\bibnamefont
  {Klein}} \emph {et~al.},\ }\bibfield  {title} {\bibinfo {title} {{Science
  with the space-based interferometer eLISA: Supermassive black hole
  binaries}},\ }\href {https://doi.org/10.1103/PhysRevD.93.024003} {\bibfield
  {journal} {\bibinfo  {journal} {Phys. Rev. D}\ }\textbf {\bibinfo {volume}
  {93}},\ \bibinfo {pages} {024003} (\bibinfo {year} {2016})},\ \Eprint
  {https://arxiv.org/abs/1511.05581} {arXiv:1511.05581 [gr-qc]} \BibitemShut
  {NoStop}%
\bibitem [{\citenamefont {Maggiore}\ \emph {et~al.}(2020)\citenamefont
  {Maggiore} \emph {et~al.}}]{Maggiore:2019uih}%
  \BibitemOpen
  \bibfield  {author} {\bibinfo {author} {\bibfnamefont {M.}~\bibnamefont
  {Maggiore}} \emph {et~al.},\ }\bibfield  {title} {\bibinfo {title} {{Science
  Case for the Einstein Telescope}},\ }\href
  {https://doi.org/10.1088/1475-7516/2020/03/050} {\bibfield  {journal}
  {\bibinfo  {journal} {JCAP}\ }\textbf {\bibinfo {volume} {03}},\ \bibinfo
  {pages} {050}},\ \Eprint {https://arxiv.org/abs/1912.02622} {arXiv:1912.02622
  [astro-ph.CO]} \BibitemShut {NoStop}%
\bibitem [{\citenamefont {Hild}\ \emph {et~al.}(2011)\citenamefont {Hild} \emph
  {et~al.}}]{Hild:2010id}%
  \BibitemOpen
  \bibfield  {author} {\bibinfo {author} {\bibfnamefont {S.}~\bibnamefont
  {Hild}} \emph {et~al.},\ }\bibfield  {title} {\bibinfo {title} {{Sensitivity
  Studies for Third-Generation Gravitational Wave Observatories}},\ }\href
  {https://doi.org/10.1088/0264-9381/28/9/094013} {\bibfield  {journal}
  {\bibinfo  {journal} {Class. Quant. Grav.}\ }\textbf {\bibinfo {volume}
  {28}},\ \bibinfo {pages} {094013} (\bibinfo {year} {2011})},\ \Eprint
  {https://arxiv.org/abs/1012.0908} {arXiv:1012.0908 [gr-qc]} \BibitemShut
  {NoStop}%
\bibitem [{\citenamefont {Littenberg}\ and\ \citenamefont
  {Cornish}(2023)}]{Littenberg:2023xpl}%
  \BibitemOpen
  \bibfield  {author} {\bibinfo {author} {\bibfnamefont {T.~B.}\ \bibnamefont
  {Littenberg}}\ and\ \bibinfo {author} {\bibfnamefont {N.~J.}\ \bibnamefont
  {Cornish}},\ }\bibfield  {title} {\bibinfo {title} {{Prototype Global
  Analysis of LISA Data with Multiple Source Types}},\ }\href@noop {} {\
  (\bibinfo {year} {2023})},\ \Eprint {https://arxiv.org/abs/2301.03673}
  {arXiv:2301.03673 [gr-qc]} \BibitemShut {NoStop}%
\bibitem [{\citenamefont {Palmese}\ \emph {et~al.}(2020)\citenamefont {Palmese}
  \emph {et~al.}}]{Palmese_2020}%
  \BibitemOpen
  \bibfield  {author} {\bibinfo {author} {\bibfnamefont {A.}~\bibnamefont
  {Palmese}} \emph {et~al.},\ }\bibfield  {title} {\bibinfo {title} {A
  statistical standard siren measurement of the hubble constant from the
  {LIGO}/virgo gravitational wave compact object merger {GW}190814 and dark
  energy survey galaxies},\ }\href {https://doi.org/10.3847/2041-8213/abaeff}
  {\bibfield  {journal} {\bibinfo  {journal} {The Astrophysical Journal}\
  }\textbf {\bibinfo {volume} {900}},\ \bibinfo {pages} {L33} (\bibinfo {year}
  {2020})}\BibitemShut {NoStop}%
\bibitem [{\citenamefont {Hunter}(2007)}]{4160265}%
  \BibitemOpen
  \bibfield  {author} {\bibinfo {author} {\bibfnamefont {J.~D.}\ \bibnamefont
  {Hunter}},\ }\bibfield  {title} {\bibinfo {title} {Matplotlib: A 2d graphics
  environment},\ }\href {https://doi.org/10.1109/MCSE.2007.55} {\bibfield
  {journal} {\bibinfo  {journal} {Computing in Science \& Engineering}\
  }\textbf {\bibinfo {volume} {9}},\ \bibinfo {pages} {90} (\bibinfo {year}
  {2007})}\BibitemShut {NoStop}%
\bibitem [{\citenamefont {Virtanen}\ \emph {et~al.}(2020)\citenamefont
  {Virtanen} \emph {et~al.}}]{2020SciPy-NMeth}%
  \BibitemOpen
  \bibfield  {author} {\bibinfo {author} {\bibfnamefont {P.}~\bibnamefont
  {Virtanen}} \emph {et~al.},\ }\bibfield  {title} {\bibinfo {title} {{{SciPy}
  1.0: Fundamental Algorithms for Scientific Computing in Python}},\ }\href
  {https://doi.org/10.1038/s41592-019-0686-2} {\bibfield  {journal} {\bibinfo
  {journal} {Nature Methods}\ }\textbf {\bibinfo {volume} {17}},\ \bibinfo
  {pages} {261} (\bibinfo {year} {2020})}\BibitemShut {NoStop}%
\bibitem [{\citenamefont {Field}\ \emph {et~al.}(2014)\citenamefont {Field},
  \citenamefont {Galley}, \citenamefont {Hesthaven}, \citenamefont {Kaye},\
  and\ \citenamefont {Tiglio}}]{Field:2013cfa}%
  \BibitemOpen
  \bibfield  {author} {\bibinfo {author} {\bibfnamefont {S.~E.}\ \bibnamefont
  {Field}}, \bibinfo {author} {\bibfnamefont {C.~R.}\ \bibnamefont {Galley}},
  \bibinfo {author} {\bibfnamefont {J.~S.}\ \bibnamefont {Hesthaven}}, \bibinfo
  {author} {\bibfnamefont {J.}~\bibnamefont {Kaye}},\ and\ \bibinfo {author}
  {\bibfnamefont {M.}~\bibnamefont {Tiglio}},\ }\bibfield  {title} {\bibinfo
  {title} {{Fast prediction and evaluation of gravitational waveforms using
  surrogate models}},\ }\href {https://doi.org/10.1103/PhysRevX.4.031006}
  {\bibfield  {journal} {\bibinfo  {journal} {Phys. Rev. X}\ }\textbf {\bibinfo
  {volume} {4}},\ \bibinfo {pages} {031006} (\bibinfo {year} {2014})},\ \Eprint
  {https://arxiv.org/abs/1308.3565} {arXiv:1308.3565 [gr-qc]} \BibitemShut
  {NoStop}%
\bibitem [{\citenamefont {Richardson}\ \emph {et~al.}(2022)\citenamefont
  {Richardson}, \citenamefont {Zanolin}, \citenamefont {Andresen},
  \citenamefont {Szczepa\'nczyk}, \citenamefont {Gill},\ and\ \citenamefont
  {Wongwathanarat}}]{Richardson:2021lib}%
  \BibitemOpen
  \bibfield  {author} {\bibinfo {author} {\bibfnamefont {C.~J.}\ \bibnamefont
  {Richardson}}, \bibinfo {author} {\bibfnamefont {M.}~\bibnamefont {Zanolin}},
  \bibinfo {author} {\bibfnamefont {H.}~\bibnamefont {Andresen}}, \bibinfo
  {author} {\bibfnamefont {M.~J.}\ \bibnamefont {Szczepa\'nczyk}}, \bibinfo
  {author} {\bibfnamefont {K.}~\bibnamefont {Gill}},\ and\ \bibinfo {author}
  {\bibfnamefont {A.}~\bibnamefont {Wongwathanarat}},\ }\bibfield  {title}
  {\bibinfo {title} {{Modeling core-collapse supernovae gravitational-wave
  memory in laser interferometric data}},\ }\href
  {https://doi.org/10.1103/PhysRevD.105.103008} {\bibfield  {journal} {\bibinfo
   {journal} {Phys. Rev. D}\ }\textbf {\bibinfo {volume} {105}},\ \bibinfo
  {pages} {103008} (\bibinfo {year} {2022})},\ \Eprint
  {https://arxiv.org/abs/2109.01582} {arXiv:2109.01582 [astro-ph.HE]}
  \BibitemShut {NoStop}%
\end{thebibliography}%
\end{document}